\pdfoutput=1
\RequirePackage{ifpdf}
\ifpdf
\documentclass[pdftex]{sigma}
\else
\documentclass{sigma}
\fi

\numberwithin{equation}{section}

\numberwithin{theorem}{section}
\numberwithin{proposition}{section}
\numberwithin{lemma}{section}
\numberwithin{corollary}{section}
\numberwithin{conjecture}{section}
\numberwithin{definition}{section}
\numberwithin{example}{section}
\numberwithin{remark}{section}
\numberwithin{note}{section}

\newcommand{\cl}{\mathrm{cl}}

\renewcommand{\and}{{\qquad {\rm and} \qquad}}

\newcommand{\dd}{\mathrm{d}}

\newcommand{\Res}{\mathop{\,\rm Res\,}}

\newcommand{\td}[1]{{\tilde{#1}}}

\newcommand{\om}{\omega}

\newcommand{\ii}{{\mathrm{i}}}

\newcommand{\ee}[1]{{{\rm e}^{#1}}}

\newcommand{\Pint}{{\int\kern -1.em -\kern-.25em}}

\renewcommand{\Im}{{\mathrm{Im}}}

\newcommand{\curve}{{\cal C}}
\newcommand{\spcurve}{{\cal S}}

\newcommand{\acycle}{{\cal A}}
\newcommand{\bcycle}{{\cal B}}
\newcommand{\bfc}{{\mathbf c}}
\newcommand{\bfu}{{\mathbf u}}

\newcommand{\bfn}{{\mathbf n}}
\newcommand{\bfm}{{\mathbf m}}
\newcommand{\genus}{{\mathfrak g}}

\newcommand{\Tau}{{\cal T}}

\begin{document}

\allowdisplaybreaks

\renewcommand{\PaperNumber}{100}

\FirstPageHeading

\ShortArticleName{Geometry of Spectral Curves and All Order Dispersive Integrable System}

\ArticleName{Geometry of Spectral Curves\\ and All Order Dispersive Integrable System}

\Author{Ga\"{e}tan BOROT~$^\dag$ and Bertrand EYNARD~$^{\ddag\S}$}

\AuthorNameForHeading{G.~Borot and B.~Eynard}

\Address{$^\dag$~Section de Math\'ematiques, Universit\'e de Gen\`eve,\\
\hphantom{$^\dag$}~2-4 rue du Li\`evre, 1211 Gen\`eve 4, Switzerland}
\EmailD{\href{mailto:gaetan.borot@unige.ch}{gaetan.borot@unige.ch}}

\Address{$^\ddag$~Institut de Physique Th\'eorique, CEA Saclay, Orme des Merisiers,\\
\hphantom{$^\ddag$}~91191 Gif-sur-Yvette, France}
\EmailD{\href{mailto:bertrand.eynard@cea.fr}{bertrand.eynard@cea.fr}}

\Address{$^\S$~Centre de Recherche Math\'ematiques de Montr\'eal, Universit\'e de Montr\'eal,\\
\hphantom{$^\S$}~P.O.~Box 6128, Montr\'eal (Qu\'ebec) H3C 3J7, Canada}

\ArticleDates{Received November 14, 2011, in f\/inal form December 11, 2012; Published online December 18, 2012}

\Abstract{We propose a def\/inition for a Tau function and a spinor kernel (closely related to Baker--Akhiezer functions), where times parametrize slow (of order $1/N$) deformations of an~algebraic plane curve. This def\/inition consists of a~formal asymptotic series in po\-wers of~$1/N$, where the coef\/f\/icients involve theta functions whose phase is linear in $N$ and therefore features generically fast oscillations when $N$ is large.
The large $N$ limit of this construction coincides with the algebro-geometric solutions of the multi-KP equation, but where the underlying algebraic curve evolves according to Whitham equations. We check that our conjectural Tau function satisf\/ies Hirota equations to the f\/irst two orders, and we conjecture that they hold to all orders. The Hirota equations are equivalent to a self-replication property for the spinor kernel. We analyze its consequences, namely the possibility of reconstructing order by order in $1/N$ an isomonodromic problem given by a Lax pair, and the relation between ``correlators'', the tau function and the spinor kernel. This construction is one more step towards a unif\/ied framework relating integrable hierarchies, topological recursion and enumerative geometry.}

\Keywords{topological recursion; Tau function; Sato formula; Hirota equations; Whitham equations}

\Classification{14H70; 14H42; 30Fxx}

\section{Introduction}

Integrable systems are nonlinear dynamical systems, and in many cases, some exact solutions can be produced in terms of algebraic geometry of Riemann surfaces.
For instance, Liouville integrable systems can be brought into the form of a linear constant motion with constant velocity in a multidimensional torus which is the Jacobian of some algebraic curve.
However, not all solutions are algebro geometric, and an important question is how to f\/ind some solutions as perturbations of algebro-geometric ones.

\subsection{Goal and motivations}

Our goal is to propose a def\/inition of a formal series for a perturbation of an algebro-geometric solution of an integrable system, in a ``small'' parameter which we call~$1/N$. Our def\/inition consists in an all order expansion in powers of~$1/N$, whose leading order is the usual algebro-geometric solution, and whose corrections to each orders contain fast oscillating terms of frequency~$N$, constructed from the invariants of~\cite{EOFg}.

The motivation for our def\/inition, is to mimick the large $N$ expansion of random $N\times N$ matrix integrals.

Indeed, it is well known \cite{AvM2, FIK90, Mehtabook, vanM} that random matrix models are particular examples of Tau-functions of some integrable systems, and also, the formal large $N$ expansion of matrix models can be obtained by formally solving Schwinger--Dyson equations (called loop equations in the context of matrix models), which leads to the invariants of \cite{EOFg}, and to an expansion of the Tau-function in terms of them in \cite{Ecv, EMhol}.

Therefore, in this paper, we propose to use the expressions introduced in \cite{Ecv, EMhol} for matrix models, in a lager context, as candidate Tau-functions associated to an arbitrary algebraic curve.
We conjecture that the expression we propose does satisfy (formally) Hirota equation to all orders in $1/N$.
We prove it to the two f\/irst non-trivial orders.

As a motivation, we recall, that it was proved in \cite{BE09} that our proposal retrieves the asymptotic expansion for the $(p,2)$ minimal model reduction of KdV. In the most general case, the matching between our construction and the asymptotic expansion of matrix models has not yet received a proof. However, in some cases like the one-hermitian matrix model with real potentials and some extra assumptions, it has been established to all orders \cite{APS01,ErcMcL} in the one-cut regime (no modulation factors, pure $1/N$ expansion) in~\cite{APS01,BG11,ErcMcL}, and has been addressed up to $o(1)$ (including the modulation by a theta function) in multi-cut regimes in the works \cite{Marcopaths, BI99,Deiftetal, DZ}.

In all examples above appeared a triple $\mathcal{S} = (\mathcal{C},X,Y)$ consisting of a Riemann surface $\mathcal{C}$ and two functions $X$ and $Y$ def\/ined on it, or some variants, like a curve and 2 meromorphic dif\/ferentials $\dd X$ and $Y\dd X$. We call this data a \emph{spectral curve}, and it plays a central role in this article.

There exists several, non tautologically equivalent approaches to integrable systems (see \cite{BBT, RevRusInt} for reviews). In this article, we take the notion of Tau function \cite{JMII,JMIII, JM81} as a starting point. It is a function of all the times which satisf\/ies Hirota bilinear relations~\cite{Hirota}; Tau functions are in correspondence with solutions of the nonlinear integrable PDE's.

Our main goal is to propose that the formal series def\/ined in Def\/inition~\ref{defTaumunu} as a functional on the space of spectral curves, is a Tau function. We hope that it will allow the prediction of the full asymptotic expansion (in the small dispersive parameter $1/N$) of some solutions of integrable PDE's in any genus $\mathfrak{g}$ regime, although we postpone precise comparisons to future works (see~\cite{Kri92b} for perturbation theory of Tau functions).

We just mention that the leading order of our construction retrieves the well-known asymptotic solutions of KdV in the genus $1$ regime. More precisely, for such a comparison, we need to consider our construction to $1$-form $Y\dd X = x \omega_{\infty,1} + 2t \omega_{\infty,3}$ def\/ined on the elliptic curve $\Upsilon^2 = \prod\limits_{i = 1}^3 (X - a_i(x,t))$, where $a_i(x,t)$ satisfy Whitham type equations \cite{Whitham}
\begin{gather*}
\partial_t a_i = \frac{\omega_{\infty,3}(a_i)}{\omega_{\infty,1}(a_i)} \partial_x a_i,
\end{gather*}
and where~$x$ (resp.~$t$) is the time associated with the $1$-form $\omega_{\infty,1}$ (resp.\ $2\omega_{\infty,3}$). We have used here the parametrization of meromorphic $1$-forms introduced in Section~\ref{poiuy},  and $\infty$ denotes the point at inf\/inity around which $\Upsilon^2 \sim X^3$). The phase $\zeta(x,t)$ coincides with $\zeta(\mathbf{t})$ of Section~\ref{sec:skBA0} and $C$ a constant depending on the initial data. $\tau(x,t)$ is the time-evolved Riemann period of the elliptic curve, and $E_2$ the second Eisenstein series. For solutions of the KdV equation
\begin{gather}
\label{kdv}u_{t} + uu_x + \frac{1}{N^2} u_{xxx} = 0
\end{gather}
with generic initial data, it has been proved \cite{Defv, FFML80,GP74, LaxLev,LaxLev2,LaxLev3,Venakdv,Venakdv2} that $u_N(x,t)$ for some time after the gradient catastrophe, $u_N(x,t)$ is asymptotic to (in a distributional sense)
\begin{gather}
\label{leads}u_N(x,t) = \frac{2\pi^2 E_2(\tau(x,t))}{3} + \frac{1}{3}\sum_{i = 1}^3 a_i(x,t) + \frac{2}{N^2} \partial_x^2\ln\theta\big(N\zeta(x,t) + C\,|\,\tau(x,t)\big),
\end{gather}
where $\theta$ is a genus $1$ Theta function. The relation between the Tau function and the solution of equation~\eqref{kdv} is $u(x,t) = 2(\ln \mathcal{T})_{xx}$, and our candidate Tau function def\/ined in Def\/inition~\ref{defTaumunu} indeed match in this setting the leading behavior equation~\eqref{leads}.

We expect our proposal to be of interest in the study of asymptotics of solutions of hierarchies known to govern those matrix models like continuous Toda or nonlinear Schr\"odinger, in any f\/inite genus regime. We also stress that matrix models are only special cases of our construction, in other words $\mathcal{T}[\mathcal{S}]$ is not in general a matrix integral. We hope that our general construction would describe the all-order asymptotics of solutions of the full dispersive hierarchies associated to Hurwitz spaces, although we do not attempt to make the comparison and do not address the hamiltonian formalism in this article.

\subsection{Outline of the article}

After a summary of algebraic geometry in Section~\ref{sec:2}, we review in Section~\ref{sec:reconstruction} the reconstruction of an isospectral Lax system from its semiclassical spectral curve (which is time-independent). The techniques for this reconstruction are closely related to those developed by Krichever \cite{KriNLS77,Kri77} to produce the algebro-geometric solutions of the Zakharov--Shabat hierarchies \cite{Zsor}. We put emphasis on the Baker--Akhiezer spinor kernel $\psi_{\cl}(z_1,z_2)$ \cite{Kor1,Kor2}, and the corresponding Tau function $\mathcal{T}_{\cl}(\mathbf{t})$ in Section~\ref{sec:tau0}. Apart from f\/ixing notations, this review is relevant to the present work, as one can illustrate in the case of KdV
where the spectral curve does not depend on the times $x$ and $t$ (i.e.~$a_i$ and~$\tau$ are assumed constant) provides an exact solution of KdV \cite{ItsMatvb, ItsMatv} which can be obtained by such a reconstruction. Whereas, if one let $a_i(x,t)$ evolve according to Whitham equations as in the second part of the paper, it also describes the leading order of a~solution of KdV in the small dispersion limit and for some time after the gradient catastrophe for generic initial data.

Then, for any spectral curve $\mathcal{S}$ whose time evolution is described by Whitham equations \cite{Kri92, Whitham} (cf.\  Section~\ref{sec:spgeom}), we shall def\/ine explicitly in Section~\ref{sec:dis} a functional $\mathcal{T}[\mathcal{S}]$ (Def\/inition~\ref{defTaumunu}) as a formal asymptotic series in a~small parameter $N$, as well as a spinor kernel $\psi(z_1,z_2)$ via a~Sato-like formula (Def\/inition~\ref{defpsis}), which plays the role of a Baker--Akhiezer spinor kernel. We also introduce in Section~\ref{sec:co} the correlators $W_n(z_1,\ldots,z_n)$ (Def\/inition~\ref{def:co}), which encode the $n$-th order derivatives of~$\mathcal{T}[\mathcal{S}]$ with respect to deformation parameters of~$\mathcal{S}$. Here, $z_1,\ldots,z_n$ denotes points on~$\mathcal{C}$.

The essential point in this article is the conjecture that $\mathcal{T}[\mathcal{S}]$ satisf\/ies a certain form of Hirota equations to all orders in $1/N$ (Conjecture~\ref{eqhys2}), and we check it holds for the two f\/irst orders (Appendix~\ref{apphirota1surN}). We present an equivalent conjecture stating that $\psi(z_1,z_2)$ is self-replicating (Conjecture~\ref{consj}). This conjecture automatically implies determinantal formulas for the correlators (Theorem~\ref{detform}), Christof\/fel--Darboux formula for the spinor kernel (Theorem~\ref{CDS}), and a Lax system satisf\/ied by the matrix $\Psi(x_1,x_2) = [\psi(z^{i}(x_1),z^{j}(x_2))]_{i,j}$, where $z^{i}(x) \in \mathcal{C}$ are the points such that $X(z^{i}(x)) = x$ (Section~\ref{sec:9}).

The coef\/f\/icient of the so-obtained Lax matrices can be computed in principle order by order in $1/N$. If our conjecture is correct, our approach describes directly a Tau function, but we do not identify the underlying nonlinear hierarchy of equations. The situation is similar to the one evoked in~\cite{DubZh}, where the dispersive hierarchy is constructed perturbatively in $1/N$, but its resummation for f\/inite $N$ is unknown~-- except in special cases.

Since our approach was strongly motivated by earlier results or heuristics in hermitian matrix models, we recapitulate their relation to the present work in Section~\ref{sec:mat}.

\section{Geometry of the spectral curve}
\label{sec:2}

We brief\/ly describe some geometric notions attached to a f\/ixed spectral curve $\mathcal{S} = (\mathcal{C},X,Y)$ \cite{Dub81, FarkasKra,Fay}. To simplify, we assume in this article that $\mathcal{C}$ is a compact Riemann surface of genus~$\mathfrak{g}$, and~$X$ and $Y$ are meromorphic functions on~$\mathcal{C}$.

\subsection{Some notations and properties}

\subsubsection{Topology and holomorphic 1-forms}

\begin{figure}
\centering
\includegraphics[width=0.4\textwidth]{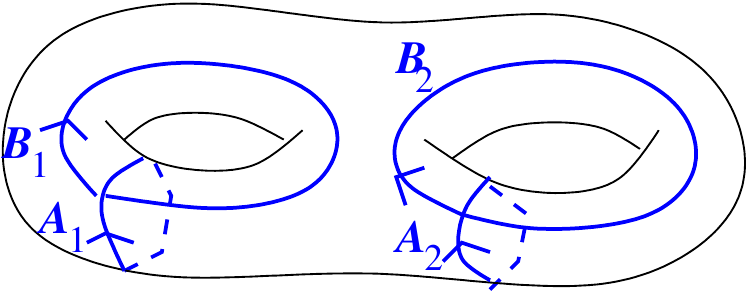}
\caption{A symplectic basis of $2\genus$ non-contractible cycles on a Riemann surface of genus $\genus$.\label{figABcycles}}
\end{figure}

The curve $\curve$ is either simply connected, and then this is the Riemann sphere $\curve = \mathbb{C}\cup\{\infty\}$, or it has genus $\genus > 0$. Then, any maximal open contractible subset of $\curve$ is called a fundamental domain. If it is of genus $\genus>0$, there exist $2\genus$ independent non-contractible cycles (see Fig.~\ref{figABcycles}), and we can choose them in such a way (but not unique) that
\begin{gather*}
\acycle_{i}\cap \bcycle_j =\delta_{i,j},\qquad
\acycle_{i}\cap \acycle_j =0,\qquad
\bcycle_{i}\cap \bcycle_j =0.
\end{gather*}
A basis satisfying these intersection relations is called ``symplectic''.

From the topological point of view, a genus $\genus > 0$ compact Riemann surface with a basis $(\mathcal{A}_i,\mathcal{B}_i)_{1 \leq i \leq \genus}$ is a $4\mathfrak{g}$ closed polygon $\Gamma$, with edges
\begin{gather*}
\big[\mathcal{A}_1,\mathcal{B}_1,\mathcal{A}_1^{-1},\mathcal{B}^{-1}_1,\ldots,\mathcal{A}_{\genus},\mathcal{B}_{\genus},
\mathcal{A}^{-1}_{\genus},\mathcal{B}_{\genus}^{-1}\big]
\end{gather*}
glued by pairs. $\mathring{\Gamma}$ is a fundamental domain of $\curve$. It is a classical result, that on a curve of genus~$\genus$, there exists~$\genus$ independent holomorphic 1-forms  $\mathrm{d}u_i$ (holomorphic means in particular having no poles), and they can be normalized on the $\acycle$-cycles
\[
\oint_{\acycle_i} \mathrm{d}u_j = \delta_{i,j}.
\]
Then, the $\genus\times\genus$ matrix $\tau_{i,j}$,
\[
\tau_{i,j} = \oint_{\bcycle_i} \mathrm{d}u_j,
\]
is known to be symmetric $\tau_{i,j}=\tau_{j,i}$ and its imaginary part is def\/inite positive
\[
\tau^t=\tau,\qquad \operatorname{Im} \tau>0.
\]
$\tau$ is called the Riemann matrix of periods of $\curve$.

\subsubsection{Theta functions}
\label{sec:theta}
Given any symmetric matrix $\tau$ such that $\Im\,\tau>0$, one can def\/ine the Riemann Theta function
\[
\theta(\bfu|\tau) = \sum_{\mathbf{n} \in \mathbb Z^\genus}  \ee{2{\rm i}\pi \mathbf{n}\cdot\bf u}  \ee{{\rm i}\pi   \mathbf{n}^t\cdot \tau\cdot\mathbf{n}}.
\]
Since $\operatorname{Im} \tau>0$, it is a well-def\/ined convergent series for all $\bfu$ in $\mathbb C^\genus$. Most often we will not write the $\tau$ dependence of the Theta function: $\theta(\bfu|\tau) \equiv \theta(\bfu)$. This function is quasi-periodic in $\bfu$: if $\bfn,\bfm \in \mathbb Z^\genus$, we have
\begin{gather}\label{eqthetaphase}
\theta(\bfu+\bfn+\tau \bfm) = \ee{-{\rm i}\pi (2 \bfm^t\cdot\bfu+ \bfm^t\cdot\tau\cdot\bfm)}\theta(\bfu).
\end{gather}
It also satisf\/ies the heat equation
\begin{gather*}
\partial_{\tau_{i,j}} \theta = \frac{1}{4\ii\pi} \partial_{u_i}\partial_{u_j} \theta.
\end{gather*}
In this equation, $\tau_{i,j}$ and $\tau_{j,i}$ are considered independent.

\subsubsection{Jacobian and Abel map}

Let us choose a generic basepoint $o\in\curve$ (it will in fact play no role). For any point $z \in \curve$, we def\/ine
\[
\forall\, i \in \{1,\dots,\genus\},\qquad u_i(z) = \int_o^z \mathrm{d}u_i,
\]
where the integration path is chosen such that it does not intersect any $\mathcal{A}$-cycle or $\mathcal{B}$-cycle.
Then we def\/ine the vector
\[
\mathbf{u}(z) = [u_i(z)]_{1 \leq i \leq \genus}\in \mathbb C^\genus.
\]
The application $z\mapsto \mathbf{u}(z)\;{\rm mod}\,(\mathbb Z^\genus + \tau\cdot\mathbb Z^\genus)$ is well-def\/ined and analytical, it maps the spectral curve into the Jacobian $\mathbb{J} = \mathbb C^\genus /(\mathbb Z^\genus + \tau\cdot\mathbb Z^\genus)$. This def\/ines the Abel map
\begin{gather*}
\curve \to \mathbb J,\\
z\mapsto \mathbf{u}(z)\; {\rm mod}\, (\mathbb Z^\genus \oplus \tau\cdot\mathbb Z^\genus).
\end{gather*}
The Jacobi inversion theorem states that every $\mathbf{w} \in \mathbb{J}$ can be represented as $\mathbf{w} = \sum\limits_{j = 1}^{\mathfrak{g}} \mathbf{u}(p_j)$ for some points $p_1,\ldots,p_{\mathfrak{g}} \in \mathcal{C}$.

The Theta function can be used with $\tau$  the Riemann matrix of periods of a Riemann surface~$\mathcal{C}$, and~$\mathbf u$ the Abel map of a point on $\curve$. In this case, it enjoys other important properties.
Its zero locus has the following description: there exists $\mathbf{k} \in \mathbb{C}^{\mathfrak{g}}$, so that $\theta(\mathbf{w}|\tau) = 0$ if\/f there exists $\mathfrak{g} - 1$ points $z_1,\ldots,z_{\mathfrak{g} - 1} \in \mathcal{C}$ satisfying $\mathbf{w} = \sum\limits_{j = 1}^{\mathfrak{g} - 1} \mathbf{u}(z_j) + \mathbf{k}$. $\mathbf{k}$ is called a ``Riemann vector of constants'', and it depends on the basepoint $o$ used to def\/ine the Abel map $\mathbf{u}$.

\subsubsection{Prime form}

An odd characteristics $\bf c$ is a vector of the form
\[
\mathbf{c} = \frac{\bfn + \tau\bfm}{2},\qquad
\bfn,\bfm \in \mathbb Z^\genus,\qquad
\bfn^t\cdot\bfm \in 2\mathbb Z+1.
\]
The Theta function vanishes at odd characteristics: $\theta({\bf c})=0$, and the following holomorphic form
\[
\mathrm{d}h_{\mathbf{c}}(z) = \sum_{i=1}^\genus \mathrm{d}u_i(z)  {\partial_{u_i} \theta(\bf c)}
\]
has only double zeroes on $\mathcal{C}$, so that we can def\/ine its squareroot, and thus one can def\/ine the prime form~\cite{MumTata,MumTata2,MumTata3}
\begin{gather*} 
E(z_1,z_2) = \frac{\theta(\bfu(z_1)-\bfu(z_2)+{\bf c})}{ \sqrt{\mathrm{d}h_{\mathbf{c}}(z_1)  \mathrm{d}h_{\mathbf{c}}(z_2)}}.
\end{gather*}
There exists choices of $\mathbf{c}$ such that $E$ is not identically $0$ (we say $\mathbf{c}$ is ``non singular''), and $E$ is in fact independent of such $\mathbf{c}$. It is a $(-\frac{1}{2},-\frac{1}{2})$-form on $\mathcal{C}\times\mathcal{C}$, and it vanishes only at $z_1=z_2$. In any local coordinate $\xi(z)$ we have
\begin{gather*} 
E(z_1,z_2) \mathop{=}_{z_1 \rightarrow z_2} \frac{\xi(z_1)-\xi(z_2)}{ \sqrt{\mathrm{d}\xi(z_1) \mathrm{d}\xi(z_2)}} + O\big((\xi(z_1)-\xi(z_2))^3\big).
\end{gather*}
Because of the Theta function, $E(z_1,z_2)$ is multivalued $\curve\times\curve$. It transforms according to equation~\eqref{eqthetaphase}.

The Theta function associated to a Riemann surface satisf\/ies a non-linear relation called Fay identity~\cite{Fay}: for any $z_1,z_2,z_3,z_4 \in \mathcal{C}$, any $\mathbf{w} \in \mathbb{C}^{\mathfrak{g}}$,
\begin{gather*}
  \theta(\mathbf{w} + \mathbf{c})\theta(\mathbf{u}_{12} + \mathbf{u}_{34} + \mathbf{w} + \mathbf{c}) \frac{E(z_1,z_3)E(z_2,z_4)}{E(z_1,z_4)E(z_2,z_3)}\,\frac{1}{E(z_1,z_2)E(z_3,z_4)} \nonumber \\
\qquad{}  =   \frac{\theta(\mathbf{w} + \mathbf{u}_{12} + \mathbf{c})}{E(z_1,z_2)}\frac{\theta(\mathbf{w} + \mathbf{u}_{34} + \mathbf{c})}{E(z_3,z_4)} - \frac{\theta(\mathbf{w} + \mathbf{u}_{14} + \mathbf{c})}{E(z_1,z_4)}\frac{\theta(\mathbf{w} + \mathbf{u}_{32} + \mathbf{c})}{E(z_3,z_2)},
\end{gather*}
where $\mathbf{u}_{jl} = \mathbf{u}(z_j) - \mathbf{u}(z_l)$.

\subsubsection{Bergman kernel}
\label{bbb}

We call Bergman kernel the ``fundamental (1,1)-form of the second kind'' \cite{Fay}, def\/ined as
\[
B(z_1,z_2) = \mathrm{d}_{z_1}\mathrm{d}_{z_2}\ln{(\theta(\bfu(z_1)-\bfu(z_2)+\bf c))}.
\]
It is independent of the choice of a non-singular, odd characteristics~$\bf c$. It is a globally def\/ined, symmetric (1,1)-form, having a double pole at $z_1=z_2$ with no residue, and no other pole.
It is normalized so that
\[
\oint_{\acycle_i} B(\cdot,z) = 0,\qquad
\oint_{\bcycle_i} B(\cdot,z) = 2{\rm i}\pi\,\mathrm{d}u_i(z).
\]
Near $z_1=z_2$, it behaves, in any local coordinate $\xi(z)$, like
\[
B(z_1,z_2) \mathop{=}_{z_1 \rightarrow z_2} \frac{\mathrm{d}\xi(z_1) \mathrm{d}\xi(z_2)}{ (\xi(z_1)-\xi(z_2))^2} + O(1).
\]

We also def\/ine the fundamental 1-form of the third kind
\[
\mathrm{d}S_{z_1,z_2}(z) = \int_{z_2}^{z_1} B(\cdot,z),
\]
where the integration contour is chosen so that it does not intersect any $\acycle$-cycle or $\bcycle$-cycle.
It is a 1-form in the variable $z$, and a function of the variable $z_1$, $z_2$, and it satisf\/ies
\[
\oint_{\acycle_j} \mathrm{d}S_{z_1,z_2} = 0,\qquad\oint_{\bcycle_j} \mathrm{d}S_{z_1,z_2} = 2{\rm i}\pi (u_j(z_1)-u_j(z_2)).
\]
It has a simple pole at $z=z_1$ with residue $+1$, a simple pole at $z=z_2$ with residue $-1$, and no other pole.
In other words, in any local coordinate $\xi(z)$
\begin{gather*}
\mathrm{d}S_{z_1,z_2}(z) \mathop{\sim}_{z \rightarrow z_1}  \frac{\mathrm{d}\xi(z)}{\xi(z)-\xi(z_1)},
\qquad
\mathrm{d}S_{z_1,z_2}(z) \mathop{\sim}_{z \rightarrow z_2}   \frac{- \mathrm{d}\xi(z)}{\xi(z)-\xi(z_2)}.
\end{gather*}
Notice that in the variable $z$ it is globally def\/ined for $z\in \curve$ (it has no monodromy if $z$ goes around a non-contractible cycle), whereas in the variable $z_1$ (resp.~$z_2$) it is def\/ined only on the fundamental domain, it has monodromies when $z_1$ (resp.~$z_2$) goes around a non-contractible cycle $\bcycle_j$
\begin{gather*}
\mathrm{d}S_{z_1+\bcycle_j,z_2}(z)  = \mathrm{d}S_{z_1,z_2}(z) + 2{\rm i}\pi \dd u_j(z),\qquad \mathrm{d}S_{z_1,z_2+\bcycle_j}(z)  = \mathrm{d}S_{z_1,z_2}(z) - 2{\rm i}\pi \dd u_j(z).
\end{gather*}

\subsubsection[Example in genus $\mathfrak{g} = 1$]{Example in genus $\boldsymbol{\mathfrak{g} = 1}$}

When $\mathfrak{g} = 1$, the Abel map is an isomorphism between $\mathcal{C}$ and $\mathbb{J} = \mathbb{C}/\mathbb{L}$ where we set $\mathbb{L} = \mathbb{Z} + \tau\mathbb{Z}$. The $\mathcal{A}$-cycle in $\mathbb{J}$ is the segment $[0,1[$, and the $\mathcal{B}$-cycle is the segment $[0,\tau[$. The Bergman kernel normalized on $\mathcal{A}$-cycles can be expressed as
\[
B(u_1,u_2) = \dd u_1\dd u_2\left(\wp(u_1 - u_2|\tau) + \frac{\pi^2E_2(\tau)}{3}\right),
\]
where $u_1,u_2 \in \mathbb{J}$, $\wp$ is the Weierstrass function and $E_2$ the second Eisenstein series
\begin{gather*}
\wp(u|\tau) = \frac{1}{u^2} + \sum_{w \in \mathbb{L}\setminus\{0\}} \left(\frac{1}{(u + w)^2} - \frac{1}{w^2}\right),\\ E_2(\tau) = \frac{3}{\pi^2}\left(\sum_{n \neq 0} \frac{1}{n^2} + \sum_{m \neq 0} \sum_{n \in \mathbb{Z}} \frac{1}{(n + m\tau)^2}\right).
\end{gather*}

\subsection{Parametrization of meromorphic 1-forms}
\label{poiuy}

\subsubsection{Sheets, ramif\/ication and branchpoints, local coordinate patches}

If $\deg X = d$, then for every value $x$, there are $d$ points $z^1(x),\dots,z^d(x)$ on the curve $\curve$ such that $X(z^i(x))=x$. $z^i(x)$ is sometimes called the preimage of $x$ in the $i$-th sheet.

\begin{definition}
We call ``ramif\/ication points of order $k$'', the zeroes of order $k \geq 1$ of the meromorphic $1$-form $\mathrm{d}X$. If $a\in\curve$ is a ramif\/ication point, the corresponding value $X(a)$ is called a branchpoint. All the other points $z \in \mathcal{C}$ at which $X(z)$ is analytical, are called ``regular points''.
\end{definition}

\begin{definition} We say that a branchpoint $x_a$ is simple if $X^{-1}(\{x_a\})$ consists in $d - 1$ points, one of them being a ramif\/ication point of order $1$, and all the remaining ones being regular points.
\end{definition}

\subsubsection{Def\/inition of local coordinates}

Near a ramif\/ication point $a$ of order $k$, $\xi_a(z) = (X - X(a))^{1/(k + 1)}$ def\/ines a local coordinate on the curve. Simple branchpoints play a special role in Sections~\ref{sec:diffsys}, \ref{sec:sympinv} and \ref{sec7}. For a simple branchpoint we have
\[
\xi_a(z) = \sqrt{X(z)-X(a)}.
\]
Since $X$ is a meromorphic function of degree $d$, it has $d$ poles with multiplicities, i.e.\ $\infty_1^{d_{\infty_1}},\ldots$, $\infty_{s}^{d_{\infty_s}}$ with $\sum_i d_{\infty_i} = d$. Near $\infty_i$, a good local variable is
\[
\xi_{\infty_i}(z) = X(z)^{-1/d_{\infty_i}}.
\]
Besides, we will need to consider also poles of a meromorphic form~$\om$.
If $p$ is a pole of $\om$, but not a pole of $X$, neither a zero of~$\mathrm{d}X$, a good local variable is
\[
\xi_p(z) = X(z)-X(p).
\]
In this case, the multiplicity of $p$ is $d_p = -1$. We shall now always use the local coordinates $\xi(z)$ def\/ined above. Notice that they depend only on the function $X(z)$.

\begin{definition}
Given a meromorphic 1-form $\om(z)$ which has no pole at ramif\/ication points, let us call
\[
\mathcal{P} = \{\mathrm{poles} \ \mathrm{of} \ \om\},\qquad \mathcal{P}_{\infty} = \{\mathrm{poles} \ \mathrm{of}\  X\},\qquad \overline{\mathcal{P}} = \mathcal{P}\cup\mathcal{P}_{\infty}.
\]
\end{definition}
To any $p \in \overline{\mathcal{P}}$, we have associated a coordinate patch $\xi_p$ on $\mathcal{C}$ centered on $p$.

\subsubsection{Poles and times, f\/illing fractions}

Following Krichever \cite{Kri92}, we def\/ine
\begin{definition} For any $p \in \overline{\mathcal{P}}$, we def\/ine the ``times'' near $p$ as the coef\/f\/icient of the negative part of the Laurent series expansion of $\om$ near $p$
\[
\om(z)  \mathop{=}_{z \rightarrow p}\;\sum_{j\geq 0} \, t_{p,j}\, (\xi_p(z))^{-(j+1)}\, \mathrm{d}\xi_p(z)  + O(1),\qquad t_{p,j} = \Res_{z\to p}  \om(z)  \xi_p(z)^j.
\]
We also write collectively $\vec t_p = [t_{p,j}]_{j\in \mathbb N }$ and $\vec t  = (\vec t_p)_{p \in \overline{\mathcal{P}}}$. We also def\/ine the ``f\/illing fractions'' (also called ``conserved quantities''), associated to non-contractible cycles, by
\[
\epsilon_i =  \frac{1}{2\ii\pi}\,\oint_{\acycle_i} \om.
\]
\end{definition}

Notice that the times $t_{p,0} = \Res_{p} \om$ are not independent, because the sum of residues of $\om$ must vanish,
\begin{gather*}
\sum_{p \in \overline{\mathcal{P}}} t_{p,0}=0.
\end{gather*}
There is a form-cycle duality \cite{Kri92}
\begin{definition}
To each time $t_{k}$, one can associate a dif\/ferential meromorphic form $\om_k(z)$, as well as a dual cycle~$\om^*_k$, and a dual orthogonal cycle $\om_k^{*\perp}$
\begin{gather} \label{eqformcycletimeduality}
t_{k}\ \longleftrightarrow\ \om_{k}(z)\ \longleftrightarrow\ \om_{k}^*\
\longleftrightarrow\ \om_{k}^{*\perp},
\end{gather}
in such a way that
\begin{gather} \label{eqdualitytkomk}
\left.\frac{\partial \om(z)}{\partial t_k}\right|_{X(z)}  = \om_k(z),\qquad
\om(z) = \sum_k t_k  \om_k(z),
\\
\om_k(z) = \int_{\om_k^*} B(\cdot,z),\qquad t_k = \int_{\om_k^{*\perp}}\om,\qquad
\om_i^* \cap \om_j^{*\perp}=\delta_{i,j}.\nonumber
\end{gather}
The symbol $\big|_{X(z)}$ means that we dif\/ferentiate keeping the local coordinates $\xi_p(z)$ f\/ixed (i.e.~$X(z)$ f\/ixed).
\end{definition}

More explicitly we have
\begin{itemize}\itemsep=0pt
\item[$\bullet$] Filling fractions $\epsilon_i \longrightarrow \om_{j} =$ f\/irst kind dif\/ferential
\[
\om_{j}(z)= 2{\rm i}\pi \mathrm{d}u_j(z) = \oint_{\bcycle_j} B(z,\cdot),\qquad \om_{j}^* = \bcycle_j,\qquad \om_{j}^{*\perp} = \frac{1}{2{\rm i}\pi} \acycle_j.
\]
\item[$\bullet$] Residues $t_{p,0} \longrightarrow \om_{p,0} =$ third kind dif\/ferential
\[
\om_{p,0}(z)= \mathrm{d}S_{p,o}(z)=\int_o^{p} B(z,\cdot),\qquad \om_{p,0}^* = [o,p],\qquad  \om_{p,0}^{*\perp} = \frac{1}{2\ii\pi }C_{p},
\]
where $o$ is an arbitrary basepoint on $\curve$, and $C_{p}$ is a small circle surrounding $p$ with index~1. As we mentioned, the $t_{p,0}$ are not independent variables, and only $(t_{p,0}-t_{p_0,0})_{p \neq p_0}$ for a~f\/i\-xed~$p_0$ are independent. As a consequence, we see that only dif\/ferences $\om_{p,0}-\om_{p',0}$ and $\om_{p,0}^*-\om_{p',0}^*$ are independent of a choice of basepoint~$o$.
\item[$\bullet$] Higher times $t_{p,j}$ with $j\geq 1 \longrightarrow \om_{p,j} =$ second kind dif\/ferential
\begin{gather*}
\om_{p,j}(z) = B_{p,j}(z) = \Res_{z'\to p}  \xi_p(z')^{-j} B(z',z),\\
\om_{p,j}^* = \frac{1}{2\ii\pi}\xi_p^{-j} C_{p},\qquad\om_{p,j}^{*\perp} = \frac{1}{2\ii\pi}\xi_p^{j+1} C_{p}.
\end{gather*}
\end{itemize}
Any meromorphic form $\om$ is a linear combination of those basis meromorphic forms, and almost by def\/inition we have
\begin{gather}\label{thydxdecomp}
\om(z) = \sum_k t_k \om_k(z) =  \sum_{i = 1}^{\mathfrak{g}} 2{\rm i}\pi \epsilon_i  \mathrm{d}u_i(z) + \sum_{p \in \overline{\mathcal{P}}} t_{p,0} \mathrm{d}S_{p,o}(z) + \sum_{p \in \overline{\mathcal{P}}, j \geq 1} t_{p,j}\,B_{p,j}(z).
\end{gather}

\subsection[$F_0$]{$\boldsymbol{F_0}$}\label{sec:prepot}

The fact that $\int_{\om_i^{*}} \int_{\om_j^{*}} B(z,z')$ is symmetric, implies that there exists a function $F_0({\vec t}\,)$ such that
\[
\frac{\partial F_0}{\partial t_i} \,\text{``=''} \, \int_{\om_i^*}\om,\qquad
\frac{\partial^2 F_0}{\partial t_i\partial t_j } \, \text{``=''}\, \int_{\om_i^*}\int_{\om_j^*} B.
\]
The problem (this is why we write quotation marks) is that those integrals are not  well-def\/ined for times associated to 3rd  kind dif\/ferentials. Such a statement is correct after an appropriate regularization. When $z$ is in the vicinity of a pole $p$, we def\/ine
\[
V_p(z) = -\sum_{j \geq 1}  \frac{t_{p,j}}{j}  \xi_p(z)^{-j},\qquad \mathrm{d}V_p(z) = \sum_{j \geq 1}  t_{p,j}  \frac{\mathrm{d}\xi_p(z)}{ \xi_p(z)^{j+1}}.
\]
It is such that $\om-\mathrm{d}V_p$ has at most a simple pole at $p$. Given an arbitrary base point $o\in \curve$, the following integral is well-def\/ined
\[
\mu_{p} =  \int^{\infty_p}_o \left(\om(z)-\mathrm{d}V_p(z)-t_{p,0} \frac{\mathrm{d}\xi_p(z)}{\xi_p(z)}\right)  - V_p(o)-t_{p,0} \ln{\xi_p(o)}.
\]
$\mu_{p}$ depends on the base point $o$, but only by an additive constant independent of $p$. Since $\sum_p t_{p,0}=0$, the sum $\sum_{p} t_{p,0} \mu_{p}$ is thus independent of $o$.
In some sense, $\mu_p$ is a regularized version of $\int_{\om_{p,0}^*} \om$ (which does not exists). Since for all the other cycles, $\int_{\om_k^*}  \om$ is well-def\/ined, we can now def\/ine $F_0$

\begin{definition}\label{defF0}
\begin{gather*}
F_0(\om) = \frac{1}{2} \left[ \sum_{p \in \overline{\mathcal{P}}} \Res_{p} V_p   \om + \sum_{p \in \overline{\mathcal{P}}} t_{p,0} \mu_p + \sum_{i = 1}^{\mathfrak{g}} \epsilon_i \oint_{\bcycle_i} \om \right].
\end{gather*}
\end{definition}

This def\/inition is closely related to that of \cite{Kri92} where~$F_0$ appears as a function of the times~$t_{p,j}$'s, but here we prefer to def\/ine it as a functional of a 1-form $\om$.

\begin{theorem}[see e.g.~\cite{Kri92}] \label{thdF0dt}
The first derivatives of $F_0$ are given by, for $j \geq 1$,
\begin{gather*}
\frac{\partial F_0}{\partial t_{p,j}} = \oint_{\om_{p,j}^*}  \om = \Res_{p} \xi_p^{-j} \om,\qquad \frac{\partial F_0}{\partial t_{p,0}}-\frac{\partial F_0}{\partial t_{p',0}} = \mu_{p}-\mu_{p'},\qquad
\frac{\partial F_0}{\partial \epsilon_i} = \oint_{\bcycle_i} \om.
\end{gather*}
\end{theorem}

The proof of this theorem has appeared in many works and contexts, initiated in~\cite{Dub92} and generalized in~\cite{Kri92}. In the context of Hurwitz spaces, this expression of $F_0$ specialized to $\omega =$ the primary dif\/ferential def\/ining the Frobenius structure, coincides with the prepotential \cite[Equation~(5.64)]{Dub92}. It follows form Theorem~\ref{thdF0dt} that
\[
F_0 = \frac{1}{2} \sum_k t_k \frac{\partial F_0}{\partial t_k},
\]
which means that $F_0$ is homogeneous of degree $2$. Another classical result is
\begin{theorem}[see e.g.~\cite{Kri92}]
\label{d2F0dti0}
The second derivatives of $F_0$ are given by
\[
\frac{\partial^2 F_0}{\partial t_{k} \partial t_{l}} = \int_{\omega_k^*}\int_{\omega_l^*} B,
\]
except for the following cases
\begin{gather*}
\frac{\partial}{\partial t_k} \left(\frac{\partial}{\partial t_{p,0}} - \frac{\partial}{\partial t_{p',0}}\right)  F_0   =  \int_{\omega_k^*}\int_{\omega_{p,0}^*} B - \int_{\omega_k^*}\int_{\omega_{p',0}^*} B,\\
\left(\frac{\partial}{\partial t_{p,0}}-\frac{\partial}{\partial t_{p',0}}\right)^2   F_0   =   - \ln{\left(E(p,p')^2 \mathrm{d}\xi_p(p) \mathrm{d}\xi_{p'}(p') \right)}, \\
\left(\frac{\partial}{\partial t_{p,0}}-\frac{\partial}{\partial t_{p',0}}\right) \left(\frac{\partial}{\partial t_{p,0}}-\frac{\partial}{\partial t_{p'',0}}\right)   F_0   =    -  \ln{\left( \frac{E(p,p')E(p,p'') \mathrm{d}\xi_p(p)}{E(p',p'')} \right)},\\
\left(\frac{\partial}{\partial t_{p,0}}-\frac{\partial}{\partial t_{p',0}}\right) \left(\frac{\partial}{\partial t_{\tilde{p},0}}-\frac{\partial}{\partial t_{\tilde{p}',0}}\right)   F_0   =   -   \ln{\left(\frac{E(p,\tilde{p}')E(p',\tilde{p}) }{E(p,\tilde{p})E(p',\tilde{p}')}\right)}.
\end{gather*}
\end{theorem}
The second derivatives of $F_0$ do not depend on the $1$-form $\om$, and thus do not depend on the times. Thus we have
\[ 
F_0 = \frac{1}{2} \sum_{k,l} t_k t_l  \frac{\partial^2 F_0}{\partial t_k\, \partial t_l}.
\]

\begin{theorem}[see e.g.~\cite{Kri92}]
\label{th23}
\[
\frac{\partial^3 F_0}{\partial t_{k} \partial t_{l} \partial t_{m}} = \sum_{a_i={\rm zeroes\ of\ dX}}  \Res_{z\to a_i}\frac{\om_k(z) \om_l(z) \om_m(z)}{dX(z) dY(z)}.
\]
\end{theorem}

\section{Reconstruction formula}
\label{sec:reconstruction}

We review in this section the reconstruction~\cite{KriNLS77, Kri77} of a Lax matrix whose evolution preserves the spectrum, and thus of an integrable system, from the spectral curve (see also the textbook~\cite{BBT} and references therein).
The only dif\/ference is that, we reformulate it intrinsically in terms of 1-forms~$\om$, instead of using time coordinates $\om=\sum_k t_k \om_k$.
For this purpose, instead of Baker--Akhiezer functions, we prefer to use a ``spinor kernel'', which is a convenient special case of Baker--Akhiezer function, which turns out to be a more intrinsic object for our formulation (see also~\cite{BertG, Kor1,Kor2}).

\subsection{Semiclassical spinor kernel}
\label{sec:skBA0}

Given a meromorphic 1-form $\om$, def\/ine the 1-form
\[
\chi(z;\om )  = \om(z) - 2{\rm i}\pi \sum_{i = 1}^{\mathfrak{g}} \epsilon_i \mathrm{d}u_i(z),
\]
which depends linearly on the times (and not on the f\/illing fractions)
\[
\chi(z;\om ) = \sum_{p \in \overline{\mathcal{P}}} t_{p,0} \mathrm{d}S_{p,o}(z) + \sum_{p \in \overline{\mathcal{P}}} \sum_{j \geq 1} t_{p,j}  \om_{p,j}(z).
\]
By construction $\chi$ is normalized on $\acycle$-cycles
\begin{gather*}\label{eqacyclechi0}
\oint_{\acycle_i}  \chi = 0.
\end{gather*}
Then we def\/ine the vector $\zeta(\om)=[\zeta_i(\om\,)]_{1 \leq i \leq \genus}$ with coordinates
\begin{gather}
\label{defzeta}
\zeta_i(\om) = \frac{1}{2{\rm i}\pi} \oint_{\bcycle_i}  \chi  = \frac{1}{2{\rm i}\pi}\left(\oint_{\bcycle_i}\om  - \sum_{j = 1}^{\mathfrak{g}} \tau_{i,j} \oint_{\acycle_i}\om\right),
\end{gather}
which we write for short as
\begin{gather*}
\zeta(\om) = \frac{1}{2{\rm i}\pi} \oint_{\bcycle-\tau\acycle}\om.
\end{gather*}
It can be decomposed as
\begin{gather*}
\zeta(\om) = \sum_{p \in \overline{\mathcal{P}}} \sum_{j \geq 0} t_{p,j}\mathbf{v}_{p,j} = \sum_{k=(p,j)} t_k \mathbf{v}_k,\qquad
\mathbf{v}_{k} = \frac{1}{2{\rm i}\pi}\oint_{\bcycle} \om_{k}.
\end{gather*}
The vector $\zeta(\om)$ is a linear function of the times $t_k$ and is independent of the f\/illing fractions~$\epsilon_i$. In other words, it follows a linear motion with constant velocity $\mathbf{v}_k$ in the Jacobian, as a~function of any of the times~$t_k$. A well-known property~\cite{BBT, Dub81, KriNLS77,Kri77} of integrable systems is that, in appropriate variables, the motion (with any of the time $t_k$) is uniform and linear. The algebraic reconstruction takes the linear evolution in the Jacobian of~$\mathcal{C}$ as starting point, and produces more complicated quantities whose evolution is described by a Lax system.

\begin{definition}\label{defpsi} We now def\/ine the \emph{semiclassical spinor kernel} as the $(1/2,1/2)$ form
\begin{gather}
\psi_{\cl}(z_1,z_2 ; \om) = \frac{\theta(\bfu(z_1)-\bfu(z_2) + \zeta(\om)+{\bf c})}{E(z_1,z_2)}  \theta(\zeta(\om)+{\bf c}) \ee{\int_{z_2}^{z_1} \chi(z;\om )},
\end{gather}
where $\mathbf{c}$ is a non-singular, odd characteristics.
\end{definition}
We write a subscript ${}_{\cl}$ to distinguish the semiclassical spinor kernel from the one proposed in the second part of the article.
This kernel was also introduced, in a similar form, in \cite{Kor1,Kor2} for solving Matrix Riemann--Hilbert problems on branched coverings of $\mathbb{CP}^1$.

\begin{proposition}
$\psi_{\cl}(z_1,z_2 ; \om)$ is a globally defined spinor in $(z_1,z_2) \in \mathcal{C}\times \mathcal{C}$, i.e.\ it is the squareroot of  a symmetric $(1,1)$-form.
\begin{itemize}\itemsep=0pt
\item[$\bullet$] It has a simple pole at $z_1 = z_2$: in any local coordinate $\xi(z)$
\begin{gather*}
\psi_{\cl}(z_1,z_2 ; \om)\,\mathop{\sim}_{z_1 \rightarrow z_2} \frac{1}{E(z_1,z_2)} \sim \frac{\sqrt{\mathrm{d}\xi(z_1) \mathrm{d}\xi(z_2)}}{\xi(z_1)-\xi(z_2)}.
\end{gather*}
\item[$\bullet$] It has essential singularities when $z_1$ $($resp. $z_2)$ approaches a pole of $\om$.
\end{itemize}
\end{proposition}

\begin{proof} The behavior at $z_1\to z_2$ is obvious, and the essential singularities at the poles of $\om$ come from the exponential term. What we need to prove, is that $\psi_{\cl}(z_1,z_2\,;\,\om)$ is unchanged when~$z_1$ (resp.~$z_2$) goes around a non-trivial cycle. When~$z_1$ (resp.~$z_2$) goes around an $\acycle$-cycle, the vector $\bfu(z_1)$ (resp. $\bfu(z_2)$) is translated by an integer vector, $\theta$ is thus unchanged, and thanks to equation~\eqref{eqacyclechi0}, $\psi_{\cl}$ is unchanged when $z_1$ (resp.~$z_2$) goes around an $\acycle$-cycle. When $z_1$ (resp.~$z_2$) goes around a $\bcycle$-cycle, the vector $\bfu(z_1)$ (resp.~$\bfu(z_2)$) is translated by a lattice vector of the form $\tau\cdot\bfn$ with $\bfn\in \mathbb Z^\genus$, and $\theta$ gets multiplied by a phase according to equation~\eqref{eqthetaphase}. Remember that the prime form $E(z_1,z_2)$ is also a $\theta$ function, and also gets a phase given by equation~\eqref{eqthetaphase}.
$\psi_{\cl}$~is thus changed by
\begin{gather*}
\psi_{\cl}(z_1+\bfn\bcycle,z_2 ; \om) \to  \psi_{\cl}(z_1,z_2 ; \om)  \ee{-2{\rm i}\pi  \bfn\cdot\zeta(\om)}  \ee{\bfn\cdot\oint_{\bcycle} \chi},
\end{gather*}
and because of equation~\eqref{defzeta}, i.e.\ $\zeta = \frac{1}{2{\rm i}\pi} \oint_{\bcycle}\chi$, we see that $\psi_{\cl}$ is unchanged when $z_1$ (resp.~$z_2$) goes around a $\bcycle$-cycle.
\end{proof}

\subsection{Duality equation}

Then we construct the following spinor matrix of size $d\times d$
\begin{gather*}
\Psi_{\cl}(x_1,x_2 ; \om) =  [\psi_{\cl}(z^i(x_1),z^j(x_2) ; \om)]_{i,j=1}^d,
\end{gather*}
where we recall that $z^i(x)$ are the $d$ preimages of $x$ on the curve $\curve$, i.e.\ $X(z^i(x))=x$, and $d=\deg X$. These preimages are distinct and this matrix is well-def\/ined when $x_1$ (or~$x_2$) is not at a branchpoint.

\begin{proposition}\label{thYB}
We have the ``duality'' equation
\[
\Psi_{\cl}(x_1,x_2 ; \om)\,\Psi_{\cl}(x_2,x_3 ; \om) = \frac{(x_1-x_3) \mathrm{d}x_2}{(x_1-x_2)(x_2-x_3)} \Psi_{\cl}(x_1,x_3 ; \om).
\]
\end{proposition}

\begin{proof}
\begin{gather*}
\frac{1}{\mathrm{d}X(z)} \psi_{\cl}(z^i(x_1),z ; \om)  \psi_{\cl}(z,z^j(x_3) ; \om)
\end{gather*}
is a meromorphic function of $z$.
Indeed, the product of two $(1/2)$-forms is a $1$-form, and when we divide by $\mathrm{d}X$, we get a function.
The essential singularities coming from the exponentials cancel in the product, so this function can only have poles, i.e.\ it is meromorphic. The only possible poles are at $z=z^i(x_1)$ or $z=z^j(x_3)$ or at the zeroes of $\mathrm{d}X(z)$. Then, summing over all sheets, we see that
\begin{gather*}
\sum_k \frac{\psi_{\cl}(z^i(x_1),z^k(x_2) ; \om)  \psi_{\cl}(z^k(x_2),z^j(x_3) ; \om)}{\dd X(z^k(x_2))}
\end{gather*}
is a symmetric sum of a meromorphic function over all sheets of $x_2$, therefore it is a meromorphic function of $x_2\in \widehat{\mathbb{C}}$, i.e.\ a rational  function of the complex variable~$x_2$.
It remains to f\/ind its poles. $1/\dd X(z^k(x_2)))$ behaves like $O(x_2-X(a_i))^{-1/2}$ at ramif\/ication points, and since a rational function of $x_2$ cannot have a singularity of power $-1/2$, this means that this rational function has no pole at branchpoints. Its only poles can then be at $x_2=x_1$ or $x_2=x_3$, and they are simple poles. The residues of the corresponding poles are easily computed and give the theorem.
\end{proof}

\begin{proposition}\label{thYBsp}
We have a refined version of the duality equation
\begin{gather*}
\psi_{\cl}(z_1,z ; \om) \psi_{\cl}(z,z_2 ; \om) = - \psi_{\cl}(z_1,z_2 ; \om)\left(\mathrm{d}S_{z_1,z_2}(z) -2{\rm i}\pi \sum_{j = 1}^{\mathfrak{g}} \alpha_j(z_1,z_2 ; \om)  \mathrm{d}u_j(z)\right),
\end{gather*}
where
\begin{gather*}
\alpha_j(z_1,z_2 ; \om) =  \frac{\theta_{u_j}(\bfu(z_1)-\bfu(z_2)+\zeta(\om) +\bfc)}{\theta(\bfu(z_1)-\bfu(z_2)+\zeta(\om) +\bfc)}-\frac{\theta_{u_j}(\zeta(\om)+\bfc)}{\theta(\zeta(\om)+\bfc)}.
\end{gather*}
\end{proposition}
This property, can be viewed as a special case of an ``addition formula'' for Baker--Akhiezer functions, found in \cite{KriBu06}. Notice that Proposition~\ref{thYB} is a corollary of Proposition~\ref{thYBsp}. Indeed the duality equation (Proposition~\ref{thYB}) can be obtained by summing the equation above on all sheets $z = z^{k}(x)$, because $\sum_{k} \mathrm{d}u_i(z^k(x))=0$ and
\begin{gather*}
\sum_k \mathrm{d}S_{z_1,z_2}(z^k(x)) = \frac{(X(z_1)-X(z_2)) \mathrm{d}X(z)}{(X(z)-X(z_1))(X(z)-X(z_2))}.
\end{gather*}

\begin{proof} Notice that $\psi_{\cl}(z_1,z;\om)\psi_{\cl}(z,z_2;\om)$ is a meromorphic 1-form in $z$, since it has no essential singularity. It has simple poles at $z=z_1$ and $z=z_2$, with residues $\mp \psi_{\cl}(z_1,z_2;\om)$, and it has no other pole. This means that $\psi_{\cl}(z_1,z;\om)\psi_{\cl}(z,z_2;\om) +\psi_{\cl}(z_1,z_2;\om)\mathrm{d}S_{z_1,z_2}(z)$ is a holomorphic 1-form, with no poles, therefore it must be a linear combination of the $\mathrm{d}u_i(z)$'s, which we choose to write
\begin{gather*}
\psi_{\cl}(z_1,z ; \om) \psi_{\cl}(z,z_2 ; \vec{t}\, ) = - \psi_{\cl}(z_1,z_2 ; \om)\left(\mathrm{d}S_{z_1,z_2}(z) - 2{\rm i}\pi\sum_{j = 1}^{\mathfrak{g}} \alpha_j(z_1,z_2 ; \om)  \mathrm{d}u_j(z) \right).
\end{gather*}
The left hand side is a well-def\/ined spinor of $z_1$ and $z_2$, whereas in the right hand side, $dS_{z_1,z_2}(z) = \int_{z_2}^{z_1} B(z,\cdot)$ gets some shifts when $z_1$ or $z_2$ go around non-trivial cycles. This implies the following relations for the coef\/f\/icients $\alpha_j(z_1,z_2;\om)$
\begin{gather*}
\alpha_j(z_1+\acycle_k,z_2;\om) = \alpha_j(z_1,z_2;\om),\qquad  \alpha_j(z_1,z_2+\acycle_k;\om) = \alpha_i(z_1,z_2;\om), \\
\alpha_j(z_1+\bcycle_k,z_2;\om) = \alpha_j(z_1,z_2;\om)-2{\rm i}\pi \delta_{j,k},\qquad \alpha_j(z_1,z_2+\bcycle_k;\om) = \alpha_j(z_1,z_2;\om)+2{\rm i}\pi \delta_{j,k}.
\end{gather*}
Moreover, we must have $\alpha_j(z_1,z_1;\om)\!=\!0$, and $\alpha_j(z_1,z_2;\om)$ may have poles when $\psi_{\cl}(z_1,z_2;\om)\!=\!0$. Apart from those poles, $\alpha_j(z_1,z_2;\om)$ has no other singularities.
The following quantity has all the required properties
\begin{gather*}
 \frac{\theta_{u_j}(\bfu(z_1)-\bfu(z_2)+\zeta+\bfc)}{\theta(\bfu(z_1)-\bfu(z_2)+\zeta+\bfc)}-\frac{\theta_{u_j}(\zeta+\bfc)}{\theta(\zeta+\bfc)}.
\end{gather*}
So, the dif\/ference of $\alpha_j$ and that quantity should be a meromorphic function of $z_1$ and $z_2$ without poles, i.e.\ a constant, and its value is zero by looking at $z_1 = z_2$.
\end{proof}

\subsection[Link with Baker-Akhiezer functions]{Link with Baker--Akhiezer functions}\label{funcin}

\subsubsection[Baker-Akhiezer functions]{Baker--Akhiezer functions}\label{fak}

The usual formulation of integrable systems \cite[Chapter~5]{BBT} is obtained by specializing one of the points to $X = \infty$.
In some sense, we would like to consider
\begin{gather*}
\psi_{\cl|i}(z) \, \text{``=''}\, \psi_{\cl}(z,\infty_i)  .
\end{gather*}
The problem is, that the expression in the right hand side is divergent, and thus we again need regularizations.

The def\/initions in this paragraph also apply to the spinor kernel constructed in Section~\ref{sec:BAsk}, so we drop here the~$_{\cl}$ index. Recall that the function~$X$ has degree $d$, so the point $X = \infty$ has~$d$ preimages $\infty_i$  (counted with multiplicities) on the curve.
We def\/ine
\begin{gather}\label{defpsiBA}
\psi_{i,0}(z) = \lim_{z_2\to\infty_i} \frac{\psi(z,z_2;\om)}{\sqrt{\mathrm{d}\xi_{\infty_i}(z_2)}} \ee{V_{\infty_i}(z_2) } (\xi_{\infty_i}(z_2))^{t_{\infty_i,0}},
\end{gather}
and if $d_{\infty_i} >1$, we def\/ine for $0 \leq j \leq (d_{\infty_i}-1)$
\begin{gather*}
\psi_{i,j}(z) = \frac{\mathrm{d}^{j}}{\mathrm{d}\xi_{\infty_i}(z_2)^{j}} \left( \frac{\psi(z,z_2;\om )}{\sqrt{\mathrm{d}\xi_{\infty_i}(z_2)}} \ee{V_{\infty_i}(z_2) }  (\xi_{\infty_i}(z_2))^{t_{\infty_i,0}} \right)_{z_2=\infty_i}.
\end{gather*}
There are $d$ pairs $I=(i,j)$ such that $0\leq j \leq d_{\infty_i}-1$, and therefore the vector $\vec\psi(z) = [\psi_{I}(z)]$ is a $d$-dimensional vector, and the matrix $\Psi(x;\om) = [\psi_I(z^k(x))]_{I,1 \leq k \leq d}$ is a $d\times d$ square matrix.

\subsubsection[Dual Baker-Akhiezer functions]{Dual Baker--Akhiezer functions}

Similarly, we would like to def\/ine $\phi_{\cl|i}(z) \, \text{``=''}\, \psi_{\cl}(\infty_i,z)$. Thus, we def\/ine the dual Baker--Akhiezer functions
\begin{gather*}
\phi_{i,0}(z) = \lim_{z_1\to\infty_i} \frac{\psi(z_1,z;\om)}{\sqrt{\mathrm{d}\xi_{\infty_i}(z_1)}}\ee{-V_{\infty_i}(z_1) } (\xi_{\infty_i}(z_1))^{-t_{\infty_i,0}},
\end{gather*}
and if $d_{\infty_i} > 1$, we def\/ine for each $0 \leq j \leq (d_{\infty_i}-1)$
\begin{gather*}
\phi_{i,j}(z) = \frac{\mathrm{d}^{j}}{\mathrm{d}\xi_{\infty_i}(z_1)^{j}}  \left(\,\frac{\psi(z_1,z;\om )}{\sqrt{\mathrm{d}\xi_{\infty_i}(z_1)}} \ee{-V_{\infty_i}(z_1) }  (\xi_{\infty_i}(z_1))^{-t_{\infty_i,0}} \right)_{z_1=\infty_i}.
\end{gather*}
There are $d$ pairs $I=(i,j)$ such that $0\leq j \leq d_i-1$, and therefore the vector $\vec\phi(z) = [\phi_{I}(z)]$ is a $d$-dimensional vector, and the matrix $\Phi(x;\om) = [\phi_I(z^k(x))]_{I,1 \leq k \leq d}$ is a $d\times d$ square matrix. From Corollary~\ref{reconsa}, one retrieves the well-known result that columns of $\Phi(x;\om)$ are eigenvectors of a Lax matrix.

\subsection[Christoffel-Darboux relations]{Christof\/fel--Darboux relations}
\label{sec:CD}

\begin{proposition}\label{thduality}
The matrix
\begin{gather*}
A_{\cl}^{-1} = \frac{1}{\mathrm{d}x} \Phi_{\cl}(x) \Psi_{\cl}^t(x)
\end{gather*}
is invertible, and independent of $x$. The matrix $A_{\cl}$ is called the Christoffel--Darboux matrix.
This can also be written $\Psi_{\cl}^t(x) A_{\cl}  \Phi_{\cl}(x)= \mathrm{d}x \mathbf{1}_{d \times d}$.
\end{proposition}

\begin{proof} This is an application of Proposition~\ref{thYB}, up to a conjugation.
Indeed
\begin{gather*}
\big(A_{\cl}^{-1}\big)_{(i,k),(i',k')}  =  \frac{\mathrm{d}^{k'-1}}{\mathrm{d}\xi_{\infty_{i'}}^{k'-1}(z_1)} \frac{\mathrm{d}^{k-1}}{\mathrm{d}\xi_{\infty_i}^{k-1}(z_2)} \Bigg[\sum_m \frac{\psi_{\cl}(z_1,z^m)\psi_{\cl}(z^m,z_2)}{\sqrt{\mathrm{d}\xi_{\infty_{i'}}(z_1) \mathrm{d}\xi_{\infty_i}(z_2)}}      \\
 \hphantom{\big(A_{\cl}^{-1}\big)_{(i,k),(i',k')}  =}{} \times \ee{V_{\infty_i}(z_2)-V_{\infty_{i'}}(z_1)} \xi_{\infty_i}(z_2)^{t_{\infty_i,0}} \xi_{\infty_{i'}}(z_1)^{-t_{\infty_{i'},0}} \Bigg]_{z_1=\infty_{i'}}^{z_2=\infty_i}   \\
\hphantom{\big(A_{\cl}^{-1}\big)_{(i,k),(i',k')}}{} = \frac{\mathrm{d}^{k'-1}}{\mathrm{d}\xi_{\infty_{i'}}^{k'-1}(z_1)} \frac{\mathrm{d}^{k-1}}{\mathrm{d}\xi_{\infty_i}^{k-1}(z_2)} \Bigg[\frac{(X(z_1) - X(z_2))\psi_{\cl}(z_1,z_2)  \ee{V_{\infty_i}(z_2)-V_{\infty_{i'}}(z_1)}}{(X(z) - X(z_1))(X(z) - X(z_2))}    \\
\hphantom{\big(A_{\cl}^{-1}\big)_{(i,k),(i',k')}=}{}
   \times \frac{\xi_{\infty_i}(z_2)^{t_{\infty_i,0}} \xi_{\infty_{i'}}(z_1)^{-t_{\infty_{i'},0}}}
{\sqrt{\mathrm{d}\xi_{\infty_{i'}}(z_1) \mathrm{d}\xi_{\infty_i}(z_2)}} \Bigg]_{z_1=\infty_{i'}}^{z_2=\infty_i}.
\end{gather*}
If $i\neq i'$, the quantity
\begin{gather*}
 \frac{\psi_{\cl}(z_1,z_2)  \ee{V_{\infty_i}(z_2)-V_{\infty_{i'}}(z_1)} \xi_{\infty_i}(z_2)^{t_{\infty_i,0}} \xi_{\infty_{i'}}(z_1)^{-t_{\infty_{i'},0}}}{\sqrt{\mathrm{d}\xi_{\infty_{i'}}(z_1)
 \mathrm{d}\xi_{\infty_i}(z_2)}}
\end{gather*}
has a well-def\/ined limit when $z_1\to\infty_{i'}$ and $z_2\to\infty_i$, and the term $\frac{1}{X(z)-X(z_1)}-\frac{1}{X(z)-X(z_2)}$ behaves like
\begin{gather*}
\frac{1}{X(z)-X(z_1)}
\mathop{\sim}_{z_1\to\infty_{i'}} \xi_{\infty_{i'}}(z_1)^{d_{\infty_{i'}}},
\end{gather*}
so we are computing the $(k'-1)$-th derivative of $O(\xi_{\infty_{i'}}(z_1)^{d_{\infty_{i'}}})$, where $k' \leq d_{\infty_{i'}}$, and therefore we get $0$, i.e.
\begin{gather*}
(A^{-1})_{\cl|(i,k),(i',k')} =0 \qquad {\rm if} \ \ i\neq i'.
\end{gather*}
If $i=i'$, we f\/irst take the limit $z_1\to \infty_i$, and again the term with $\frac{1}{X(z)-X(z_1)}$ vanishes.
Then, remember that $\psi_{\cl}(z_1,z_2)$ has a simple pole at $z_1=z_2$, and thus the derivative with respect to~$z_1$, can generate a pole of degree~$k'$ at $z_2=\infty_i$.
Therefore, we are computing the $(k-1)$-th derivative of $O(\xi_{\infty_i}(z_2)^{d_{\infty_i}-k'})$.
We get zero if~$k+k'\leq d_{\infty_i}$, and therefore
\begin{gather*}
(A^{-1})_{\cl|(i,k),(i,k')} =0 \qquad {\rm if} \ \ k+k'\leq d_{\infty_i}.
\end{gather*}
If $i=i'$ and $k+k'=d_{\infty_i}+1$, the only non-vanishing contribution is
\begin{gather*}
  \frac{1}{(k' - 1)!} \big(A^{-1}\big)_{\cl|(i,k),(i,k')}  \\
\qquad{} =  \frac{\mathrm{d}^{k-1}}{\mathrm{d}\xi_{\infty_i}^{k-1}(z_2)}
 \lim_{z_1\to\infty_i}\bigg[\frac{E(z_1,z_2) \psi_{\cl}(z_1,z_2)}{(\xi_{\infty_i}(z_2)-\xi_{\infty_i}(z_1))^{k'}} \ee{V_{\infty_i}(z_2)-V_{\infty_i}(z_1)}
 \\
 \qquad\quad\qquad{}\times
 \xi_{\infty_i}(z_2)^{t_{\infty_i,0} + d_{\infty_i}} \xi_{\infty_i}(z_1)^{-t_{\infty_i,0}} \bigg]_{z_2=\infty_i}   \\
 \qquad{}  =  \frac{\mathrm{d}^{k-1}}{\mathrm{d}\xi_{\infty_i}^{k-1}(z_2)}\lim_{z_1\to\infty_i}\Big[\psi_{\cl}(z_1,z_2)E(z_1,z_2) \ee{V_{\infty_i}(z_2)-V_{\infty_i}(z_1)}\\
 \qquad\quad\qquad{}\times
 \xi_{\infty_i}(z_2)^{t_{\infty_i,0} + d_{\infty_i} - k'} \xi_{\infty_i}(z_1)^{-t_{\infty_i,0}}\Big]_{z_2=\infty_i}   \\
\qquad{} =
 \frac{\mathrm{d}^{k-1}}{\mathrm{d}\xi_{\infty_i}^{k-1}(z_2)} \left[ \xi_{\infty_i}(z_2)^{d_{\infty_i} - k'}\right]_{z_2=\infty_i}
  =   \frac{\mathrm{d}^{k-1}}{\mathrm{d}\xi_{\infty_i}^{k-1}(z_2)} \left[ \xi_{\infty_i}(z_2)^{k-1}\right]_{z_2=\infty_i}
  =  (k-1)! \neq 0.
\end{gather*}
The matrix $A_{\cl}^{-1}$ has thus typically the shape
\begin{gather*}
A_{\cl}^{-1} =  \begin{pmatrix}
& &  * & & \\
& * & * & &  \\
* & * & * & & \\
& & & & * \\
& & & * & * \end{pmatrix}
\end{gather*}
it is made of (inverted) triangular blocks. Since the diagonal of each triangle is non-zero,
 this proves that the matrix $A_{\cl}^{-1}$ is invertible.

Then, if $i=i'$ and $k+k'\geq d_{\infty_i}+1$, we write that
\begin{gather*}
\frac{1}{X(z)-X(z_1)} = - \frac{1}{X(z_1)} + O\big(1/X(z_1)^2\big),
\end{gather*}
and we see that the leading term $\frac{1}{X(z_1)}$ is independent of $X(z)$, and the part which depends on $X(z)$ is $O(1/X(z_1)^2)=O(\xi_{\infty_i}(z_1)^{2d_{\infty_i}})$.
A non vanishing contribution to the part which depends on $X(z)$ could occur only if $k+k' > 2d_{\infty_i}$, which can never happen since we assumed $k,k'\leq d_{\infty_i}$. This proves that $A_{\cl}$ is independent of $X(z)$.
\end{proof}

\begin{corollary}
The matrices $\Psi_{\cl}(x;\om)$ and $\Phi_{\cl}(x;\om)$ are invertible.
\end{corollary}
As a consequence, $\psi_{\mathrm{cl}}(z_1,z_2;\om)/\sqrt{\dd x(z_1)\dd x(z_2)}$ can be identif\/ied with an integrable kernel in the sense of \cite{IIKS}, i.e.\ we have

\begin{proposition} Christoffel--Darboux relation:
\begin{gather*}
\psi_{\cl}(z_1,z_2;\om) = - \frac{\sum_{I,J} \psi_{\cl|I}(z_1)  A_{\cl|I,J}  \phi_{\cl|J}(z_2)}{X(z_1)-X(z_2)}.
\end{gather*}
\end{proposition}

\begin{proof}
This is an application of Property~\ref{thduality} and Proposition~\ref{thYB}.
Indeed, the very def\/inition of the $\psi_{\cl|I}$'s, means exactly that there exists a matrix $C_{\cl}(x)$ such that
\begin{gather*}
\Psi_{\cl}^t(X(z_1)) = \lim_{x\to\infty} \Psi_{\cl}(X(z_1),x)  C_{\cl}(x),
\end{gather*}
and similarly, there exists a matrix $\td{C}_{\cl}(x)$ such that
\begin{gather*}
\Phi_{\cl}(X(z_2)) = \lim_{x\to\infty} \td C_{\cl}(x) \Psi_{\cl}(x,X(z_2)).
\end{gather*}
When $z_1=z_2$, we have $\Psi_{\cl}(x,X(z_1))\Psi_{\cl}(X(z_1),x)=-\frac{\mathrm{d}x \mathrm{d}X(z_1)}{(x-X(z_1))^2} \mathbf{1}_{d \times d}$, which implies
\begin{gather*}
- \lim_{x\to\infty} \frac{\mathrm{d}x \mathrm{d}X(z_1)}{(x-X(z_1))^2} \td C_{\cl}(x)   C_{\cl}(x) = \Phi_{\cl}(X(z_1))  \Psi_{\cl}^t(X(z_1)) = {A_{\cl}^{-1}  \mathrm{d}X(z_1)},
\end{gather*}
and therefore
\begin{gather}\label{eqACCId}
\lim_{x\to\infty}  A_{\cl}  \td C_{\cl}(x) C_{\cl}(x)  \frac{\mathrm{d}x}{x^2} = -\mathbf{1}_{d \times d}.
\end{gather}
Then, we have from the duality equation, for any $x$
\begin{gather*}
  (X(z_1)-X(z_2))\,\Psi_{\cl}(X(z_1),X(z_2))   \\
\qquad{}  =   \Psi_{\cl}(X(z_1),x)\Psi_{\cl}(x,X(z_2))  \frac{ (X(z_1)-x)(X(z_2)-x)}{\mathrm{d}x} \\
 \qquad{} =    \Psi_{\cl}(X(z_1),x) C_\cl(x)  C_\cl^{-1}(x)\Psi_{\cl}(x,X(z_2))  \frac{ (X(z_1)-x)(X(z_2)-x)}{\mathrm{d}x},
\end{gather*}
and in particular, we may take the limit $x\to\infty$ and insert  equation~\eqref{eqACCId}
\begin{gather*}
  - (X(z_1)-X(z_2)) \Psi_{\cl}(X(z_1),X(z_2))   \\
 =   \lim_{x\to\infty} \Psi_{\cl}(X(z_1),x) C_\cl(x) A_{\cl}  \td C_{\cl}(x) C_{\cl}(x)  C_\cl^{-1}(x)\Psi_{\cl}(x,X(z_2))  \frac{ (X(z_1)-x)(X(z_2)-x)}{x^2}   \\
 =   \lim_{x\to\infty} \Psi_{\cl}(X(z_1),x) C_\cl(x) A_{\cl}  \td C_{\cl}(x) \Psi_{\cl}(x,X(z_2))
 =  \Psi_{\cl}^t(X(z_1))  A_{\cl}   \Phi_{\cl}(X(z_2)).\tag*{\qed}
\end{gather*}
\renewcommand{\qed}{}
\end{proof}

\subsection{Lax matrix}
\label{sec:recon}

As corollary of Proposition~\ref{thYB}
\begin{corollary}
\label{duall} If $x_1$ and $x_2$ are not branchpoints, and if $x_2\neq x_1$, then the matrix $\Psi_{\cl}(x_1,x_2)$ is invertible, and
\begin{gather*}
\Psi_{\cl}(x_1,x_2 ; \om)\Psi_{\cl}(x_2,x_1 ; \om) = - \frac{\mathrm{d}x_1 \mathrm{d}x_2}{(x_1-x_2)^2} \mathbf{1}_{d \times d}.
\end{gather*}
\end{corollary}

\begin{proof}
Take $x_3=x_1$ in the duality relation.
\end{proof}

\begin{corollary}[Reconstruction formula]\label{reconsa}
Let
\begin{gather*}
\td L(x) = \operatorname{diag}(Y_1(x),\dots,Y_d(x))= {\rm diag}\big(Y(z^1(x)),Y(z^2(x)),\dots,Y(z^d(x))\big).
\end{gather*}
For every $x_1$, the matrix
\begin{gather*}
L_{x_1}(x;\vec t\, ) = \Psi_{\cl}(x_1,x ; \om)  \td L(x)  \Psi_{\cl}^{-1}(x_1,x;\om) = -\frac{(x_1-x)^2}{\mathrm{d}x_1 \mathrm{d}x} \Psi_{\cl}(x_1,x;\om)  \td L(x)  \Psi_{\cl}(x,x_1;\om)
\end{gather*}
$($it depends on times $\vec t$ through $\om = \sum_k t_k \om_k$ as in equation~\eqref{thydxdecomp}$)$
is a rational function of~$x$. Its characteristic polynomial is independent of the times $\vec t$, and its zero locus defines the semiclassical spectral curve $\det{\left(y-L_{x_1}(x;\vec t\,)\right)} = \det(y-\td L(x)) = 0$. Changing~$x_1$ just amounts to a~conjugation
\begin{gather*}
L_{x'_1}(x;\vec t\,) = \Psi_{\cl}(x'_1,x_1;\om) \td L_{x_1}(x;\vec t\,) \Psi_{\cl}^{-1}(x'_1,x_1;\om).
\end{gather*}
\end{corollary}

\begin{proposition}
For any of the times $t_{p,j}$ with $j \geq 1$, the matrix $L_{x_1}(x;\vec t\,)$ obeys the Lax equation
\begin{gather}
\frac{\partial}{\partial t_{(p,j)}}  L_{x_1}(x;\vec t\,) = [M_{(p,j);x_1}(x;\vec t\,),L_{x_1}(x;\vec t\,)],
\end{gather}
where the matrix $M_{(p,j);x_1}$ is $M_{(p,j);x_1}(x;\vec{t}\, ) = \partial_{t_{(p,j)}}\Psi_{\cl}(x_1,x;\om) \Psi_{\cl}^{-1}(x_1,x;\om)$.
\end{proposition}

\subsection{Dif\/ferential systems}

\label{sec:diffsys}
Baker--Akhiezer functions satisfy simultaneously several f\/irst order dif\/ferential systems with respect to the spectral parameter and the times \cite{Kor1,Kor2}.

\begin{proposition}
\label{thx}
The matrix $\Psi_\cl(x_1,x;\om)$ is solution of a linear ODE with respect to the spectral parameter $x$
\[
\left(\frac{\mathrm{d}}{\mathrm{d}x} +\frac{\mathbf{1}_{d \times d}}{x-x_1} - M_{x;x_1}(x ; \vec{t}\, )\right) \frac{\Psi_{\cl}(x_1,x ; \om)}{\sqrt{\mathrm{d}x \mathrm{d}x_1}} = 0,
\]
where $M_{x;x_1}$ is a rational function of $x$ having poles only on $\overline{\mathcal P}$, with the same order as those of $\om(z)/\mathrm{d}X(z)$.
\end{proposition}

\begin{proof}
Write $x = X(z)$, we have
\begin{gather*}
\mathrm{d}X(z) \big[ M_{x;x_1}(X(z);\vec t\, )\big]_{i,j} -\delta_{i,j} \frac{\mathrm{d}X(z)}{(X(z)-x_1)}  \\
\qquad{} = -(X(z)-x_1)^2 \sum_{k} \mathrm{d}\left( \frac{\psi_\cl\big(z^i(x_1),z^k\big)}{\sqrt{\mathrm{d}X(z) \mathrm{d}x_1}}\right) \frac{\psi_\cl\big(z^k,z^j(x_1)\big)}{\sqrt{\mathrm{d}X(z) \mathrm{d}x_1}}.
\end{gather*}
In the right hand side, the essential singularities cancel, and only meromorphic singularities remain. Since we perform the sum over all sheets, the result is necessarily a rational function of~$X(z)$.
Poles could occur at singularities of~$\psi$, or also at zeroes of $\sqrt{\mathrm{d}X(z)}$, or at~$X(z)=x_1$.

$\psi_\cl$ has a simple pole at $z^k=z^i(x_1)$ for some $k$. Taking the derivative yields a double pole, and multiplying by $(X(z)-x_1)^2$ cancels the double pole. If $i=j$, there is also a simple pole coming from $\psi_\cl(z^k,z^j(x_1))$, and if $i\neq j$ there is no pole.
It is easy to check that the residue of the pole at $x=x_1$ is $-\delta_{i,j}$.
This implies that $M_{x;x_1}(x)$ has no pole at $x=x_1$. From its def\/inition, $M_{x,x_1}(X(z))$ must behave as $o\big((X(z) - X(a))^{-1/2}\big)$ at a ramif\/ication point $a$, and since it is a rational fraction of $X(z)$, this must actually be $O(1)$, meaning that $M_{x;x_1}(x)$ has no poles when $x$ approaches a branchpoint.

Near $z\to p\in \bar{\cal P}$, the only singularity comes from the exponential term and we have
\begin{gather*}
\big[M_{x;x_1}(X(z);\vec t\,)\big]_{i,j}
= (X(z)-x_1)^2 \sum_{k} \om(z^k)  \frac{\psi_\cl\big(z^i(x_1),z^k\big) \psi_\cl\big(z^k,z^j(x_1)\big)}{(\mathrm{d}X(z))^2 \mathrm{d}x_1} + O(1).
\end{gather*}
This shows that $\mathrm{d}x\,M_{x;x_1}(x;\vec t\, )$ has poles in $\bar{\cal P}$, of order at most that already present in~$\om/dX$.
\end{proof}

\begin{proposition}
\label{tht}
We have with respect to the times $t_{p,j}$ ($j \geq 1$)
\begin{itemize}\itemsep=0pt
\item[$(i)$] $\left(\partial_{t_{p,j}} - M_{t_{p,j};x_1}(x ; \vec{t}\, )\right)\Psi_{\cl}(x_1,x ; \om) = 0$.
\item[$(ii)$] $M_{t_{p,j};x_1}$ is a rational function of $x$, with possible poles only at $x = X(p)$.
\end{itemize}
If $X(p) \neq \infty$, the pole is of order $j$. If $X(p) = \infty$, the pole is of degree $1 + \lfloor (j-1)/d_p \rfloor$.
\end{proposition}
The method of the proof is similar to that of Proposition~\ref{thx}. There also exists an analog theorem for $\partial_{t_{p,0}} - \partial_{t_{p',0}}$. We retrieve after sending~$x_1$ to $\infty$ the usual formulation of Lax systems for $\Psi_{\cl}(x;\om)$ and $\Phi_{\cl}(x;\om)$ introduced in Section~\ref{fak}.

\section{Semiclassical Tau function}
\label{sec:tau0}

For later convenience, we def\/ine $\td F_0$ as a shifted version of $F_0$, as follows
\begin{gather*}
\tilde{F_0}(\om)   :=   F_0(\om) - \sum_{i = 1}^{\mathfrak{g}} \epsilon_i \frac{\partial F_0(\om)}{\partial\epsilon_i} + {\rm i}\pi\,\sum_{j,j' = 1}^{\mathfrak{g}} \epsilon_j\epsilon_{j'}\tau_{j,j'} \\
\hphantom{\tilde{F_0}(\om)}{} =   \frac{1}{2}\sum_{k,l = (p,j)} t_kt_l \frac{\partial^2F_0(\om)}{\partial t_k t_l} + \frac{1}{2}\sum_{j,j'  = 1}^{\mathfrak{g}} 2{\rm i}\pi \epsilon_j\epsilon_{j'}\tau_{j,j'},
\end{gather*}
where $\epsilon_i = \frac{1}{2{\rm i}\pi}\oint_{\acycle_i} \om$, and $\om=\sum_k \om_k t_k$.

\begin{definition}
\label{dsa}We def\/ine
\begin{gather*}
\Tau_{\cl}(\mathcal{C},\om) = \ee{\tilde{F_0}(\om)} \theta(\zeta(\om)+\bfc),
\end{gather*}
where we recall that $\zeta(\om)=\frac{1}{2{\rm i}\pi} \oint_{\bcycle-\tau\acycle}\om=\sum_k v_k t_k$ depends linearly on the times. We shall also sometimes use as notation $\Tau_{\cl}(\mathcal{C},\om) \equiv \Tau_{\cl}(\vec t)$ when there is no ambiguity.
In this article, we call~$\Tau_{\mathrm{cl}}$ the ``semiclassical'' Tau function.
\end{definition}

$\Tau_{\cl}(\curve,\om)$ depends as well on the data of a non-singular odd characteristics~$\bfc$ for the Theta function. It is the Tau function \cite{JMII,JMIII,JM81} associated here to the solution of the problem of isospectral evolution described in Section~\ref{sec:recon}. We call Tau function, any function satisfying Hirota bilinear equation, and related to Baker--Akhiezer functions by Sato relation. Sections~\ref{sec:Satocl} and~\ref{sec:Hirotabeq} review the fact that~$\Tau_{\cl}$ f\/its in this def\/inition.

\subsection{Sato relation}
\label{sec:Satocl}
Sato relation \cite{Sato2} means that the Baker--Akhiezer kernel $\psi_{\cl}(z_1,z_2;\om)$ is obtained from the Tau function by a Schlesinger transformation. This can be formulated intrinsically in terms of 1-forms, as

\begin{theorem}\label{thSato}
\begin{gather*}
\frac{\Tau_{\cl}(\curve,\om+\mathrm{d}S_{z_1,z_2})}{\Tau_{\cl}(\curve,\om)} \sqrt{\mathrm{d}X(z_1) \mathrm{d}X(z_2)}  = \psi_{\cl}(z_1,z_2 ; \om).
\end{gather*}
\end{theorem}

\begin{proof} We add two simple poles at $z_1$ and $z_2$ by considering
\begin{gather*}
\om_\lambda(z) = \om(z)+ \lambda\,\mathrm{d}S_{z_1,z_2}(z).
\end{gather*}
According to Theorem~\ref{thdF0dt} we have
\begin{gather}
\label{reg9}
\left.\frac{\partial F_0}{\partial \lambda}\right|_{\lambda=0} = \mu_{z_1}-\mu_{z_2} = \int_{z_2}^{z_1}\om,
\end{gather}
and according to Theorem~\ref{d2F0dti0} we have
\begin{gather}
\label{reg10}
\left.\frac{\partial^2 F_0}{\partial \lambda^2}\right|_{\lambda=0} = - \ln{\left(E(z_1,z_2)^2 \mathrm{d}X(z_1) \mathrm{d}X(z_2)\right)}.
\end{gather}
Since $F_0$ is a quadratic functions of the times, and thus a degree $2$ polynomial in $\lambda$, we have
\begin{gather*}
F_0(\om_\lambda)
 =  F_0(\om) + \lambda\left.\frac{\partial F_0}{\partial \lambda}\right|_{\lambda=0}+ \frac{\lambda^2}{2} \left.\frac{\partial^2 F_0}{\partial \lambda^2}\right|_{\lambda=0}  \\
\hphantom{F_0(\om_\lambda)}{}
 =  F_0(\om) +\lambda\int_{z_2}^{z_1}\om - \lambda^2 \ln{\left(E(z_1,z_2) \sqrt{\mathrm{d}X(z_1) \mathrm{d}X(z_2)}\right)}.
\end{gather*}
It is also easy to see that
\begin{gather*}
\zeta(\om+dS_{z_1,z_2}) = \zeta(\om)  +\frac{\lambda}{2{\rm i}\pi} \oint_{\bcycle-\tau\acycle}  \mathrm{d}S_{z_1,z_2}
 = \zeta(\om) + \lambda(\mathbf{u}(z_1)-\mathbf{u}(z_2)).
\end{gather*}
Taking $\lambda =1$ implies the theorem.
\end{proof}

\subsection{Expansion near poles}

We claim that Theorem~\ref{thSato} (which we wrote intrinsically in terms of 1-forms on $\curve$) is equivalent to Sato's formula (which is written in terms of times $t_k$'s). In order to see that, we need to expand for $z_1,z_2 \in U_p$ near a pole $p \in \mathcal{P}$, where $U_p$ denote an open neighborhood of $p$ on which the local coordinate $\xi_p$ is well-def\/ined. The Schlesinger transformation
\begin{gather*}
\omega \rightarrow \omega + \dd S_{z_1,z_2}
\end{gather*}
can also be written by Taylor expansion, assuming $||\xi_{p}(z)| > \mathrm{max}(|\xi_p(z_1)|,|\xi_p(z_2)|)$
\begin{gather*}
\mathrm{d}S_{z_1,z_2}(z) = \sum_{j \geq 1} \frac{\xi_p(z_1)^j}{j} \omega_{p,j}(z) - \frac{\xi_p(z_2)^j}{j} \omega_{p,j}(z).
\end{gather*}
Thus, the times associated to $p$ after Schlesinger transformation are
\begin{gather*}
\forall \, j \geq 1 \qquad t_{p,j} \longrightarrow  t_{p,j} + \left(\frac{\xi_p(z_1)^j}{j} - \frac{\xi_p(z_2)^j}{j}\right).
\end{gather*}
The usual notation for this special inf\/inite collection of times is
\begin{gather}\label{defsatonotation}
[z_1]_p = \left(\frac{\xi_p(z_1)^j}{j}\right)_{j \geq 1}.
\end{gather}

Thus  Theorem~\ref{thSato}, together with equation~\eqref{defpsiBA} gives the Sato formula in its usual presentation involving an inf\/inite set of times
\begin{gather*}
\psi_{\cl|(p,0)}(z_1) = \frac{\Tau_{\cl}(\vec{t} + [z_1]_p)}{\Tau_{\cl}(\vec{t})}.
\end{gather*}

\subsection{Hirota bilinear equation}
\label{sec:Hirotabeq}

\begin{definition}
For any point $z \in \mathcal{C}$, we def\/ine the insertion operator~$\delta_z $, which acts on functions of a meromorphic $1$-form $\om$ on $\mathcal{C}$,  by considering the one parameter deformation consisting in adding a Bergman kernel $B(z,\cdot)/\mathrm{d}X(z)$ (the dot $\cdot$ means the variable on which $\om=\om(\cdot)$ depends)
\begin{gather}\label{defdeltaz}
\delta_z  f(\om) = \mathrm{d}X(z) \frac{\partial}{\partial \lambda}  f\left.\left(\om+ \lambda  \frac{B(z,\cdot)}{\dd X(z)}\right) \right|_{\lambda=0}.
\end{gather}
$\delta_z$ is a derivation (it satisf\/ies the chain rule $\delta_z (fg) = g \delta_z f+ f \delta_z g$).
In terms of times coordinates
\begin{gather*}
\delta_z f(\vec t) = dX(z)  \partial_{t_{z,1}} f(\vec t\, ).
\end{gather*}
\end{definition}
If $z$ is near a pole $p$, in the local coordinates we have
\begin{gather}\label{defdeltazlocal}
\delta_{z} \equiv \sum_{j =1}^\infty \xi_p(z)^{j - 1} \mathrm{d}\xi_p(z) \partial_{t_{p,j}}.
\end{gather}
Again, we prefer equation~\eqref{defdeltaz}, which is intrinsic and does not require the introduction of an inf\/inite number of times. We write equation~\eqref{defdeltazlocal} just to make the contact with usual presentation (given for instance in~\cite{BBT}).

\begin{proposition}
For any two $1$-forms $\om$ and $\hat\om$ defined on the same Riemann surface $\mathcal{C}$, we have
\begin{gather*}
\Res_{z'\to z} \psi_{\cl}(z,z' ; \om) \psi_{\cl}(z',z ; \hat{\om}) = -\delta_z \ln\frac{\Tau_\cl(\om)}{\Tau_{\cl}(\hat\om)}.
\end{gather*}
\end{proposition}

\begin{proof}
 $\psi_{\cl}(z,z';\om)\psi_{\cl}(z',z ; \hat{\om})$ has a double pole at $z=z'$, and thus evaluating the residue computes a derivative.
\end{proof}

 As a corollary, we get

\begin{theorem}\label{thHirotacl} For any data $(\mathcal{C},\omega)$, the Baker--Akhiezer kernel is self-replicating
\begin{gather*}
\delta_z \psi_{\cl}(z_1,z_2;\om) = - \psi_{\cl}(z_1,z;\om) \psi_{\cl}(z,z_2;\om).
\end{gather*}
\end{theorem}

\begin{proof}
We choose $\hat\om = \om + \mathrm{d}S_{z_1,z_2}$. Then, $\psi_{\cl}(z,z' ; \om) \psi_{\cl}(z',z ; \om + \mathrm{d}S_{z_1,z_2})$ is a meromorphic form in~$z'$ (it has no essential singularity). Its only poles are located at $z'=z$ and $z'=z_2$. Moving the integration contour, we f\/ind
\begin{gather*}
\delta_z \ln\frac{\Tau_\cl(\om+dS_{z_1,z_2})}{\Tau_{\cl}(\om)}
 = \Res_{z'\to z} \psi_{\cl}(z,z' ; \om) \psi_{\cl}(z',z ; \om + \mathrm{d}S_{z_1,z_2}) \\
 \hphantom{\delta_z \ln\frac{\Tau_\cl(\om+dS_{z_1,z_2})}{\Tau_{\cl}(\om)}}{}
 =   - \Res_{z'\to z_2} \psi_{\cl}(z,z' ; \om) \psi_{\cl}(z',z ; \om + \mathrm{d}S_{z_1,z_2}) .
\end{gather*}
We have
\begin{gather*}
\psi_{\cl}(z',z ; \om + \mathrm{d}S_{z_1,z_2})
 =
\frac{E(z',z_1)E(z,z_2) \ee{\int_{z}^{z'}\om + \int_{z_2}^{z_1}\om}}{E(z',z) E(z',z_2) E(z_1,z) E(z_1,z_2)} \\
\hphantom{\psi_{\cl}(z',z ; \om + \mathrm{d}S_{z_1,z_2})=}{}
\times \frac{\theta(\zeta(\om)+\mathbf u(z')-\mathbf u(z)+\mathbf u(z_1)-\mathbf u(z_2)+\mathbf c)}{\theta(\zeta(\om)+\mathbf u(z_1)-\mathbf u(z_2)+\mathbf c)}
\end{gather*}
and thus near $z'\to z_2$
\begin{gather*}
\psi_{\cl}(z',z ; \om + \mathrm{d}S_{z_1,z_2})
 \sim  - \frac{1}{E(z',z_2)}
\frac{\ee{\int_{z}^{z_1}\om }}{E(z_1,z)} \frac{\theta(\zeta(\om)+\mathbf u(z_1)-\mathbf u(z)+\mathbf c)}{\theta(\zeta(\om)+\mathbf u(z_1)-\mathbf u(z_2)+\mathbf c)}   \\
\hphantom{\psi_{\cl}(z',z ; \om + \mathrm{d}S_{z_1,z_2})}{}
\sim  - \frac{1}{E(z',z_2)} \ee{\int_{z_2}^{z_1} \om}
\frac{\psi_{\cl}(z_1,z ; \om )}{\psi_\cl(z_1,z_2;\om)}.
\end{gather*}
Therefore
\begin{gather*}
\delta_z \ln\frac{\Tau_\cl(\om+\dd S_{z_1,z_2})}{\Tau_{\cl}(\om)}   =   - \Res_{z'\to z_2} \psi_{\cl}(z,z' ; \om) \psi_{\cl}(z',z ; \om + \mathrm{d}S_{z_1,z_2}) \\
\hphantom{\delta_z \ln\frac{\Tau_\cl(\om+\dd S_{z_1,z_2})}{\Tau_{\cl}(\om)}}{}
 =   - \psi_{\cl}(z_1,z ; \om) \psi_{\cl}(z,z_2 ; \om).\tag*{\qed}
\end{gather*}
  \renewcommand{\qed}{}
\end{proof}

\noindent This self-replication property is the analog of the Ricatti equation in~\cite{BBT}. Notice that it can also be obtained by a straightforward computation of $\delta_z \ln\psi_{\cl}(z_1,z_2;\om)$ from Def\/inition~\ref{defpsi}) and comparison with the ref\/ined duality equation (Proposition~\ref{thYBsp}).

\subsubsection*{Hirota equation in terms of times}

Theorem \ref{thHirotacl} is written intrinsically in terms of forms~$\om$, but to make contact with usual notations in the literature, let us translate it in terms of times~$t_k$'s. Given any derivation~$\partial_t$, one def\/ines a Hirota operator $D_t$ \cite{Hirota} acting on two functions $f(t)$, $g(t)$, such that
\begin{gather*}
(D_t f \cdot g)(t) \equiv \partial_{u} f(t + u)g(t - u) \big|_{u = 0} =  g^2(t)\partial_t(f/g).
\end{gather*}
In particular, there is a Hirota operator $D_z$ associated to derivation $\delta_z = \partial_{t_{z,1}}$. This allows us to reformulate the self-replication property as
\begin{proposition}
\label{Hirohiro} For $\Tau_{\cl}=\Tau_{\cl}(\vec t\, )$ written as a function of times, for any pole~$p$, the self-replication property is equivalent to the Hirota bilinear difference equation
\begin{gather*}
D_{z} \Tau_{\cl}(\vec{t} + [z_1]_p - [z_2]_p) \cdot \Tau_{\cl}(\vec{t}) = - \Tau_{\cl}(\vec{t} + [z_1]_p - [z]_p)\Tau_{\cl}(\vec{t} + [z]_p - [z_2]_p).
\end{gather*}
\end{proposition}

Actually, this property for~$\Tau_{\mathrm{cl}}$ is equivalent to the Fay identity (see Section~\ref{sec:theta}) satisf\/ied by the Theta function $\theta(\cdot|\tau)$. The Hirota equation can also be written in a more symmetric way by setting
\begin{gather*}
\vec{t} \longleftarrow \vec{t} - \frac{[z_1]_p}{2} + \frac{[z_2]_p}{2}.
\end{gather*}
Namely
\begin{gather*}
  D_{z} \Tau_{\cl}\left(\vec{t} + \frac{[z_1]_p}{2} - \frac{[z_2]_p}{2}\right) \cdot \Tau_{\cl}\left(\vec{t} - \frac{[z_1]_p}{2} + \frac{[z_2]_p}{2}\right) \nonumber \\
 \qquad{} =   - \Tau_{\cl}\left(\vec{t} - [z]_p + \frac{[z_1]_p}{2} + \frac{[z_2]_p}{2}\right) \Tau_{\cl}\left(\vec{t} + [z]_p - \frac{[z_1]_p}{2} - \frac{[z_2]_p}{2}\right).
\end{gather*}
The procedure to translate it into a set of dif\/ferential equations (with respect to the times) is well-known, it is merely obtained by Taylor expansion in $\xi_p(z)$, $\xi_p(z_1)$ and $\xi_p(z_2)$ (see e.g.~\cite{JM1}). This gives an inf\/inite set of partial dif\/ferential equations involving derivatives of the Tau function with respect to times $t_{p,j}$, which are equations of a KP hierarchy. These equations are equivalent to Hirota bilinear equation. The fact that Theta functions via Def\/inition~\ref{dsa} provides solutions to the KP hierarchy (the so-called algebro-geometric solutions) was discovered by~\cite{ItsMatv,Kri77}. The notion of Tau function was only introduced later in~\cite{JM81}, but its expression for the algebro-geometric solution is a straightforward reformulation of those results, and coincides with what we call semiclassical tau function~$\mathcal{T}_{\cl}(\vec{t})$.

\section{Proposal for a new tau function and spinor kernel}
\label{sec:dis}
So far, the spectral curve $(\curve,X,Y)$ was f\/ixed once for all, and was equipped with a 1-form~$\om$ depending linearly on times ($\om= \sum_k t_k\om_k$).
Now, we shall let the spectral curve itself change around~$(\curve,X,Y)$, and in particular we may vary the complex structure of the curve $\curve$. $Y\mathrm{d}X$ shall play the role of the 1-form.

\subsection{More geometry: symplectic invariants}
\label{sec:sympinv}

For any spectral curve $\spcurve=(\curve,X,Y)$, a sequence of symplectic invariants $F_g[\spcurve]$, and of symplectic covariant forms $\om_n^{(g)}(\spcurve)$ was def\/ined in~\cite{EOFg}.
Let us recall their def\/inition and main properties.

\subsubsection{Topological recursion}
\label{sec:toporec}
Let $a_i$ be the ramif\/ication points of $\spcurve$, i.e. the zeroes of $\mathrm{d}X$. We assume that the spectral curve~$\spcurve$ is regular, i.e.~$a_i$ are simple zeroes of $\mathrm{d}X$ and $\mathrm{d}Y(a_i) \neq 0$. Then, $Y$ behaves like a~squareroot near ramif\/ication points
\begin{gather*}
Y(z) = Y(a_i)+Y'(a_i) \sqrt{X(z)-X(a_i)}+ O\big(X(z) - X(a_i)\big),
\end{gather*}
and there is a unique other point~$\overline{z} \neq z$ such that $X(z) = X(\overline{z})$, at least for $z$ in a neighborhood~$U_{a_i}$ of~$a_i$. Then we def\/ine the recursion kernel~\cite{EOFg}
\begin{gather*}
K(z_0,z) = -\frac{1}{2}\frac{\int_{\overline{z}}^z  B(z_0,\cdot)}{(Y(z)-Y(\overline{z}))  \mathrm{d}X(z)}.
\end{gather*}
In the variable $z_0$, it is a meromorphic 1-form def\/ined globally for $z_0\in \curve$, and in the variable~$z$, it is the inverse of a 1-form, only def\/ined in $\bigcup_i U_{a_i}$.

\begin{definition}
\label{defspinv}
The ``symplectic covariant forms'' $\om_n^{(g)}(z_1,\dots,z_n)$ are meromorphic $\underbrace{(1,\ldots,1)}_{n \ \mathrm{times}}$-forms def\/ined by the following recursion
\begin{gather*}
\om_1^{(0)}(z) = Y(z) \mathrm{d}X(z),\qquad \om_2^{(0)}(z_1,z_2) = B(z_1,z_2),
\end{gather*}
and for $J = \{z_2,\ldots,z_n\}$
\begin{gather*}
\om_{n}^{(g)}(z_0,J) = \sum_i \Res_{z \rightarrow a_i}  K(z_0,z) \left(\om_{n+1}^{(g-1)}(z,\overline{z},J) + \sum_{0 \leq h \leq g, I\subseteq J}' \! \om_{1+|I|}^{(h)}(z,I)  \om_{1+n-|I|}^{(g-h)}(\overline{z},J\setminus I) \right),
\end{gather*}
where $\sum'$ means that terms containing a $\omega_1^{(0)}$ factor are excluded.
\end{definition}

One can prove that
\begin{theorem}[\cite{EOFg}]
For $2-2g-n<0$,
$\om_n^{(g)}(z_1,\dots,z_n)$ is symmetric in its $n$ variables, and it is a meromorphic form in each variable, having poles only at ramification points, with vanishing residues. The poles are of order at most $6g+2n-4$.
\end{theorem}

For instance, for $(g,n) = (0,3)$, this def\/inition yields
\begin{gather}
\omega_3^{(0)}(z_0,z_1,z_2)   =   \sum_{i} \Res_{z \rightarrow a_i} \left[-\frac{1}{2}\frac{\int_{\overline{z}}^{z} B(z_0,\cdot)}{(Y(z) - Y(\overline{z}))\dd X(z)} \big(B(z,z_1)B(\overline{z},z_2) + B(\overline{z},z_1)B(z,z_2)\big)\right] \nonumber \\
\hphantom{\omega_3^{(0)}(z_0,z_1,z_2)}{}
  =   \sum_{i} \Res_{z \rightarrow a_i} \frac{B(z_0,z)B(z_1,z)B(z_2,z)}{\dd X(z) \dd Y(z)}.\label{W30ex}
\end{gather}
We shall comment this expression in Section~\ref{sec:spgeom} below.

\begin{definition}
The ``symplectic invariants'' $F_g$ are numbers associated to $\mathcal{S}$, as follows:
\begin{itemize}\itemsep=0pt
\item[$\bullet$] For $g=0$, we def\/ine $F_0$ as in Def\/inition~\ref{defF0}, using $\om=YdX$, i.e., $F_0(\spcurve):=F_0(YdX)$.
\item[$\bullet$] For $g=1$, $F_1$ is def\/ined in terms of the Bergman Tau function $\Tau_{B,X}$ introduced by Kokotov and Korotkin~\cite{KoKo}
\begin{gather*}
F_1(\spcurve) := -\frac{1}{2}\ln(\Tau_{B,X}) - \frac{1}{24}\ln\left(\prod_{i} Y'(a_i)\right),
\end{gather*}
where $Y'(a_i) = \lim\limits_{z\to a_i} (Y(z)-Y(a_i))/\sqrt{X(z)-X(a_i)}$.
$F_1$ is related to the logarithm of the determinant of the Laplacian on $\curve$ with metrics $|y\mathrm{d}x|^2$.
\item[$\bullet$] For $g\geq 2$
\begin{gather*}
F_g(\spcurve) := \frac{1}{2-2g} \sum_i \Res_{z\to a_i} \Phi(z) \om_1^{(g)}(z),
\end{gather*}
where $\Phi(z)$ is any primitive\footnote{Since $\omega_1^{(g)}(z)$ for $g \geq 2$ has no residues, $F_g$ is independent of the choice of $\Phi$.} of $Y\mathrm{d}X$, i.e.\ $\mathrm{d}\Phi=Y\mathrm{d}X$.
\end{itemize}
As a notation we write $\om_0^{(g)} \equiv F_g$.
\end{definition}

The name ``symplectic invariants'' comes from the important property, proved in~\cite{EOFg, EO2MM}, that $F_g$ and the cohomology class of $\omega_n^{(g)}$ are invariant under the following transformations of spectral curves, each of which let the symbolic symplectic form $\dd X\wedge \dd Y$ invariant
\begin{gather*}
(X,Y) \rightarrow (-Y,X),\qquad (X,Y) \rightarrow (X,Y + R(X)),\qquad (X,Y) \rightarrow (\lambda X,Y/\lambda),
\end{gather*}
where $\mathrm{R}$ is any rational function and $\lambda\in \mathbb C^*$.

\subsubsection{Inf\/initesimal deformations: special geometry}
\label{sec:spgeom}

The times $t_{k}$'s of the 1-form $Y\dd X$ provide locally a set of coordinates of the moduli space of spectral curves $\spcurve=(\curve,X,Y)$~\cite{Kri92}. Inf\/initesimal deformations of the spectral curves are given by f\/lows associated to the 1-forms $\om_{k}$ dual to the times~$t_k$'s. The following theorem holds for the invariants $\om_{n}^{(g)}$

\begin{theorem}[\cite{EOFg}]
\label{thspecialgeo}
For any of the times $t_k$, we have
\begin{gather*}
\forall \, n,g \geq 0,\qquad \frac{\partial}{\partial t_k}\Big|_{X(z_i)\ {\rm f\/ixed}} \om_n^{(g)}(z_1,\dots,z_n)  = \int_{\om_k^*}  \om_{n+1}^{(g)}(\cdot,z_1,\dots,z_n),
\end{gather*}
where $\om_k^*$ is the dual cycle associated to the time $t_k$ in equation~\eqref{eqformcycletimeduality}.
\end{theorem}
This set of relations is called ``special geometry'' by physicists in the context of string theory. Let us give some examples:
\begin{itemize}\itemsep=0pt
\item[$\bullet$] First kind deformations/addition of holomorphic forms
\begin{gather*}
\frac{\partial \om_n^{(g)}(z_1,\dots,z_n)}{\partial \epsilon_i} = \oint_{{\cal B}_i}  \om_{n+1}^{(g)}(\cdot,z_1,\dots,z_n). \nonumber
\end{gather*}
\item[$\bullet$] Second kind deformations/addition of a double pole
\begin{gather*}
\frac{\partial\om_n^{(g)}(z_1,\dots,z_n)}{\partial t_{p,1}} = \frac{\omega_{n + 1}^{(g)}(p,z_1,\dots,z_n)}{\mathrm{d}\xi_p(p)}.
\end{gather*}
If $p$ is a pole of $Y\mathrm{d}X$, this does not apply to $(g,n) = (0,0)$.
\item[$\bullet$] Third kind deformations/addition of simple poles
\begin{gather*}
\frac{\partial \om_n^{(g)}(z_1,\dots,z_n)}{\partial t_{p,0}}- \frac{\partial \om_n^{(g)}(z_1,\dots,z_n)}{\partial t_{p',0}} = \int_{p'}^p \om_{n+1}^{(g)}(\cdot,z_1,\dots,z_n).
\end{gather*}
\item[$\bullet$] $(g,n) = (0,0)$. We retrieve Theorem~\ref{thdF0dt}  (Section~\ref{sec:prepot}) for the derivatives of~$F_0$
\begin{gather*}
\frac{\partial F_0}{\partial t_k} = \oint_{\om_k^*}  Y\mathrm{d}X.
\end{gather*}
\item[$\bullet$] $(g,n)=(0,1)$. We retrieve the def\/inition of the form-cycle duality equation~\eqref{eqdualitytkomk}
\begin{gather*}
\frac{\partial Y\mathrm{d}X}{\partial t_k}(z) = \om_k(z) = \oint_{\om_k^*}  B(\cdot,z).
\end{gather*}
\item[$\bullet$] $(g,n) = (0,2)$. Thanks to equation~\eqref{W30ex},
\begin{gather*}
\frac{\partial B(z_1,z_2)}{\partial t_k} = \sum_{i} \Res_{z \rightarrow a_i} \frac{B(z,z_1)B(z,z_2)\omega_k(z)}{\dd X(z) \dd Y(z)}.
\end{gather*}
If we compare it to Rauch variational formula~\cite{Rauchvar}, we retrieve that the variation of the complex structure of $\mathcal{C}$ is such that
\begin{gather}
\label{varco}
\frac{\partial X(a_i)}{\partial_{t_k}} = \frac{\omega_k(a_i)}{\dd Y(a_i)}.
\end{gather}
In particular, we deduce that the time evolution of the spectral curve that we consider obeys the Whitham equations~\cite{Kri92, Whitham}
\begin{gather*}
(\partial_{t_k} X(a_i)) \omega_l(a_i) = (\partial_{t_l} X(a_i)) \omega_k(a_i).
\end{gather*}
Furthermore, if we integrate $z_1$ and $z_2$ over the dual cycles $\omega^*_l$ and $\omega^*_m$, we retrieve Theorem~\ref{th23} (see also~\cite{Kri92}).
\end{itemize}
Although we make use of the meromorphic deformations (listed above) in this article, the special geometry relations may remain valid for deformations along certain multivalued $1$-forms, namely the primaries and their descendents listed in~\cite{Dubrovin} in the context of Hurwitz spaces, or the $1$-forms coupled to the Whitham times of \cite[Section~7]{Kri92} for a slightly dif\/ferent moduli space.

\subsubsection{Finite deformations}
\label{dowu}

Instead of inf\/initesimal deformations of the spectral curve, one may consider f\/inite deformations: $t_k\to t_k+\frac{1}{N} c_k$ (where $N$ is a formal parameter, which will serve as a formal expansion parameter for Taylor series).
In other words, to a spectral curve $\spcurve$ and a meromorphic $1$-form $\Omega=\sum_k c_k \om_k$, we shall associate a new spectral curve, denoted $\spcurve+\frac{1}{N}\Omega$, as push-forward of~$\mathcal{S}$ by the f\/low def\/ined by $\Omega$. This convenient notation should not hide the fact that the conformal structure of~$\mathcal{C}$ changes under such global deformations (see equation~\eqref{varco}). A more detailed example of computations with global deformation is provided in the proof of Lemma~\ref{qpqpq} below.

\begin{definition}
For any meromorphic 1-form $\Omega$ on $\curve$, written with its times coordinates $c_k$ (as in equation~\eqref{eqformcycletimeduality})
\begin{gather*}
\Omega = \sum_k c_k \om_k
\qquad {\rm with \ dual \  cycle}\quad
\Omega^* = \sum_k c_k \om_k^*,
\end{gather*}
and for any functional $f[\mathcal{S}]\equiv f(\vec t\, )$ depending on a spectral curve $\mathcal{S}$, i.e.\ on the times $t_k$'s, we take as a def\/inition\footnote{With an appropriate regularization  when $\Omega$ has simple poles.} of $f[\spcurve+\frac{1}{N}\Omega]$ the following formal series in $N^{-1}$
\begin{gather*}
f\left[\mathcal{S}+\frac{1}{N}\Omega\right] := \sum_{k \geq 0} \frac{N^{-k}}{k!}\,
\left(\frac{\partial^k}{\partial\lambda^{k}} f[\vec t+\lambda\vec c\,]\Big|_{\lambda=0} \right).
\end{gather*}
\end{definition}
If the functional $f$ is $F_g$ or $\omega_n^g$, the coef\/f\/icients in the Taylor series can be computed by the relations of special geometry Theorem~\ref{thspecialgeo}, namely
\begin{gather*}
\omega_n^{(g)}\!\left[\mathcal{S}+\frac{1}{N}{\Omega}\right]\!(z_1,\dots,z_n)   =   \omega_n^{(g)}[\mathcal{S}](z_1,\dots,z_n)  + \!\sum_{k \geq 1}\! \frac{N^{-k}}{k!}\!\underbrace{\int_{\Omega^*}\!\cdots\!\int_{\Omega^*}}_{k \ \mathrm{times}}\! \omega_{k + n}^{(0)}[\mathcal{S}](z_1,\ldots,z_n,\bullet), \\
F_g[\mathcal{S}+\frac{1}{N}{\Omega}]  =   F_g[\mathcal{S}] + \sum_{k \geq 1} \frac{N^{-k}}{k!}\underbrace{\int_{\Omega^*}\cdots\int_{\Omega^*}}_{k \ \mathrm{times}} \omega_{k}^{(0)}[\mathcal{S}],
\end{gather*}
so that the computation of $F_g$'s or $\om_n^{(g)}$ for ${\spcurve+\frac{1}{N}\Omega}$ does not involve derivatives with respect to times, but only integrals at f\/ixed times on~$\spcurve$.

\subsection{Tau function}

\subsubsection{Preliminaries}

We need Theta functions, and we shall introduce an appropriate notation for our purposes. For any $\mathbf w\in \mathbb C^\genus$, for any $\mathbf\epsilon\in \mathbb C^\genus$, and for any $\genus\times\genus$ symmetric matrix $\tau$ of positive imaginary part $\operatorname{Im} \tau>0$, and for any $(\mu,\nu)\in \mathbb C^\genus \times \mathbb C^\genus$ we def\/ine the $\Theta_{[\mu,\nu]}$ function as
\begin{gather*}
\Theta_{[\mu,\nu]}(\mathbf{w}|\tau;\mathbf\epsilon) := \sum_{\mathbf{p} \in \mathbb{Z}^{\mathfrak{g}}} e^{{\rm i}\pi(\mathbf{p} + \mu - N\epsilon)\cdot\tau\cdot(\mathbf{p} + \mu - N\epsilon) + (\mathbf{p} + \mu - N\epsilon)\cdot\mathbf{w} + 2{\rm i}\pi\mathbf{p}\cdot\nu}.
\end{gather*}
It is closely related to the usual Siegel theta function with characteristics $[\mu,\nu]$
\begin{gather*}
\Theta_{[\mu,\nu]}(\mathbf{w}|\tau;\mathbf\epsilon) = e^{{\rm i}\pi N^2 \epsilon\cdot\tau\cdot\epsilon - N\epsilon\cdot \mathbf{w} - 2{\rm i}\pi\mu\cdot\nu} \vartheta\bigl[{}^{\mu}_{\nu}\bigr](\mathbf{w}/2{\rm i}\pi - N\tau\cdot\epsilon|\tau),
\end{gather*}
where
\begin{gather*}
\vartheta\bigl[{}^{\mu}_{\nu}\bigr](\mathbf{w}|\tau)= \sum_{\mathbf{p}\in {\mathbb Z}^{\genus}}
\exp\bigl[ {\rm i}\pi (\mathbf{p}+\mu) \tau (\mathbf{p}+\mu) + 2{\rm i}\pi  (\mathbf{p}+\mu)(\mathbf{w}+\nu)\bigr],
\end{gather*}
$[\mu,\nu]$ plays the role of a characteristics, although we do not require it to be half-integer here. Most often, we shall omit to write the dependance in $[\mu,\nu]$, as well as the dependence in~$\tau$,~$\mathbf\epsilon$, and we shall use the notation $\Theta', \Theta'', \dots, \Theta^{(k)}$ for the tensor of derivatives with respect to~$\mathbf{w}$. For instance
\begin{gather*}
\Theta' = \left(\frac{\partial \Theta}{\partial w_1},\ldots,\frac{\partial \Theta}{\partial w_\genus}\right)^t.
\end{gather*}
This $\Theta_{[\mu,\nu]}$ function satisf\/ies the heat equation
\begin{gather*}
\frac{1}{2}\left(\partial_{\tau_{i,j}} + \partial_{\tau_{j,i}}\right) \Theta_{[\mu,\nu]}(\mathbf{w}) = {\rm i}\pi \Theta''(\mathbf{w})_{i,j},
\end{gather*}
where, in this equation, $\tau_{i,j}$ and $\tau_{j,i}$ are considered independent.

\subsubsection{Def\/inition and comments}

\begin{definition}[\cite{Ecv,EMhol}]\label{defTaumunu}
For any $[\mu,\nu] \in \mathbb{C}^{\mathfrak{g}}\times \mathbb{C}^{\mathfrak{g}}$ and any spectral curve $\spcurve=(\mathcal{C},X,Y)$ of genus~$\genus$, we def\/ine
\begin{gather*}
\Tau_{[\mu,\nu]}[\spcurve] = \exp\left(\sum_{g \geq 0} N^{2-2g} F_g\right)\left\{\sum_{k \geq 0} \sum_{l_i > 0} \sum_{h_i > 1 - l_i/2} \frac{N^{\sum_i (2-2h_i-l_i)}}{k!   l_1!\cdots l_k!}\,F_{h_1}^{(l_1)}\cdots F_{h_k}^{(l_k)}\,\Theta^{(\sum_i l_i)}\right\},
\end{gather*}
where we def\/ined
\begin{gather*}
F_h^{(k)} = \overbrace{\oint_{\bcycle}\dots\oint_{\bcycle}}^{k \ {\rm times}}  \om_k^{(h)}(\spcurve),
\qquad
\Theta^{(k)} = \nabla_{\mathbf{w}}^{\otimes k}\Theta_{[\mu,\nu]}(\mathbf{w} = \mathbf{w}_0 \,|\,\tau;\mathbf\epsilon),
\\
\epsilon = \frac{1}{2\ii\pi}\oint_{\mathcal{A}}\om_1^{(0)},\qquad \mathbf{w}_0 = NF_0' = N\oint_{\mathcal{B}}\om_1^{(0)},\qquad \tau = \frac{1}{2{\rm i}\pi}F_0'' = \frac{1}{2\ii\pi}\oint_{\mathcal{B}}\oint_{\mathcal{B}} \om_2^{(0)}.
\end{gather*}
$\Tau$ is def\/ined formally order by order in $N$ (in the coef\/f\/icient of~$1/N^k$, each term $\Theta^{(j)}$ is considered formally of order~1). It can be seen as a genuine asymptotic series when $\spcurve$ is such that~$ \mathbf w_0$ is of order~$O(1)$.
\end{definition}

To the f\/irst few orders in $1/N$ we have explicitly
\begin{gather*}
\Tau_{[\mu,\nu]}[\spcurve]
  =   \ee{N^2 F_0}  \ee{F_1} \biggl\{\Theta + \frac{1}{N}\left(\Theta' F_1' + \frac{1}{6} \Theta''' F_0'''\right)+ \frac{1}{N^2}\bigg(\Theta\,F_2+\frac{1}{2}\Theta'' F_1''   \\
 \hphantom{\Tau_{[\mu,\nu]}[\spcurve]=}{} + \frac{1}{2}\Theta'' F_1'^2 + \frac{1}{24}\Theta^{(4)} F_0'''' + \frac{1}{6}\Theta^{(4)} F_0''' F_1' + \frac{1}{72}\Theta^{(6)} F_0'''^2 \bigg)  + o\big(1/N^2\big) \biggr\}.
\end{gather*}
Let us emphasize that this def\/inition of $\Tau$ does not need require the computation of derivatives with respect to times, all the terms consist in cycle integrals on $\curve$ at a given time.

Notice that when $\mu + \tau\nu$ is a non-singular half-integer odd characteristics, the leading term when $N \rightarrow \infty$ of $\Tau(\spcurve)$ coincides with the semiclassical Tau function of Section~\ref{sec:tau0} computed for the dif\/ferential form $\omega = N Y\dd X$, up to an exponential factor
\begin{gather*}
\Tau[\spcurve] \sim  e^{N^2F_0} \Theta(\mathbf{w}_0|\tau) \sim e^{N^2\tilde{F_0}}\,\theta(\zeta + \nu + \mu\cdot\tau) e^{2{\rm i}\pi\mu\cdot\zeta} \sim \Tau_{\mathrm{cl}}(NY\dd X) e^{2{\rm i}\pi\mu\cdot\zeta}.
\end{gather*}
Notice that the Fay identity for $\theta$ presented in Section~\ref{sec:theta} is also true if we multiply $\theta$ by an exponential factor of its argument. Therefore, Theorem~\ref{Hirohiro} ensures that the large~$N$ limit of~$\Tau[\mathcal{S}]$ satisf\/ies the Hirota bilinear equation (Theorem.~\ref{Hirohiro}), i.e.\ is a~Tau function to leading order. This is also true for arbitrary characteristics $[\mu,\nu]$, although the times have to be shifted by a~(maybe complex) constant in this case. We conjecture (see Section~\ref{sec:Hi}) that $\Tau$ is actually a~Tau function, i.e.\ satisf\/ies Hirota equations to all orders.

We also mention that under a modular transformation (i.e.\ a change of choice for the~$\acycle$ and~$\bcycle$ cycles), $\Tau$ changes like the Siegel Theta function of characteristics $[\mu,\nu]$ (see~\cite{EMhol})

\begin{proposition}[\cite{EMhol}]\label{modularZ}
Under a modular ${\rm Sp}({2\genus},\mathbb Z)$  transformation $\tau\to\td\tau= (\alpha\tau+\beta)(\gamma\tau+\delta)^{-1}$, the characteristics $[\mu,\nu]$ changes as $\mu\to \td\mu= \delta\mu-\gamma\nu+\frac{1}{2}(\gamma \delta^t)_{\rm diag}$, $\nu\to \td\nu=-\beta\mu+\alpha\nu+\frac{1}{2}(\alpha \beta^t)_{\rm diag}$, the Tau function $\mathcal{T} \equiv \mathcal{T}_{[\mu,\nu]}$ transforms as
\begin{gather*}
\Tau_{[\mu,\nu]} \to \zeta_{[\mu,\nu]}(\alpha,\beta,\gamma,\delta) \Tau_{[\td\mu,\td\nu]},
\end{gather*}
where $\zeta_{[\mu,\nu]}(\alpha,\beta,\gamma,\delta)$ is the phase factor, independent of the spectral curve.
\end{proposition}

\subsubsection{Heuristic motivation for the def\/inition}

Let us consider
\begin{gather*}
Z_N[\mathcal{S},\mathbf{n}] = \exp\left(\sum_{g = 0}^{\infty} N^{2-2g}\,F_g(\mathcal{S},\mathbf{n})\right),
\end{gather*}
where we emphasize the dependence of $F_g(\spcurve)=F_g(\mathcal{S},\mathbf{n})$ in the vector $\mathbf n=(n_1,\dots,n_\genus)$ of f\/illing fractions $n_k=\frac{1}{2{\rm i}\pi}\oint_{\acycle_k} Y \dd X$.
Let $F_g(\mathcal{S},\mathbf{n})$ be the symplectic invariants associated to the curve~$\mathcal{S}$ whose vector of f\/illing fractions is $\mathbf{n} = (n_1,\ldots,n_{\genus}) \in \mathbb{C}^{\mathfrak{g}}$.
$\ln Z[\mathcal{S},\mathbf{n}]$  is a formal Laurent series in $N$ (in particular we emphasize that it contains no oscillatory terms, by opposition with the def\/inition to come). The relevance of $Z[\mathcal{S},\mathbf{n}]$ in topological strings has been pointed out in the work of Dijkgraaf and Vafa~\cite{DVphase}.
Def\/inition~\ref{defTaumunu} gives a precise meaning to the sum over all shifts of f\/illing fractions by integers
\begin{gather}
\label{eq:taumean}
\Tau_{\mu,\nu}[\spcurve] \, \text{``=''}\, \sum_{\mathbf{n'} \in \mathbb Z^\genus}  e^{2{\rm i}\pi\mathbf{n'}\cdot\nu} Z[\mathcal{S}',\mathbf{n}+\mathbf{n'} + \mu].
\end{gather}
Such sums over a lattice have also been considered in the context of the small dispersion limit of KdV in the genus $\mathfrak{g}$ regime~\cite{Venakdv2}. In the context of hermitian matrix integrals, it has been used in~\cite{BDE,Ecv} to arrive to the formal asymptotic series of Def\/inition~\ref{defTaumunu}. This series has then been proposed to describe non-perturbative ef\/fects in topological strings~\cite{EMhol}.

\subsection[Baker-Akhiezer spinor kernel]{Baker--Akhiezer spinor kernel}\label{sec:BAsk}

We now def\/ine the spinor kernel $\psi(z_1,z_2 ; \mathcal{S})$ through Sato's relation

\begin{definition}
\label{defpsis}
\begin{gather*}
\psi(z_1,z_2 ; \mathcal{S})  =   \frac{\Tau(\spcurve ; \spcurve+\frac{1}{N}\,\mathrm{d}S_{z_1,z_2})}{\Tau(\spcurve)}   \sqrt{\mathrm{d}X(z_1) \mathrm{d}X(z_2)}   =   \frac{\Tau(\spcurve+[z_1]-[z_2])}{\Tau(\spcurve)}  \sqrt{\mathrm{d}X(z_1) \mathrm{d}X(z_2)},
\end{gather*}
where $[z_1]-[z_2]=\frac{1}{N} dS_{z_1,z_2}$ is Sato's notation, see equation~\eqref{defsatonotation}.
\end{definition}

$\psi(z_1,z_2)$ is again def\/ined formally, order by order in $1/N$. The leading order coincides with $\psi_{\mathrm{cl}}$ introduced in Def\/inition~\ref{defpsi}. Let us give the f\/irst few orders
\begin{gather*}
  \psi(z_1,z_2 ; \mathcal{S})
  =   \frac{e^{N\int_{z_2}^{z_1}Y\mathrm{d}X}}{E(z_1,z_2)} \frac{\Theta\big(\mathbf{w}_0 + 2{\rm i}\pi(\mathbf{u}(z_1) - \mathbf{u}(z_2))\big)}{\Theta\big(\mathbf{w}_0\big)}\Bigg\{ 1+ \frac{1}{N}\Bigg[\frac{1}{6}\int_{z_2}^{z_1}\int_{z_2}^{z_1}\int_{z_2}^{z_1} \omega_3^{(0)} \nonumber \\
  \hphantom{\psi(z_1,z_2 ; \mathcal{S}) =}{}
+ \int_{z_2}^{z_1} \omega_1^{(1)}  + \frac{1}{2}\frac{\Theta'\big(\mathbf{w}_0 + 2{\rm i}\pi(\mathbf{u}(z_1) - \mathbf{u}(z_2)))}{\Theta\big(\mathbf{w}_0 + 2{\rm i}\pi(\mathbf{u}(z_1) - \mathbf{u}(z_2))\big)} \int_{z_2}^{z_1}\int_{z_2}^{z_1}\oint_{\mathcal{B}} \omega_3^{(0)} \nonumber \\
\hphantom{\psi(z_1,z_2 ; \mathcal{S}) =}{}
  + \frac{1}{2}\frac{\Theta''\big(\mathbf{w}_0 + 2{\rm i}\pi(\mathbf{u}(z_1) - \mathbf{u}(z_2))\big)}{\Theta\big(\mathbf{w}_0 + 2{\rm i}\pi(\mathbf{u}(z_1) - \mathbf{u}(z_2))\big)} \int_{z_2}^{z_1} \oint_{\mathcal{B}}\oint_{\mathcal{B}} \omega_3^{(0)} \nonumber \\
\hphantom{\psi(z_1,z_2 ; \mathcal{S}) =}{}
  + \frac{1}{6}\left(\frac{\Theta'''\big(\mathbf{w}_0 + 2{\rm i}\pi(\mathbf{u}(z_1) - \mathbf{u}(z_2))\big)}{\Theta\big(\mathbf{w}_0 + 2{\rm i}\pi(\mathbf{u}(z_1) - \mathbf{u}(z_2))\big)} - \frac{\Theta'''\big(\mathbf{w}_0\big)}{\Theta\big(\mathbf{w}_0\big)}\right)F_0''' \nonumber \\
\hphantom{\psi(z_1,z_2 ; \mathcal{S}) =}{}
  + \left(\frac{\Theta'\big(\mathbf{w}_0 + 2{\rm i}\pi(\mathbf{u}(z_1) - \mathbf{u}(z_2))\big)}{\Theta\big(\mathbf{w}_0 + 2{\rm i}\pi(\mathbf{u}(z_1) - \mathbf{u}(z_2))\big)} - \frac{\Theta'\big(\mathbf{w}_0\big)}{\Theta\big(\mathbf{w}_0\big)}\right)F_1'\Bigg]
+ o(1/N)\Bigg\}.
\end{gather*}

\begin{lemma}
\label{qpqpq}
$\psi(z_1,z_2;\mathcal{S})$ is a well-defined spinor in $z_1$ and $z_2$. Furthermore
\begin{itemize}\itemsep=0pt
\item[$\bullet$] $\psi(z_1,z_2)$ has a simple pole at $z_1=z_2$
\[
\psi(z_1,z_2 ; \mathcal{S}) \mathop{\sim}_{z_1 \rightarrow z_2}  \frac{\sqrt{\mathrm{d}X(z_1) \mathrm{d}X(z_2)}}{X(z_1)-X(z_2)}.
\]
\item[$\bullet$] It has an essential singularity near any pole $p$ of $Y\mathrm{d}X$, of the form $\ee{N\int_{z_2}^{z_1} Y\mathrm{d}X}$.
\item[$\bullet$] At all orders in $1/N$ $($except at leading order$)$, $\psi(z_1,z_2)$ has poles at the ramification points~$a_i$. Their order increase with the order of $1/N$.
\end{itemize}
\end{lemma}

\begin{proof} When $z_1$ goes around an $\mathcal{A}$-cycle, $\mathrm{d}S_{z_1,z_2}$ is unchanged.
When~$z_1$ goes around a cycle~$\bcycle_j$, $\mathrm{d}S_{z_1,z_2}$ is shifted by a holomorphic form $\mathrm{d}S_{z_1,z_2}\to \mathrm{d}S_{z_1,z_2}+2{\rm i}\pi \mathrm{d}u_j$, which is dual to a $\partial/\partial \epsilon_j$, and since the Tau function is background independent (it was proved in \cite{EMhol} that $\partial \Tau/\partial \epsilon_i=0$), then it is unchanged. This shows that $\psi(z_1,z_2;\mathcal{S})$ is a well-def\/ined spinor.

Then we compute each term of $\Tau(\spcurve+[z_1]-[z_2])$ by writing the Taylor expansion (Section~\ref{dowu})
\begin{gather*}
F_g\left[\spcurve +\frac{\lambda}{N}\mathrm{d}S_{z_1,z_2}\right] = \sum_{n \geq 0}  \frac{\lambda^n}{n!  N^n}  \left. \frac{\partial^nF_g}{\partial \lambda^n}\right|_{\lambda=0},
\end{gather*}
which we need to evaluate at $\lambda=1$. The $n$-th derivatives of $F_g$ at $\lambda=0$ are computed by the special geometry relations Theorem~\ref{thspecialgeo}, using the dual cycle $(\mathrm{d}S_{z_1,z_2})^* = [z_2,z_1]$
\begin{gather*}
\left. \frac{\partial^n F_g}{\partial \lambda^n}\right|_{\lambda=0} = \int_{z_2}^{z_1}\cdots\int_{z_2}^{z_1}\om_n^{(g)}.
\end{gather*}
All the $\om_n^{(g)}$ with $2-2g-n<0$ are meromorphic, and have poles only at ramif\/ication points, without residues. This implies that their contribution to $\psi(z_1,z_2)$ provides only poles at ramif\/ication points. The only terms involving $\om_n^{(g)}$ with $2-2g-n\geq 0$, are $\partial_\lambda F_0$, $\partial_\lambda^2 F_0$ and $\partial_\lambda F_0'$.
\begin{itemize}\itemsep=0pt
\item[$\bullet$] $\left. \frac{\partial F_0}{\partial \lambda}\right|_{\lambda=0} = \int_{z_2}^{z_1}Y\mathrm{d}X$, which contributes to $\psi$ as the essential singularity $\ee{N\int_{z_2}^{z_1} Y\mathrm{d}X}$.
\item[$\bullet$] $\left. \frac{\partial^2 F_0}{\partial \lambda^2}\right|_{\lambda=0} = -\ln{(E(z_1,z_2)^2 \mathrm{d}X(z_1) \mathrm{d}X(z_2))}$
which contributes to $\psi$ as $1/E(z_1,z_2)$.
\item[$\bullet$] $\left. \frac{\partial F'_0}{\partial \lambda}\right|_{\lambda=0} = 2{\rm i}\pi (\mathbf{u}(z_1)-\mathbf{u}(z_2))$
which does not yield any singularity.
\end{itemize}
All the other terms have $2-2g-n<0$, and contribute order by order, only as combinations of meromorphic forms and derivatives of Theta functions, having poles at ramif\/ication points conveyed by the $\omega_n^{(g)}$'s with $2 - 2g - n < 0$.
\end{proof}

Notice that as a corollary of Proposition~\ref{modularZ}, $\psi$ has nice modular properties:
\begin{corollary}
Under a modular ${\rm Sp}({2\genus},\mathbb Z)$  transformation with the notations of Proposition~{\rm \ref{modularZ}}, the spinor kernel $\psi_{[\mu,\nu]}(z_1,z_2)$ transforms as
\begin{gather*}
\psi_{[\mu,\nu]}(z_1,z_2) \to  \psi_{[\td\mu,\td\nu]}(z_1,z_2).
\end{gather*}
\end{corollary}

\section{Correlators}
\label{sec:co}

\subsection[Second kind deformations of $\mathcal{S}$]{Second kind deformations of $\boldsymbol{\mathcal{S}}$}

\label{sec7}
Let us recall the def\/inition of the insertion operator $\delta_z$, already encountered in Section~\ref{sec:Hirotabeq} and adapted now for varying spectral curves.
\begin{definition} \label{def:delta} We def\/ine the insertion operator~$\delta_z$, acting on a functional $f(\spcurve)$ of a spectral curve $\spcurve$, as follows
\begin{gather*}
\delta_z f =\left. \mathrm{d}X(z) \frac{\partial}{\partial \lambda} f(\spcurve_\lambda)\right|_{\lambda=0},
\end{gather*}
where the family of spectral curves $\spcurve_\lambda=\spcurve+\lambda B(z,\cdot)/\dd X(z)$ is such that
\begin{gather*}
 (Y\mathrm{d}X)_\lambda = Y\mathrm{d}X + {\lambda} \frac{B(z,\cdot)}{\mathrm{d}X(z)}.
\end{gather*}
In other words $\delta_z=\dd X(z) \partial/\partial t_{z,1}$ as in Section~\ref{sec:Hirotabeq}.
\end{definition}

The dual cycle of $B(z,\cdot)/\mathrm{d}X(z)$ is the contour surrounding $z$ with index~$1$
\begin{gather*}
B(z,\cdot) =  \Res_{z'\to z}  B(z',\cdot) \frac{\mathrm{d}X(z)}{(X(z')-X(z))}.
\end{gather*}
Then, the relations of special geometry (Theorem~\ref{thspecialgeo}) for $\omega_n^{(g)}$ imply, for any $n,g \geq 0$
\begin{gather*}
\delta_z \om_n^{(g)}(z_1,\dots,z_n)   =   \Res_{z'\to z}  \om_{n+1}^{(g)}(z',z_1,\dots,z_n) \frac{\mathrm{d}X(z)}{(X(z')-X(z))}
  =   \om_{n+1}^{(g)}(z,z_1,\dots,z_n).
\end{gather*}
For instance
\begin{gather*}
\delta_z F_0 = \om_1^{(0)}(z) = Y(z) \mathrm{d}X(z) ,\qquad \delta_z F_g = \om_1^{(g)}(z) ,\qquad \delta_z \om_1^{(0)}(z') = B(z,z'),
\end{gather*}
and the conformal structure of $\mathcal{C}$ changes such that
\begin{gather*}
\delta_{z} X(a_i) = \frac{B(z,a_i)}{\dd Y(a_i)}.
\end{gather*}

\begin{definition}
\label{def:co}For $n$ positive integer, we def\/ine the correlators $W_n(z_1,\ldots,z_n)$ and the disconnected correlators $\overline{W}_n(z_1,\ldots,z_n)$ as
\begin{gather*}
W_n(z_1,\ldots,z_n) = N^{-n} \delta_{z_1}\cdots\delta_{z_n} \ln {\Tau(\spcurve)},\qquad \overline{W}_n(z_1,\ldots,z_n) = \frac{N^{-n} \delta_{z_1}\cdots\delta_{z_n} \Tau(\spcurve)}{\Tau(\spcurve)}.
\end{gather*}
\end{definition}
$W_n(z_1,\dots,z_n)$ and $\overline{W}_n(z_1,\ldots,z_n)$ are $(1,\ldots,1)$-forms ($n$~times), symmetric in their~$n$ variables. Each coef\/f\/icient, order by order in $1/N$, is a meromorphic form with poles at ramif\/ication points.

\subsection{Examples}
\label{exex}
For instance the three f\/irst orders of $W_1$ are
\begin{gather*}
W_1(z)  =   N Y(z) \mathrm{d}X(z) + (\ln \Theta)'\cdot 2{\rm i}\pi\mathrm{d}\mathbf{u}(z) +\frac{1}{N}\Bigg\{\om_1^{(1)}(z) + \frac{\Theta''}{\Theta}\cdot (i\pi\delta_z\tau)\\
\hphantom{W_1(z)  =}{}
  + \left[F'_1 \left(\frac{\Theta''}{\Theta}-\frac{\Theta'^2}{\Theta^2}\right) + \frac{F_0'''}{6} \left(\frac{\Theta^{''''}}{\Theta}-\frac{\Theta''' \Theta'}{\Theta^2}\right)\right]\cdot 2{\rm i}\pi \mathrm{d}\mathbf{u}(z)\Bigg\}
  + o(1/N),
\end{gather*}
where we recall that
\begin{gather*}
\delta_z \tau_{j,k}   =   \frac{\delta_z F_0''}{2{\rm i}\pi} = \frac{1}{2{\rm i}\pi}\oint_{\mathcal{B}_j}\oint_{\mathcal{B}_k} \omega_3^{(0)}(\cdot,\cdot,z) \\
\hphantom{\delta_z \tau_{j,k}}{}
 =   4 i\pi  \sum_l \Res_{z'\to a_l} K(z,z') \mathrm{d}u_j(z')\mathrm{d}u_k(\overline{z'}) = 2{\rm i}\pi  \sum_l \Res_{z'\to a_l} \frac{B(z,z') \mathrm{d}u_j(z')\mathrm{d}u_k(z')}{\mathrm{d}X(z') \mathrm{d}Y(z')}, \\
\om_1^{(1)}(z_1)   =   \sum_l  \Res_{z\to a_l} K(z_1,z) B(z,\overline{z}).
\end{gather*}
For the 2-point correlator, the three f\/irst orders are
\begin{gather*}
W_2(z_1,z_2)   =   B(z_1,z_2) + (\ln \Theta)''\cdot 2{\rm i}\pi \mathrm{d}\mathbf{u}(z_1) \otimes 2{\rm i}\pi \mathrm{d}\mathbf{u}(z_2)
  + \frac{1}{N}\Bigg\{\frac{\Theta'}{\Theta}\cdot\int_{\mathcal{B}}\omega_3^{(0)}(\cdot,z_1,z_2) \nonumber \\
\hphantom{W_2(z_1,z_2)   =}{}
 + \left(\frac{\Theta''}{\Theta}\right)'\cdot\big[i\pi(\delta_{z_1}\tau) \otimes 2{\rm i}\pi\dd \mathbf{u}(z_2) + 2{\rm i}\pi\dd \mathbf{u}(z_1)\otimes i\pi(\delta_{z_2}\tau)\big]
  \\
\hphantom{W_2(z_1,z_2)   =}{}
  +\left[(F_1)' \left(\frac{\Theta''}{\Theta} - \frac{\Theta'^2}{\Theta^2}\right)' + \frac{F_0'''}{6}\left(\frac{\Theta''''}{\Theta} - \frac{\Theta'''\Theta'}{\Theta^2}\right)'\right]\cdot 2{\rm i}\pi\dd \mathbf{u}(z_1)\otimes 2{\rm i}\pi\dd\mathbf{u}(z_2)\Bigg\}  \\
\hphantom{W_2(z_1,z_2)   =}{}
  + o(1/N).
\end{gather*}
For $n \geq 3$, the leading order of the $n$-point correlator is a $O(1)$, and is obtained by successive applications of $\delta_z$ to the $(\ln \Theta)''$ term
\begin{gather*}
W_n(z_1,\dots,z_n)   =   (\ln \Theta)^{(n)}\cdot \bigotimes_{j = 1}^n 2{\rm i}\pi \dd\mathbf{u}(z_j) + \frac{1}{N}\Bigg\{\delta_{n,3}\omega_3^{(0)}(z_1,z_2,z_3) \\
\hphantom{W_n(z_1,\dots,z_n)   =}{}
  + \sum_{j = 1}^n \left(\frac{\Theta''}{\Theta}\right)^{(n - 1)}\cdot (i\pi\delta_{z_j}\tau)\otimes\bigotimes_{k \neq j} 2{\rm i}\pi\dd\mathbf{u}(z_k)  \\
\hphantom{W_n(z_1,\dots,z_n)   =}{}
  + \sum_{1 \leq j < k \leq n} (\ln \Theta)^{(n - 1)}\cdot\left(\int_{\mathcal{B}} \omega^{(0)}_3(\cdot,z_j,z_k)\right)\otimes\bigotimes_{l \neq j,k} 2{\rm i}\pi\dd\mathbf{u}(z_l)   \\
\hphantom{W_n(z_1,\dots,z_n)   =}{}
  + \left[(F_1)' \!\left(\frac{\Theta''}{\Theta} - \frac{\Theta'^2}{\Theta^2}\right)^{(n - 1)}\!\! + \frac{F_0'''}{6}\!\left(\frac{\Theta''''}{\Theta} - \frac{\Theta'''\Theta'}{\Theta^2}\right)^{(n - 1)}\!\right]\!\cdot\bigotimes_{j = 1}^n 2{\rm i}\pi\dd\mathbf{u}(z_j)\Bigg\}    \\
\hphantom{W_n(z_1,\dots,z_n)   =}{}
  + o(1/N).
\end{gather*}

\subsection{Loop equations}

\begin{theorem}\label{llop}
The dispersive Tau function obeys the loop equations. Namely, let us denote by~$\Gamma$ a contour separating the poles of~$Y\dd X$, from the set of preimages of a point $x\in \mathbb{C}$
\begin{alignat*}{3}
& (i)   \quad &&   \oint_{z \in \Gamma} \frac{\delta_{z} \ln\Tau[\mathcal{S}] - N (Y\dd X)(z)}{X(z) - x} = 0, &  \\
& (ii) \quad && \oint_{z \in \Gamma} \Res_{z' \rightarrow z} \frac{1}{(X(z) - x)(X(z') - X(z))}\left(\frac{1}{\Tau[\spcurve]} \delta_{z}\delta_{z'} \Tau[\mathcal{S}] - \frac{\dd X(z) \dd X(z')}{(X(z) - X(z'))^2} \right) = Q(x),  &
\end{alignat*}
where $Q$ is a rational function of $x$, whose only poles are those of $Y\mathrm{d}X$, and with degree one less than that of $Y\mathrm{d}X$.
\end{theorem}

Those loop equations can be written in terms of correlators by applying $\delta_{z_2}\cdots\delta_{z_n}$.

\begin{theorem}
Let $J=\{z_2,\dots,z_n\}$. The correlators satisfy
\begin{itemize}\itemsep=0pt
\item[$(i)$] The linear loop equations. For all $n \geq 1$
\begin{gather*}
\oint_{z \in \Gamma} \frac{1}{X(z) - x}\left(W_{n}(z,J) -\delta_{n,1}\,N\,(Y\dd X)(z) - \delta_{n,2} \frac{\mathrm{d}X(z) \mathrm{d}X(z_2)}{(X(z)-X(z_2))^2}\right) = 0.
\end{gather*}
\item[$(ii)$] The quadratic loop equations. For all $n \geq 1$
\begin{gather*}
  \oint_{z \in \Gamma}\Res_{z' \rightarrow z} \frac{1}{(X(z) - x)(X(z') - X(z))}\Bigg\{\sum_{I\subseteq J} W_{1+|I|}(z,I)\,W_{n-|I|}(z',J\setminus I) \\
\qquad\quad{} +\frac{1}{N^2} W_{n+1}(z,z',J) + \dd X(z) \dd X(z') \sum_{z_k\in J}\mathrm{d}_{z_k}\left(\frac{W_{n - 1}(J)}{(x-X(z_k)) \mathrm{d}X(z_k)} \right)\Bigg\}\\
\qquad{} = Q_n(x;J)
\end{gather*}
defines a quantity $Q_n(x;J)$ which is a rational function of~$x$, whose only poles are located at those of~$Y\mathrm{d}X$.
\end{itemize}
\end{theorem}
The important information in loop equations, is that those particular combinations of $W_n$'s have no monodromies in the variable $x$ around the branchpoints. Since every $W_n$ has poles at ramif\/ication points to all orders in $1/N$, this is a highly non-trivial property.

\begin{proof} The $F_g$ were precisely introduced such that for any $\mathbf{n}$
\begin{gather*}
Z[\mathcal{S},\mathbf{n}] = \exp\left(\sum_g N^{2-2g}  F_g(\mathcal{S},\mathbf{n} + \mu)\right)
\end{gather*}
is a solution of the loop equations. Since $\mathcal{\tau}[\spcurve]$ is constructed formally as a linear combination (see equation~\eqref{eq:taumean}) of such objects, $\Tau[\spcurve]$ satisfy the same loop equations. As a matter of fact, $\mathcal{\Tau}[\mathcal{S}]$ was introduced in~\cite{Ecv} so as to solve those loop equations while preserving modularity.
\end{proof}

\section{Hirota equations}
\label{sec:Hi}

We mentioned in Theorem~\ref{Hirohiro} that the self-replication property of the Baker--Akhiezer spinor kernel~$\psi_{\mathrm{cl}}$ is equivalent to an inf\/initesimal Fay identity for the semiclassical spectral curve, which is known in turn to be equivalent to Hirota equations. In a similar way, we conjecture here
\begin{conjecture}
\label{consj}$\psi$ is self-replicating
\begin{gather*}
\frac{1}{N} (\delta_z\psi)(z_1,z_2) = -\psi(z_1,z)\psi(z,z_2).
\end{gather*}
\end{conjecture}

We have not been able to prove this conjecture. We prove in Appendix~\ref{apphirota1surN} that it holds up to $o(1/N)$. We argue in Section~\ref{sec:mat} that it is compatible with what is known for spectral curves coming from the one matrix model (hyperelliptic curves), or the two matrix model. Besides, these matrix models do not allow to reach all plane curves~$\mathcal{S}$.
The dif\/f\/iculty in f\/inding a proof of Conjecture~\ref{consj} comes from the singularities at ramif\/ication points. For instance, one can always write
\begin{gather*}
\frac{1}{N} (\delta_z\psi)(z_1,z_2 ; \mathcal{S}) = \Res_{z' \rightarrow z} \psi(z,z' ; \mathcal{S}) \psi(z',z ; \mathcal{S} + [z_1] - [z_2]),
\end{gather*}
where $\mathcal{S} + [z_1] - [z_2]=\spcurve + \frac{1}{N} \dd S_{z_1,z_2}$. Since the integrand is a dif\/ferential form on the Riemann surface $\curve$ underlying $\mathcal{S}$, we can move the contour to the poles at the ramif\/ication points, and the pole at $z' = z_2$
\begin{gather*}
  \frac{1}{N} (\delta_z\psi)(z_1,z_2 ; \mathcal{S})
  =   - \psi(z_1,z ; \mathcal{S})\psi(z,z_2 ; \mathcal{S}) - \sum_{i} \Res_{z' \rightarrow a_i} \psi(z,z' ; \mathcal{S}) \psi(z',z ; \mathcal{S} + [z_1] - [z_2]).
\end{gather*}
Then, it remains to show that the sum of residues at ramif\/ication points vanishes.
So, Conjectu\-re~\ref{consj} is equivalent to
\begin{conjecture}\label{consjai}
\begin{gather*}
\sum_{i} \Res_{z' \rightarrow a_i} \psi(z,z' ; \mathcal{S}) \psi(z',z ; \mathcal{S} + [z_1] - [z_2]) =0.
\end{gather*}
\end{conjecture}

In Appendix~\ref{apphirota1surN}, we check that this residue at each~$a_i$ is~$o(1/N)$. This involves already non-trivial identities between Theta functions associated to a complex curve, like Fay identity, its degenerations and dif\/ferentiations with respect to the moduli of the curve, and involves the precise expression of~$\om_3^{(0)}$ (equation~\eqref{W30ex}). We have not been able yet to f\/ind a general way to show that this residue is~$0$ to all orders in~$1/N$.

In terms of the $\mathcal{T}$ function, Conjecture~\ref{consj} can be rephrased
\begin{conjecture}
\label{eqhys}
$\mathcal{T}$ satisfy an infinitesimal version of Hirota equations
\begin{gather*}
  \Tau\big[\mathcal{S}\big] (\delta_z\Tau)\big[\mathcal{S} + [z_1] - [z_2]\big] - (\delta_z\Tau)\big[\mathcal{S}\big] \Tau\big[\mathcal{S} + [z_1] - [z_2]\big] \nonumber \\
\qquad{} =   - N \Tau\big[\mathcal{S} + [z_1] - [z]\big] \Tau\big[\mathcal{S} + [z] - [z_2]\big].
\end{gather*}
\end{conjecture}
There is also a global version of the former conjecture. First, notice from our def\/inition in Section~\ref{sec:dis} that
\begin{gather*}
\Tau\big[(\mathcal{S} + [z_1] - [z_2]) + [z_3] - [z_4]\big] = \Tau\big[(\mathcal{S} + [z_3] - [z_4]) + [z_1] - [z_2]\big],
\end{gather*}
so omitting the parentheses makes sense, but
\begin{gather*}
\Tau\big[\mathcal{S} + [z_1] - [z_4] + [z_3] - [z_2]\big] = - \Tau\big[\mathcal{S} + [z_1] - [z_2] + [z_3] - [z_4]\big].
\end{gather*}
This sign comes from the fact that the def\/inition of $\Tau$ contains a regularization procedure (for $\int_{z_i}^{z_j}\int_{z_i}^{z_j} B$), whose result depends on the way we form the pairs of simple poles to add to~$\mathcal{S}$.

\begin{conjecture}\label{eqhys2}
$\Tau$ satisfies Hirota equations
\begin{gather*}
  \Tau\big[(\mathcal{S} + [z_1] - [z_2]) + [z_3] - [z_4]\big]\,\Tau\big[\mathcal{S}\big]  \nonumber \\
\qquad{} =   \Tau\big[\mathcal{S} + [z_1] - [z_2]\big]\,\Tau\big[\mathcal{S} + [z_3] - [z_4]\big] - \Tau\big[\mathcal{S} + [z_3] - [z_2]\big]\,\Tau\big[\mathcal{S} + [z_1] - [z_4]\big].
 \end{gather*}
\end{conjecture}
Provided our conjectures hold, $\Tau$ and $\psi$ are actually the Tau function and the ``wave function'' of a dispersive integrable system.

\begin{proof}[Proof of equivalence of Conjectures~\ref{eqhys} and \ref{eqhys2}.]
 We can obtain Conjecture~\ref{eqhys} from \linebreak Conjectu\-re~\ref{eqhys2} by letting $z_1$ and $z_2$ merge to a point $z$. In the other direction, we use shorter notations
\begin{gather*}
\Tau_{ijkl} = \Tau\big[\mathcal{S} + [z_i] - [z_j] + [z_k] - [z_l]\big],\qquad \Tau_{ij} = \Tau\big[\mathcal{S} + [z_i] - [z_j]\big],\qquad \Tau = \Tau[\mathcal{S}],
\end{gather*}
and we apply Conjecture~\ref{eqhys} to the spectral curve $\mathcal{S} + [z_3] - [z_4]$
\begin{gather*}
 \delta_z \Tau_{1234} = \frac{\delta_z\Tau_{34}}{\Tau_{34}} \Tau_{1234} - N \frac{ \Tau_{1z34}\Tau_{z234}}{\Tau_{34}} = \frac{\delta_z\Tau}{\Tau} \Tau_{1234} - N \frac{\Tau_{3z}\Tau_{z4}\Tau_{1234}}{\Tau \Tau_{34}} - N \frac{ \Tau_{1z34}\Tau_{z234}}{\Tau_{34}}.
\end{gather*}
Exchanging the roles of $z_1\leftrightarrow z_3$ and $z_2\leftrightarrow z_4$ also gives
\begin{gather*}
 \delta_z \Tau_{1234}
= \frac{\delta_z\Tau}{\Tau} \Tau_{1234} - N \frac{\Tau_{1z}\Tau_{z2}\Tau_{1234}}{\Tau \Tau_{12}} - N \frac{ \Tau_{3z12}\Tau_{z412}}{\Tau_{12}},
\end{gather*}
and comparing the two, we may get rid of the terms involving $\delta_z$
\begin{gather}
\label{hulo}
\Tau_{12}\Tau_{3z}\Tau_{z4}\Tau_{1234} + \Tau\Tau_{12}\Tau_{1z34}\Tau_{z234} =\Tau_{34}\Tau_{1z}\Tau_{z2}\Tau_{1234} + \Tau\Tau_{34}\Tau_{3z12}\Tau_{z412}.
\end{gather}
Let us def\/ine
\begin{gather*}
U_{1234} = \Tau\Tau_{1234} - \Tau_{12}\Tau_{34}+\Tau_{14}\Tau_{32},
\end{gather*}
which is the quantity which should vanish at the end of our computation. Notice that
\begin{gather*}
\lim_{z_1\to z_2} \mathcal{U}_{1234}=0.
\end{gather*}
Indeed
\begin{gather*}
\mathcal{U}_{1234} \mathop{\sim}_{z_1\to z_2}\left(
\frac{\Tau\Tau_{34}}{E_{12}}+\Tau\delta_1\Tau_{34} - \frac{\Tau\Tau_{34}}{E_{12}} - \delta_1\Tau \Tau_{34} + \Tau_{14}\Tau_{31}\right)\\
\hphantom{\mathcal{U}_{1234}}{}
 \mathop{\sim}_{z_1\to z_2}
\big(\Tau\delta_1\Tau_{34}  - \delta_1\Tau \Tau_{34} + \Tau_{14}\Tau_{31}\big),
\end{gather*}
where $E_{ij} = E(z_i,z_j)$, and this expression vanish by application of Conjecture~\ref{eqhys} to the spectral curve $\mathcal{S}$. Notice also that the remark about $\Tau$ made above Conjecture~\ref{eqhys2} implies $\mathcal{U}_{ijkl} = \mathcal{U}_{klij} = -\mathcal{U}_{ilkj}$. Let us rewrite equation~\eqref{hulo} in terms of $\mathcal{U}_{ijkl}$ only
\begin{gather} \Tau_{12} \mathcal{U}_{1z34}\,\mathcal{U}_{z234} - \Tau_{34} \mathcal{U}_{3z12} \mathcal{U}_{z412} + \big(\Tau_{12}\Tau_{3z}\Tau_{z4} - \Tau_{1z}\Tau_{z2}\Tau_{34}\big)\mathcal{U}_{1234} \nonumber \\
\qquad{} + \big(\Tau_{1z}\Tau_{34} - \Tau_{14}\Tau_{3z}\big)\Tau_{12} \mathcal{U}_{z234}  + \big(\Tau_{z2}\Tau_{34} - \Tau_{32}\Tau_{z4}\big)\Tau_{12} \mathcal{U}_{1z34} \nonumber \\
\qquad{} + \big(\Tau_{32}\Tau_{1z} - \Tau_{12}\Tau_{3z}\big)\Tau_{34} \mathcal{U}_{z412} + \big(\Tau_{z2}\Tau_{14} - \Tau_{12}\Tau_{z4}\big)\Tau_{34} \mathcal{U}_{3z12}
  =   0.\label{yyy}
\end{gather}
The left hand side may have simple poles when $z_i \rightarrow z_j$ due to $\Tau_{ij}$, but not higher degree poles since we know that $\mathcal{U}_{ijkl}$ is actually regular when $z_i \rightarrow z_j$ or $z_l$. For equation~\eqref{yyy} to hold, in particular, the coef\/f\/icient of the simple pole when $z_2 \rightarrow z_3$ must vanish
\begin{gather*}
-\Tau_{13}\Tau_{z4} \mathcal{U}_{1z34} + \Tau_{1z}\Tau_{34}\,\mathcal{U}_{z413} = 0,
\end{gather*}
which we can also write after reindexing the points and using the symmetries of~$\mathcal{U}$
\begin{gather}
\label{opop}
\Tau_{31}\Tau_{z4} \mathcal{U}_{1z34} = \Tau_{3z}\Tau_{14} \mathcal{U}_{z134}.
\end{gather}
Similarly, the coef\/f\/icient of the simple pole when $z_1 \rightarrow z_2$ must vanish
\begin{gather*}
\mathcal{U}_{1z34} \mathcal{U}_{z134}
 + (\Tau_{z1}\Tau_{34}-\Tau_{z4}\Tau_{31})\mathcal{U}_{1z34}
 + (\Tau_{1z}\Tau_{34}-\Tau_{14}\Tau_{3z})\mathcal{U}_{z134} = 0.
\end{gather*}
Now, we may combine the latter with equation~\eqref{opop}, set $z = z_2$ for convenience, and isolate~$\mathcal{U}_{1234}$
\begin{gather*}
\frac{\Tau_{31}\Tau_{24}}{\Tau_{32}\Tau_{14}} \mathcal{U}_{1234}^2 + \left((\Tau_{21}\Tau_{34} - \Tau_{24}\Tau_{31}) + \frac{\Tau_{31}\Tau_{24}}{\Tau_{32}\Tau_{14}}(\Tau_{12}\Tau_{34} - \Tau_{14}\Tau_{32})\right)\mathcal{U}_{1234} = 0.
\end{gather*}
If $\mathcal{U}_{1234}$ were not identically zero, we would have (by continuity of all the coef\/f\/icients of the series) for any points $z_1$, $z_2$, $z_3$, $z_4$ on the curve
\begin{gather*}
\mathcal{U}_{1234} =  \frac{\Tau_{32}\Tau_{14}}{\Tau_{31}\Tau_{24}}(\Tau_{24}\Tau_{31} - \Tau_{21}\Tau_{34}) + (\Tau_{14}\Tau_{32} - \Tau_{12}\Tau_{34}).
\end{gather*}
But matching the coef\/f\/icient of the simple pole when $z_1 \rightarrow z_4$ in this equation yields
\begin{gather*}
0 = \frac{\Tau_{32}}{\Tau_{31}\Tau_{21}}(\Tau_{21}\Tau_{31} - \Tau_{21}\Tau_{31}) + \Tau_{32} = \Tau_{32},
\end{gather*}
which is not true. Therefore, $\mathcal{U}_{1234} \equiv 0$.
\end{proof}

\section{Consequences}
\label{sec:more}

\subsection{Exponential formula}

We start with a remark which does not rely on the conjectures of Section~\ref{sec:Hi},
but which is natural to present now. Recall that the kernel is def\/ined by Sato's formula
\begin{gather*}
\psi(z_1,z_2 ; \mathcal{S}) = \frac{\Tau(\spcurve+\frac{1}{N} \mathrm{d}S_{z_1,z_2})}{\Tau(\spcurve)}  \sqrt{\mathrm{d}X(z_1) \mathrm{d}X(z_2)}.
\end{gather*}
Adding a double pole can be realized by adding two simple poles and taking the limit where the two simple poles collapse. In other words, we can write
\begin{gather*}
\mathrm{d}S_{z_1,z_2} = \int_{z_2}^{z_1} B = (X(z_1)-X(z_2)) B(\cdot,z_1)/\dd X(z_1) + O((X(z_1)-X(z_2)))^2,
\end{gather*}
and then express $\psi$ in terms of second kind deformations of $\Tau$, i.e.\ in terms of the correlators~$W_n$. However, when we substitute second kind deformations instead of third kind deformations, we must pay attention to regularization for the term $\int_{z_2}^{z_1}\int_{z_2}^{z_1} B$, like in Sections~\ref{sec:Satocl} and~\ref{sec:BAsk}. Besides, for any~$n$, $W_n$ has a f\/inite number of terms which are not~$O(1/N)$. For this reason we def\/ine the following quantities, which are~$O(1/N)$
\begin{gather*}
\widehat{W}_1(z)   =   W_1(z) - N\,Y\dd X(z) - (\ln \Theta)'\cdot 2{\rm i}\pi \dd\mathbf{u}(z), \\
\widehat{W}_2(z_1,z_2)   =   W_2(z_1,z_2) - B(z_1,z_2) - (\ln \Theta)''\cdot 2{\rm i}\pi \dd\mathbf{u}(z_1)\otimes 2{\rm i}\pi \dd\mathbf{u}(z_2),
\end{gather*}
and for $n \geq 3$
\begin{gather*}
\widehat{W}_n(z_1,\ldots,z_n) = W_n(z_1,\ldots,z_n) - (\ln \Theta)^{(n)}\cdot\bigotimes_{j = 1}^n 2{\rm i}\pi \dd\mathbf{u}(z_j).
\end{gather*}
Then, we have the exponential formula
\begin{proposition}\label{psop}
\begin{gather*}
\psi(z_1,z_2) =  \frac{e^{N\int_{z_2}^{z_1}Y\dd X}}{E(z_1,z_2)} \frac{\Theta_{12}}{\Theta} \exp\left(\sum_{n \geq 1} \frac{1}{n!} \int_{z_2}^{z_1} \cdots \int_{z_2}^{z_1} \widehat{W}_n\right),
\end{gather*}
where $\Theta_{12}=\Theta\big(\mathbf{w}_0 + 2{\rm i}\pi(\mathbf{u}(z_1) - \mathbf{u}(z_2))\big)$ and $\Theta=\Theta\big(\mathbf{w}_0\big)$. This formula is an equality if we collect on both sides all the terms of the same order.
\end{proposition}

\begin{proof} The Taylor formula allows to express $\ln\Tau\big[\mathcal{S} + [z_1] - [z_2]\big] - \ln\Tau\big[\mathcal{S}\big]$ to all orders in $1/N$. For the f\/irst and second order, we use equations~\eqref{reg9} and~\eqref{reg10}
\begin{gather*}  \ln\Tau\big[\mathcal{S} + [z_1] - [z_2]\big] - \ln\Tau\big[\mathcal{S}\big]
  =   \int_{z_2}^{z_1} N^{-1}\delta_{\zeta}(\ln \Tau) + \frac{1}{2}\,\text{``}
  \int_{z_2}^{z_1}\int_{z_2}^{z_1} N^{-2}\delta_{\zeta}\delta_{\zeta'}(\ln \Tau)\big[\mathcal{S}\big]\,\text{''}    \\
\qquad\quad{} + \sum_{n \geq 3} \frac{N^{-n}}{n!} \int_{z_2}^{z_1} \underbrace{\delta_{\zeta_1} \cdots\int_{z_2}^{z_1} \delta_{\zeta_n}}_{n\ \mathrm{times}} (\ln \Tau)\big[\mathcal{S}\big]   \\
\qquad {} =  \int_{z_2}^{z_1} W_1 + \frac{1}{2}\,\text{``} \int_{z_2}^{z_1}\int_{z_2}^{z_1} W_2\,\text{''} + \sum_{n \geq 3} \frac{1}{n!} \underbrace{\int_{z_2}^{z_1}\cdots\int_{z_2}^{z_1}}_{n\ \mathrm{times}} W_n   \\
\qquad{}  =   N\int_{z_2}^{z_1} Y\dd X - \frac{1}{2}\ln\big(\big(E(z_1,z_2)\big)^2\dd X(z_1)\dd X(z_2)\big)  \\
\qquad\quad{}
 + \sum_{n \geq 1} \frac{1}{n!}(\ln \Theta)^{(n)}\cdot \big(2{\rm i}\pi (\mathbf{u}(z_1) - \mathbf{u}(z_2))\big)^{\otimes n}
  + \sum_{n \geq 1} \frac{1}{n!} \underbrace{\int_{z_2}^{z_1}\cdots\int_{z_2}^{z_1}}_{n\ \mathrm{times}} \widehat{W}_n.
\end{gather*}
In the last step, we have used the expression for the leading order of $W_n$ found in Section~\ref{exex}. Then, the second line can be resummated into~$\ln \Theta_{12}$, and the whole result exponentiated leads to the announced formula.
\end{proof}

Notice that, when expanding the exponential, to any given order $O(N^{-k})$, all the $W_n$ give a~contribution involving derivatives of theta functions contracted with tensor products $\big(\mathbf{u}(z_1) - \mathbf{u}(z_2)\big)^{\otimes n}$. These contributions have to be resummated into a single theta function (or derivatives thereof) with argument shifted by $\mathbf{u}(z_1) - \mathbf{u}(z_2)$, then producing an expression at the order sought involving only a f\/inite number of terms.

\subsection{Determinantal formulas}
\label{detto}
Conversely, the correlators $W_n(z_1,\ldots,z_n)$ can be retrieved from the spinor kernel $\psi(z_1,z_2)$, they are the determinantal correlation functions built with~$\psi(z_i,z_j)$. Let us consider f\/irst the case of~$W_1$, which does not rely on the conjectures of Section~\ref{sec:Hi}.
\begin{lemma}\label{lW1}
\begin{gather*}
W_1(z)  =  N\,Y\dd X(z) + \lim_{z'\to z}\left(\psi(z',z)e^{-N\int_{z_2}^{z_1} Y{\mathrm d}X}-\frac{\sqrt{\mathrm{d}X(z') \mathrm{d}X(z)}}{X(z')-X(z)}\right).
\end{gather*}
\end{lemma}

\begin{proof}
First, we notice that adding a double pole can be realized by adding two simple poles and take the limit where the two simple poles collapse. More precisely, when $z'\to z$, we have
\begin{gather*}
\mathrm{d}S_{z',z}(z_0) \mathop{\sim}_{z' \rightarrow z} (X(z')-X(z)) \frac{B(z,z_0)}{\mathrm{d}X(z)}.
\end{gather*}
We can thus use the def\/inition of $\delta_z$ with $\lambda = X(z')-X(z)$.
For any regular functional $f[\spcurve]$ of the spectral curve, we thus have
\begin{gather*}
f\left[\spcurve+\frac{1}{N}\mathrm{d}S_{z',z}\right]=f[\spcurve]+\frac{X(z')-X(z)}{N \mathrm{d}X(z)} \delta_z f[\spcurve] + O\big((X(z')-X(z))^2\big).
\end{gather*}
In some sense, we trade a variation of $\spcurve$ by a second kind dif\/ferential with a variation with a~third kind dif\/ferential.

In particular, $F_g$ with $g\geq 1$, or every $\Theta$-term in Def\/inition~\ref{defTaumunu} are regular functionals of the spectral curve.
We just have to pay attention to the $F_0$ term, because the derivative of $F_0$ with respect to third kind dif\/ferentials involves a regularization procedure, whereas the derivative with respect to second kind dif\/ferentials does not.
We have, by Taylor expansion, and computing all derivatives from special geometry (Theorem~\ref{thspecialgeo})
\begin{gather*}
F_0\big[\spcurve+\lambda\,\mathrm{d}S_{z',z}\big]
 =  F_0[\spcurve]+\sum_{n=1}^\infty  \frac{\lambda^n}{n!} \frac{\partial^n}{\partial\lambda^n}  F_0\big[\spcurve+\lambda \mathrm{d}S_{z',z}\big]\big|_{\lambda=0}   \\
\hphantom{F_0\big[\spcurve+\lambda\,\mathrm{d}S_{z',z}\big]}{}
=  F_0[\spcurve]+\lambda\int_{z}^{z'} Y\mathrm{d}X  -  \frac{\lambda^2}{2}\ln{(E(z,z')^2\mathrm{d}X(z)\mathrm{d}X(z'))} \\
\hphantom{F_0\big[\spcurve+\lambda\,\mathrm{d}S_{z',z}\big]=}{}
  + \sum_{n=3}^{\infty} \frac{\lambda^n}{n!} \int_{z}^{z'}\cdots \int_{z}^{z'}\, W_n^{(0)}   \\
\hphantom{F_0\big[\spcurve+\lambda\,\mathrm{d}S_{z',z}\big]}{}
 =  F_0(\spcurve)+\lambda\int_{z}^{z'} Y\mathrm{d}X -  \frac{\lambda^2}{2}\ln\big(E(z,z')^2\mathrm{d}X(z)\mathrm{d}X(z')\big)    \\
\hphantom{F_0\big[\spcurve+\lambda\,\mathrm{d}S_{z',z}\big]=}{}
 + O\big((X(z')-X(z))^2\big).
\end{gather*}
Taking $\lambda=1/N$ gives
\begin{gather*}
  F_0\left[\spcurve+\frac{1}{N} \mathrm{d}S_{z',z}\right]
+  \frac{1}{N^2}\ln{(E(z',z) \sqrt{\mathrm{d}X(z)\mathrm{d}X(z')})}   \\
\qquad{}=  F_0[\spcurve]+\frac{X(z')-X(z)}{N\,\mathrm{d}X(z)}  \delta_z F_0[\spcurve] + O\big((X(z')-X(z))^2\big).
\end{gather*}
Finally we have
\begin{gather*}
\psi(z',z) \ee{-N\int_z^{z'}Y\dd X} E(z',z) = 1+\frac{X(z')-X(z)}{N \mathrm{d}X(z)} \delta_z \big(\ln\Tau-N^2F_0\big) + O\big((X(z')-X(z))^2\big),
\end{gather*}
and thus
\begin{gather*}
\psi(z',z) \ee{-N\int_z^{z'}Y\dd X}-\frac{1}{E(z',z)} = \frac{X(z')-X(z)}{N\,\mathrm{d}X(z) E(z',z)} \delta_z \big(\ln\Tau-N^2 F_0\big) + O(X(z')-X(z)).
\end{gather*}
Taking the limit $z'\to z$, and noticing that
\[
\frac{1}{E(z',z)}={\sqrt{\mathrm{d}X(z) \mathrm{d}X(z')}}/(X(z')-X(z)) + O(X(z')-X(z)),
\] we f\/ind
\begin{gather*}
\frac{1}{N} \delta_z\big(\ln\mathcal{T}-N^2 F_0\big) = W_1(z)-N Y(z)\dd X(z)\\
 \hphantom{\frac{1}{N} \delta_z\big(\ln\mathcal{T}-N^2 F_0\big)}{}
 = \lim_{z'\to z} \left(\psi(z',z) \ee{-N\int_z^{z'}Y\dd X}-\frac{\sqrt{\mathrm{d}X(z')\mathrm{d}X(z)}}{X(z')-X(z)}\right),
\end{gather*}
hence the lemma.
\end{proof}

\begin{theorem}
\label{detform} If Conjecture~{\rm \ref{consj}} holds, then
\begin{gather*}
\forall\, n \geq 2,\qquad W_n(z_1,\dots,z_n) = (-1)^{n+1}  \sum_{\sigma \ {\rm cyclic\  perm.}} \;  \prod_{i=1}^n \psi(z_i,z_{\sigma(i)}).
\end{gather*}
Equivalently
\begin{gather*}
\overline{W}_n(z_1,\dots,z_n) = \text{``$\det$''}\, \psi(z_i,z_j),
\end{gather*}
where ``$\det$'' means that when we decompose the determinant as a sum of permutations, each factor $\psi(z_i,z_i)$ should be replaced by $W_1(z_i)$.
\end{theorem}

\begin{proof}
The formula for $W_1$ is proved in Lemma~\ref{lW1}. Then, we get the formula for $W_n$ by recursively applying $\delta_{z_i}$ and using the self-replication of $\psi$.
\end{proof}

  It is thus clear that this determinantal structure relies on the (conjectured) existence of Hirota equations for~$\Tau$, i.e.\ on integrability.

\subsection[Baker-Akhiezer functions]{Baker--Akhiezer functions}
\label{BAfunk}
We recall the notations of Section~\ref{sec:reconstruction}. With the kernel $\psi(z_1,z_2)$ that we have just constructed, we introduce a $d \times d$ matrix $\Psi(x_1,x_2) = \psi(z^{i}(x_1),z^{j}(x_2))_{i,j=1,\dots,d}$ for $x_1,x_2 \in \mathbb{C}$, where $z^j(x)$ are the $d$ preimages of $x$ on the curve. This def\/inition has to be regularized when $x_1$ or $x_2$ is equal to $X(p)$ where $p$ is a pole of~$X$ or~$Y$. In particular, it was explained in Section~\ref{funcin} how to take~$x_1$ or~$x_2 \rightarrow \infty$
\begin{gather*}
\Psi(x) \,\text{``=''}\,  \big[\Psi\big(z^j(x),\infty_{I}\big)\big]_{I,1 \leq j \leq d},
\qquad\Phi(x) \,\text{``=''}\, \big[\Psi\big(\infty_{I},z^j(x)\big)\big]_{I,1 \leq j \leq d}.
\end{gather*}

\subsection{Duality equation}
\label{duadau}
\begin{theorem}
\label{th:YB1}If Conjecture~{\rm \ref{consj}} holds, we have for any spectral curve $\mathcal{S}$
\begin{gather*}
\Psi(x_1,x_2)\Psi(x_2,x_3) = \frac{(x_1 - x_3) \mathrm{d}x_2}{(x_2 - x_1)(x_3 - x_1)}\Psi(x_1,x_3).
\end{gather*}
\end{theorem}

\begin{proof}
For $\mathcal{S} = (\mathcal{C},x,y)$, we introduce an auxiliary spectral curve
\begin{gather*}
\hat{\mathcal{S}}_{ij} = \spcurve + \frac{1}{N}\,\mathrm{d}S_{z^i(x_1),z^j(x_3)}.
\end{gather*}
To compute the matrix element, we use the self-replication of~$\psi$
\begin{gather*}
[\Psi(x_1,x_2)\Psi(x_2,x_3)]_{ij} = \sum_m \psi\big(z^{i}(x_1),z^{m}(x_2)\big)\psi\big(z^{m}(x_2),z^{j}(x_3)\big)  \\
\hphantom{[\Psi(x_1,x_2)\Psi(x_2,x_3)]_{ij}}{}= - \sum_m  \delta_{z^m(x_2)} \ln\left(\frac{\Tau[\hat{\mathcal{S}}_{ij}]}{\Tau[\mathcal{S}]}\right).
\end{gather*}
Now, we use the fact that $\delta_z$ is a derivation, and by def\/inition, $\delta_z \ln\Tau[\hat{\mathcal{S}}_{ij}] = W_1[z ; \hat{\mathcal{S}}_{ij}]$
\begin{gather*}
[\Psi(x_1,x_2)\Psi(x_2,x_3)]_{ij} \nonumber \\
= - \sum_m W_1\big[z^m(x_2) ; \hat{\mathcal{S}}_{ij}\big] \psi\big(z^i(x_1),z^j(x_3)\big) + \sum_m W_1\big[z^m(x_2) ; \mathcal{S}\big] \psi\big(z^i(x_1),z^j(x_3)\big).
\end{gather*}
The linear loop equation (Theorem~\ref{llop}) tells us the sum over sheets of $W_1[z^m(x_2) ; \hat{\mathcal{S}}_{ij}]$
\begin{gather*}
  [\Psi(x_1,x_2)\Psi(x_2,x_3)]_{ij}   =    [\Psi(x_1,x_3)]_{ij}\\
  \qquad{}
 \times  \left(-\sum_m N Y\mathrm{d}X\big(z^m(x_2)\big) - \mathrm{d}S_{z^i(x_1),z^j(x_3)}\big(z^m(x_2)\big) + N Y\mathrm{d}X\big(z^m(x_2)\big)\right) \\
 \qquad{}
 =  - [\Psi(x_1,x_3)]_{ij} \sum_m \mathrm{d}S_{z^i(x_1),z^j(x_3)}\big(z^m(x_2)\big)
 =  \frac{(x_1 - x_3) \mathrm{d}x_2}{(x_1 - x_2)(x_2 - x_3)} [\Psi(x_1,x_3)]_{ij}.\tag*{\qed}
\end{gather*}
\renewcommand{\qed}{}
\end{proof}

The Baker--Akhiezer spinor kernel $\psi_{\cl}(z_1,z_2)$ of Section~\ref{sec:reconstruction} was regular at ramif\/ication points. So, we could f\/ind a formula for $\psi_{\cl}(z_1,z)\psi_{\cl}(z,z_2)$ even before summing over the sheets where~$z$ is located (ref\/ined duality equation, Proposition~\ref{thYBsp}). Here, the spinor kernel $\psi(z_1,z_2)$ does have, order by order in~$1/N$, poles at ramif\/ication points. So, we do not have a simple expression for $\psi(z_1,z)\psi(z,z_2)$. However, Theorem~\ref{th:YB1} shows all contributions from the ramif\/ication points cancel in the sum over sheets.

\subsection[Christoffel-Darboux relations]{Christof\/fel--Darboux relations}

The matrix $\Psi(x_1,x_2)$ is invertible, since it is def\/ined as a series for which the leading term coincides with $\Psi_{\mathrm{cl}}(x_1,x_2)$ which is invertible (see Lemma~\ref{duall}). We also have a duality relation to express the inverse

\begin{corollary}\label{Cfd}
If Conjecture~{\rm \ref{consj}} holds,
\begin{gather*}
\Psi^{-1}(x_1,x_2) = - \frac{(x_1 - x_2)^2}{\mathrm{d}x_1\mathrm{d}x_2} \Psi(x_2,x_1).
\end{gather*}
\end{corollary}

\begin{theorem}
\label{CDS}If Conjecture~{\rm \ref{consj}} holds, the matrices $\Psi(x)$ and $\Phi(x)$ are invertible, and we have the Christoffel--Darboux relation
\begin{gather*}
\psi(z_1,z_2) = \frac{\sum_{I,J}\psi_I(z_1)A_{I,J}\phi_J(z_2)}{X(z_1) - X(z_2)},
\end{gather*}
where the matrix $A$ is invertible, independent of $x$ and given by
\begin{gather*}
A^{-1} = \frac{1}{\mathrm{d}x} \Phi(x)\Psi^{t}(x).
\end{gather*}
\end{theorem}
The proofs can be copied from the semiclassical case (Section~\ref{sec:CD}), because they are based only on the duality equation.

\subsection{Dif\/ferential systems}
\label{sec:9}

$\Psi(x_1,x_2)$ is the solution of a system of dif\/ferential equations with respect to the positions of the poles $X(p)$ ($x = \infty$ is a f\/ixed pole) and the times $t_{p,j}$.

\begin{theorem}
If Conjecture~{\rm \ref{consj}} holds, then for any deformation parameter $\lambda = t_{p,j}$ or $\lambda = X(p) \neq \infty$, there exists a $d\times d$ matrix $M_{\lambda}(x_1,x_2)$, such that
\begin{itemize}\itemsep=0pt
\item[$(i)$] $\left(\partial_{\lambda} - M_\lambda\right)\Psi(x_1,x_2) = 0$;
\item[$(ii)$] $M_{\lambda}$ is a rational function of $x_2$;
\item[$(iii)$] $M_{\lambda}$ has no pole at branchpoints.
\end{itemize}
These deformations are compatible $($since $\Psi$ is invertible$)$ and isomonodromic.
\end{theorem}
Implicitly, for $j = 0$, only deformations $\partial_{\lambda} = \partial_{t_{p,0}} - \partial_{t_{p',0}}$ are considered.

\begin{proof} By def\/inition
\begin{gather*}
M_{\lambda} = \partial_{\lambda}\Psi(x_1,x_2)\,\Psi^{-1}(x_1,x_2) = - \frac{(x_1 - x_2)^2}{\mathrm{d}x_1\mathrm{d}x_2}\partial_{\lambda}\Psi(x_1,x_2)\,\Psi(x_2,x_1).
\end{gather*}
When $\lambda$ is a time $t_{p,k}$ ($k \geq 1$), we have by self-replication
\begin{gather*}
\partial_{t_{p,k}} \psi(z_1,z_2) = - \int_{\omega_{p,k}^*} \psi(z_1,\cdot)\psi(\cdot,z_2).
\end{gather*}
We compute
\begin{gather*}
[M_{t_{p,k}}]_{ij}   =    \frac{(x_1 - x_2)^2}{\mathrm{d}x_1\mathrm{d}x_2}\sum_{m} \int_{\omega_{p,k}^*} \psi\big(z^i_1,z\big)\psi\big(z,z^m_2\big)\psi\big(z^m_2,z^j_1\big) \nonumber \\
\hphantom{[M_{t_{p,k}}]_{ij}}{}
 =
  \frac{1}{\mathrm{d}x_1}\int_{\omega_{p,k}^*} \frac{(X(\cdot) - x_1)(x_2 - x_1)}{(X(\cdot) - x_2)} \psi\big(z^i_1,\cdot\big)\psi\big(\cdot,z^j_1\big).
\end{gather*}
It is clear that $M_{t_{p,k}}(x_1,x_2)$ is a rational function of $x_2$ without poles at branchpoints. If we assume $X(p) \neq \infty$, it has a pole only at $x_2 = X(p)$, and this pole is of order $k + 1$. If we assume $X(p) = \infty$, it has a pole only at $x_2 = \infty$, and this pole is of order $1 + \lfloor k/d_p\rfloor$ (we recall that~$d_p$ is the multiplicity of order of the pole~$p$ of~$X$). It is straightforward to adapt this proof for $k = 0$, namely $\partial_{\lambda} = \partial_{t_{p,0}} - \partial_{t_{p',0}}$.

Now, we turn to $\lambda = X(p) \neq \infty$. We have assumed that $\mathrm{d}X(p) \neq 0$, so the preimages $p^1,\ldots,p^d$ of $X(p)$ are distinct, and $(X(z) - X(p))$ is a local coordinate near each $p^m$. The Laurent expansion of $Y\mathrm{d}X(z)$ when $z \rightarrow p^m$ is
\begin{gather*}
Y\mathrm{d}X(z) = \sum_{k \geq 0} t_{p,k}\frac{\mathrm{d}X(z)}{(X(z) - X(p))^{k + 1}} + O(1),
\end{gather*}
where only a f\/inite number of $t_{p,k'} = 0$ are non zero. If we perform $X(p) \rightarrow X(p) + \lambda$ while the times are f\/ixed, we change $Y\mathrm{d}X$ to $(Y\mathrm{d}X)_{\lambda}$, with Laurent expansion at $p^m$
\begin{gather*}
(Y\mathrm{d}X)_{\lambda} = Y\mathrm{d}X + \lambda\left(\sum_{k \geq 1} \frac{kt_{p,k - 1}}{(X(z) - X(p))^{k + 1}}\right) + O(1).
\end{gather*}
So, we can identify
\begin{gather*}
\frac{\partial}{\partial X(p)} \longrightarrow \sum_{k \geq 1} kt_{p,k - 1}\frac{\partial}{\partial t_{p,k}},
\end{gather*}
which is a f\/inite sum. Hence
\begin{gather*}
[M_{X(p)}]_{ij} = -\frac{1}{\mathrm{d}x_1}\sum_m \Res_{z \rightarrow z^m(X(p))} \frac{\mathrm{d}Y(z)}{\mathrm{d}X(z)} \frac{(X(z) - x_1)(x_2 - x_1)}{(X(z) - x_2)}  \psi\big(z^i_1,z\big)\psi\big(z,z^j_1\big).
\end{gather*}
This is a rational fraction of $x_2$, with a pole at
$x_2 = X(p)$ of order
\begin{gather*}
1 + \lfloor(\max_m \operatorname{ord}\limits_{z^m(p)} Y\mathrm{d}X)/d_p\rfloor. \tag*{\qed}
\end{gather*}
\renewcommand{\qed}{}
\end{proof}

One could also def\/ine a matrix $L$ with the reconstruction formula presented in Section~\ref{sec:recon}. But now, the function~$Y$ depends on the spectral curve, the evolution of $L$ under the f\/lows generated by the times is not isospectral. However, since the position and the order of the poles is f\/ixed, the deformation parameters~$t_{p,j}$ preserve the monodromies of~$\Psi$. Such a reconstruction has also been performed at leading order by Bertola and Gekhtman~\cite{BertG}, and our expressions up to $o(1)$ match their results.

\section{Dictionary for matrix models}
\label{sec:mat}

We recall here the correspondence between the def\/initions in integrability (and in this article), and the observables def\/ined in matrix models. Let $M$ be a random square matrix of size~$N\times N$, diagonalizable by a unitary conjugation, with eigenvalues restricted to some contour $\Gamma$ in the complex plane. The probability measure $\mathrm{d}\mu(M)$ is a data of the matrix model. Its normalization def\/ines the partition function $Z_N = \int \mathrm{d}\mu(M)$. For a function~$f$, we note $\langle f(M)\rangle$ the expectation value of the random variable~$f(M)$
 \begin{gather*}
 \langle f(M)\rangle = \frac{\int \mathrm{d}\mu(M) f(M)}{\int \mathrm{d}\mu(M)}.
\end{gather*}

\subsection{Examples of integrable matrix models}\label{secexrmt}

Important examples of integrable matrix models, described in details in \cite{Mehtabook}, include:

{\bf The one hermitian matrix model.}
$M\in H_N$ is a $N\times N$ hermitian matrix, with eigenvalues on the real axis
\begin{gather*}
{\mathrm d}\mu(M) = \ee{-N \operatorname{Tr} V(M)} {\mathrm d}M,\qquad
{\mathrm d}M = \prod_{i=1}^N {\mathrm d}M_{i,i}\prod_{1 \leq i<j \leq N} {\mathrm d}\operatorname{Re} M_{i,j}{\mathrm d}\operatorname{Im} M_{i,j},
\end{gather*}
and $V(M)$ is a ``semiclassical potential'' (see \cite{Marcopaths}), i.e.\ $V'(x)$ is a rational function of its variable~$x$, chosen such that $\int_{\mathbb R}  \ee{-V(x)} {\mathrm d}x$ is absolutely convergent.

{\bf The one normal matrix model, with eigenvalues on a contour $\boldsymbol{\Gamma}$.}
We def\/ine
\begin{gather*}
H_N(\Gamma)  = \big\{ U\operatorname{diag}(x_1,\dots,x_N) U^\dagger, \   U\in U(N) \ \mathrm{and} \ (x_1,\dots,x_N)\in \Gamma^N\big\}
\end{gather*}
equipped with a ``measure'' (not necessarily real or normalized)\footnote{For a given $M\in H_N(\Gamma)$, there are many choices of~$U$ and~$x_i$'s (in fact~$U$ can be multiplied by any element of~$U(1)^N$ and the~$x_i$'s can be permuted, and $\dd U$ is in fact the measure on the quotient $U(N)/\mathfrak{S}_N\times U(1)^N$ by the Haar measure on $U(N)$.}
\begin{gather*}
{\mathrm d}M = {\mathrm d}U \prod_{i<j}(x_i-x_j)^2 \prod_{i=1}^N {\mathrm d}x_i,
\end{gather*}
where $\dd U$ is the Haar measure on the unitary group $U(N)$ and $\dd x_i$ is the curvilinear\footnote{If $\Gamma\subset \mathbb C$ is a path in the complex plane parametrized by a $C^1$ function $\gamma:\mathbb R\to \Gamma\subset\mathbb C$, i.e.\ $\Gamma=\{\gamma(s), s\in\mathbb R \}$, at $x=\gamma(s)$ we def\/ine the curvilinear measure  $\dd x = \gamma'(s) \dd s$ where $\dd s$ is the Lebesgue measure on $\mathbb R$.} measure along~$\Gamma$. This measure is always invariant under unitary transformations. We then def\/ine the measure ${\mathrm d}\mu(M)$ on $H_N(\Gamma)$
\begin{gather*}
{\mathrm d}\mu(M) = \ee{-N\operatorname{Tr} V(M)} {\mathrm d}M,
\end{gather*}
where $V'(x)\in \mathbb C(x)$, i.e.~$V'(x)$ is a rational function of its variable $x$, chosen such that $\int_{\Gamma}  \ee{-V(x)} {\mathrm d}x$ is absolutely convergent.
This means that for a given potential $V(x)$, $\Gamma$ must go to $\infty$ or to the poles of $V'$, only in sectors where $\operatorname{Re}  V(x) \to +\infty$. When $\Gamma=\mathbb R$, this def\/inition is the same as the hermitian matrix model $H_N(\mathbb R)=H_N$, with the usual Lebesgue measure on $H_N$. When~$\Gamma$ is the unit circle~$\mathbb{S}_1$, this def\/inition is that of the unitary matrix model $H_N(\mathbb{S}_1)= U(N)$.

{\bf The two normal matrices model.}
Given two paths $\Gamma_1$, $\Gamma_2$, we def\/ine a measure on $H_N(\Gamma_1)\times H_N(\Gamma_2)$
\begin{gather*}
{\mathrm d}\mu_{2}(M_1,M_2) = \ee{-N\operatorname{Tr} [V_1(M_1)+V_2(M_2) - M_1 M_2 ]}{\rm d}M_1{\rm
d}M_2,
\end{gather*}
where $V_1$ and $V_2$ are semiclassical potentials ($V'_1$ and $V'_2$ are rational functions), chosen such that integrals on $\Gamma_1\times\Gamma_2$ are absolutely convergent. Upon integration on $M_2$, this measure induces a measure ${\mathrm d}\mu(M)$ on~$M_1$ that we rename~$M$
\begin{gather*}
{\rm d}\mu(M_1) = \int_{M_2\in H_N(\Gamma_2)}{\mathrm d}\mu_{2}(M,M_2).
\end{gather*}

{\bf The chain of  matrices.}
This is the natural generalization of the case of two matrices. Consider $k$ paths $\Gamma_1,\dots, \Gamma_k$, and $k$ semiclassical potentials $V_1,\dots, V_k$, and def\/ine the measure on $H_N(\Gamma_1)\times\dots\times H_N(\Gamma_k)$ as
\begin{gather*}
{\mathrm d}\mu_k(M_1,\dots,M_k) = \ee{-N\operatorname{Tr} \big[\sum\limits_{i=1}^k V_i(M_i) - \sum\limits_{i=1}^{k-1} M_i M_{i+1} \big]}
{\rm d}M_1\cdots{\rm d}M_k.
\end{gather*}
For any $i\in \{1,\dots,k\}$, we integrate on $M_j$'s with $j\neq i$, and renaming $M=M_i$, this measure induces a measure ${\mathrm d}\mu(M)$:
\begin{gather*}
{\mathrm d}\mu(M_i) = \int_{M_j\in H_N(\Gamma_j), j\neq i}{\mathrm d}\mu_k(M_1,\dots,M_k).
\end{gather*}

\subsection{Correspondences}

\subsubsection{Partition function}

All the listed matrix models above have the property to be integrable, in a sense explained below. There is a huge literature on the subject, let us mention among others the early works in physics \cite{Karch,kostovhirota}, and in mathematics \cite{AvM2,HTW,vanM}. For the one and two matrices models, it is established that the partition function $Z_N =\int {\mathrm d}\mu(M)$ is an isomonodromic Tau function \cite{BEHtau,BEH04,BMtau}, but such a result is not known at present for the chain of matrices. The so-called double scaling limit of matrix models has been also intensively from the point of view of isomonodromic deformations \cite{doug,FIKN, FIK90,moore}.

\subsubsection{Correlators, spectral curves, loop equations}
For all the matrix models above, the correlators are
\begin{gather*}
{\cal W}_n(x_1,\ldots,x_n) = N^{-n} \Big\langle\prod_{i = 1}^n \operatorname{Tr} \frac{1}{x_i - M}\Big\rangle_C,\qquad \overline{{\cal W}}_n(x_1,\ldots,x_n) = N^{-n} \Big\langle\prod_{i = 1}^n\operatorname{Tr} \frac{1}{x_i - M}\Big\rangle,
\end{gather*}
where $C$ means `cumulant'. ${\cal W}_n$ is often called $n$-point function, or {\em connected} $n$-point function, and $\overline{{\cal W}}_n$ is called disconnected $n$-point function.

The one point function ${\cal W}_1(x)$ (also called \emph{resolvent}) plays an important role. In all the matrix models above, under reasonable assumptions, it has a large $N$ limit denoted
\begin{gather*}
\frac{1}{N} {\cal W}_1(x) \mathop{{\sim}}_{N\to\infty} {\cal W}_1^{(0)}(x) = {\cal Y}(x),
\end{gather*}
which is furthermore an algebraic function of $x$, i.e.\ there exists a polynomial ${\cal E}(x,y)$ such that ${\cal E}(x,{\cal Y}(x))=0$. Hence, there exists a spectral curve $\mathcal{S}_{\mathrm{MM}} = (\mathcal{C},X,Y)$, such that
\begin{gather*}
\forall\, z\in \curve,\qquad
{\cal Y}(X(z)) = Y(z).
\end{gather*}
If we def\/ine
\begin{gather*}
W_n(z_1,\dots,z_n) =  N^n\,{\cal W}_n(X(z_1),\dots,X(z_n)) \bigotimes_{i = 1}^n\dd X(z_i) + \delta_{n,2}  \frac{\dd X(z_1)\otimes \dd X(z_2)}{(X(z_1)-X(z_2))^2},
\end{gather*}
it is a classical result of random matrix theory (for instance it can be proved by integration by parts in the matrix integral) that, for all matrix models listed above, $W_n$ satisfy the loop equations of Theorem~\ref{llop}.

\subsubsection[Large $N$ asymptotic expansion]{Large $\boldsymbol{N}$ asymptotic expansion}

It is conjectured that the large $N$ asymptotic expansion of the partition function $Z_N$ matches with that of $\mathcal{T}[\mathcal{S}_{\mathrm{MM}}]$ function for the spectral curve $\spcurve$ introduced in Def\/inition~\ref{defTaumunu}, i.e.\ has an asymptotic expansion
\begin{gather*}
\ln Z_N = N^2 F_0 + F_1 + \ln\Theta + \frac{1}{N}\left(F'_1 \frac{\Theta'}{\Theta} + F'''_0 \frac{\Theta'''}{\Theta}\right) + o(1/N).
\end{gather*}
The characteristics $[\mu,\nu]$ of the Theta function $\Theta$ is determined by the choice of the integration contour~$\Gamma$. This conjecture was derived heuristically in~\cite{BDE,Ecv}.

When the semiclassical spectral curve is of genus $0$, there is no Theta function and the expansion involves only powers of $1/N$ (in fact, powers of $1/N^2$). This happens for the so-called ``one-cut regime'', and for the one matrix model, the existence of such an expansion has been proved for the one matrix model for real-valued, analytic potential \cite{APS01}, and then the coef\/f\/icients are necessarily given by the symplectic invariants of the semiclassical spectral curve~\cite{E1MM}. Beyond the one-cut regime, the Riemann--Hilbert steepest descent analysis~\cite{DZ, FIK90} has been applied to f\/ind explicitly the asymptotics up to $o(1)$ in the one hermitian matrix model with real-valued, polynomial potential~\cite{BI99,Deiftetal}. It features in general a pseudo-periodic behavior with $N$, encoded in the Theta function, but the one-cut regime can also be retrieved with this Riemann--Hilbert method \cite{ErcMcL}. These results have later been extended to the one normal matrix model with complex-valued polynomial potential \cite{Be}. This method can be used in principle to f\/ind recursively the subleading orders, although it does not allow to write the answer a priori explicitly to all orders. Def\/inition~\ref{defTaumunu} is expected to give the correct answer to all orders.

\subsubsection[Baker-Akhiezer spinor kernel]{Baker--Akhiezer spinor kernel}

The spinor kernel is related to the expectation value of ratios of characteristic polynomials
\begin{gather*}
\psi(z_1,z_2) =  \Big\langle\frac{\det(X(z_1) - M)}{\det (X(z_2) - M)}\Big\rangle \frac{\sqrt{\mathrm{d}X(z_1)\mathrm{d}X(z_2)}}{X(z_1)-X(z_2)}.
\end{gather*}
The Baker--Akhiezer spinor kernel $\psi_{\cl}(z_1,z_2)$ is the large $N$ limit of $\psi(z_1,z_2)$.
When sending $X(z_2)\to\infty$, one gets, after proper renormalization
\begin{gather*}
\psi(z_1) = \text{``}\psi(z_1,\infty)\text{''} =  \big\langle{\det (X(z_1) - M)}\big\rangle,
\end{gather*}
which is clearly a polynomial in $X(z_1)$ of degree $N$. It is a classical result of random matrix theory~\cite{Mehtabook} that this expectation value of the characteristic polynomial of $M$ is the orthogonal polynomial of degree $N$, for the orthogonality given by the measure $\dd\mu$ for matrices of size~$1$.
The dual function for matrix models is
\begin{gather*}
\phi(z_1) = \text{``}\psi(\infty,z_1)\text{''} =  \Big\langle{\frac{1}{\det (X(z_1) - M)}}\Big\rangle.
\end{gather*}
It can be derived from orthogonality relations of the orthogonal polynomials  that those spinor kernels satisfy Hirota equations, also called ``determinantal formula'', for instance,
\begin{gather*}
  \frac{(x_1-x_2)(x_3-x_4)}{(x_1-x_3)(x_1-x_4)(x_2-x_3)(x_3-x_4)}\Big\langle\frac{\det(x_1-M) \det(x_2-M)}{\det(x_3-M) \det(x_4-M)}\Big\rangle \\
\qquad{}  =  \frac{1}{(x_1-x_3)(x_2-x_4)} \Big\langle \frac{\det(x_1-M)}{\det(x_3-M)}\Big\rangle\Big\langle\frac{\det(x_2-M)}{\det(x_4-M)}\Big\rangle \\
\qquad\quad{}  - \frac{1}{(x_1-x_4)(x_2-x_3)}\Big\langle\frac{\det(x_1-M)}{\det(x_4-M)}\Big\rangle\Big\langle\frac{\det(x_2-M)}{\det(x_3-M)}\Big\rangle.
\end{gather*}
In that case, $N$ is an integer (the size of the matrices), and Hirota equation is an equality between sequences indexed by $N$, not only of formal asymptotic series.
This relations have been proved in \cite{FyodStra} for the one matrix model, and in~\cite{Bergere2004,AkePottier} for the two matrices model, from which the case of chain of matrices can be deduced. It shows that, modulo the heuristic derivation of asymptotics of matrix integrals, Hirota equations (as stated as in Conjecture~\ref{eqhys2}) hold for the semiclassical spectral curves of matrix models (however, not all algebraic curves can be reached in this way).

\subsubsection{Dif\/ferential systems}

The orthogonal polynomials, as well as their duals, in all cases above, do satisfy ODE's of order $d$ ($d$ depends on the degrees of potentials). As we have seen in Section~\ref{sec:more}, the proof is a consequence of Hirota equations. For instance for the one matrix model, this is a second order ODE ($d = 2$). For the two matrices model with potentials~$V_1$ and~$V_2$, we always have $d=1+\deg V'_2$ (in case~$V'_2$ is a~rational function, $\deg V'_2$ is the sum of degrees of all poles).
In all cases we have a~$d$-dimensional vector
\begin{gather}\label{defeqdpsicalDMM}
\vec\Psi(z)  =  \begin{pmatrix}\psi(z_1) \\
\vdots \end{pmatrix},
\end{gather}
where the f\/irst entry is $\psi(z_1)$. The other entries are obtained from $\psi(z_1)$ with a procedure described in~\cite{BEH} and very similar to Section~\ref{funcin}. This vector satisf\/ies an ODE
\begin{gather*}
\frac{\mathrm d}{{\mathrm d} X(z)} \vec \Psi(z)  =  {\cal D}(X(z))  \vec \Psi(z),
\end{gather*}
where ${\cal D}(x)$ is a $d\times d$ matrix, whose entries are rational functions of~$x$, that depend implicitly on~$N$, on the coef\/f\/icients of the potentials, and on the choice of integration contour $\Gamma$. The locus of eigenvalues of ${\cal D}(x)$, i.e.\ the polynomial equation
\begin{gather}
{\cal E}_N(x,y) = \det{(y - {\cal D}(x))} = 0
\end{gather}
def\/ines a spectral curve for any f\/inite $N$. The semiclassical spectral curve (i.e.\ its large $N$ limit) coincides with the spectral curve $\mathcal{S}_{\mathrm{MM}}$
\begin{gather*}
{\cal E}(x,y) = \lim_{N\to\infty} {\cal E}_N(x,y).
\end{gather*}
There are also $d\times d$ dif\/ferential systems for derivatives with respect to all coef\/f\/icients of the potentials, and there is also a linear recursion relation on~$N \rightarrow N + 1$ (see~\cite{Ismachen} for the one matrix model). All these systems are compatible as shown in~\cite{BEHmm} in full generality for the chain of matrices.

\subsubsection{Symplectic invariance}

Notice that, in the two matrices model
\begin{gather*}
Z= \int_{H_N(\Gamma)\times H_N(\td\Gamma)}\dd\mu(M_1,M_2),
\end{gather*}
we have def\/ined our semiclassical spectral curve $\spcurve=(\curve,X,Y)$ from the large $N$ limit of the resolvent ${\cal W}_1(x) = \langle \operatorname{tr} \frac{1}{x-M_1}\rangle$ associated to the matrix $M_1$.
Since $M_1$ and $M_2$ play a symmetric role, it is clear that we would have obtained the same partition function, starting from the semiclassical spectral curve $\td\spcurve=(\td\curve,\td X,\td Y)$ associated to the resolvent of matrix $M_2$, and thus we must have
\begin{gather*}
\Tau[\spcurve] = \Tau[\td\spcurve].
\end{gather*}
One can easily f\/ind that the two spectral curves $\spcurve_{\mathrm{MM}}=(\curve,X,Y)$ and $\td\spcurve_{\mathrm{MM}}=(\td\curve, \td X,\td Y)$ are related by $\curve=\td\curve$ and $\td X=Y$, $\td Y=X$, in other words they arse symplectically equivalent. Hence, the fact that $\Tau[\spcurve_{\mathrm{MM}}] = \Tau[\td\spcurve_{\mathrm{MM}}]$ can be seen at all order in the large $N$ expansion as a consequence of the symplectic invariance of the~$F_g$'s. In fact, this is a manifestation at large~$N$ of an exact result for f\/inite~$N$. The orthogonal polynomials $\vec\Psi(x)$ associated to matrix~$M_1$ satisfy an ODE of some order~$d$ (see equation~\eqref{defeqdpsicalDMM})
\begin{gather*}
\frac{\mathrm d}{{\mathrm d} X(z)} \vec{\Psi}(z)  =  {\cal D}(X(z))  \vec{\Psi}(z),
\end{gather*}
whereas the orthogonal polynomials $\vec{\td{\Psi(y)}}$ associated to matrix $M_2$ satisfy another ODE of some order $\td d$ (in general $\td d\neq d$)
\begin{gather*}
\frac{\mathrm d}{{\mathrm d} Y(z)} \vec{\td{\Psi}}(z)  =  \td{\cal D}(Y(z))  \vec{\td\Psi}(z).
\end{gather*}
It was discovered in \cite{BEHduality} that
\begin{gather*}
\det \big(y\,\mathbf{1}_{d\times d}-{\cal D}(x)\big) = \det \big(x\,\mathbf{1}_{\td d\times \td d}-\td {\cal D}(y)\big).
\end{gather*}
This also implies that the semiclassical spectral curve def\/ined from~$M_1$ or from~$M_2$ are related as explained above.

\section{Conclusion}

For integrable systems with a small dispersive parameter~$1/N$, using the theory of symplectic invariants~\cite{EOFg}, we have introduced a formal object~$\Tau$, which is conjectured to be a Tau function in the sense that it satisf\/ies Hirota equations. It is challenging to f\/ind a full proof of Conjecture~\ref{eqhys2}, and that would certainly prove that a certain quantity constructed out of $\Tau$ (like $u(x,t) = 2(\ln \Tau)_{xx}$ for KdV) provides the all-order asymptotics of solutions of nonlinear integrable PDE's in the small dispersion limit.

One can wonder how to generalize the construction. First, to algebraic curves in $\mathbb{C}^*\times\mathbb{C}^*$, having in view the mirror curves appearing in Gromov--Witten theory, which all are of the form $\mathrm{Pol}(e^x,e^y) = 0$. In a recent work, we have described an application of the expressions for $\Tau$ and $\psi$ for such curves to the computation of perturbative knot invariants~\cite{BEknot}. This explicit example suggests that the construction remains meaningful in this context. Second, it should be possible to extend our construction to the bundles appearing in generalized matrix models like the $O(n)$ model, which might be related to isomonodromic deformations on Riemann surfaces of positive genus and to Hitchin systems \cite{Hitchin}. And, even further to $\mathcal{D}$-modules, for which an adapted topological recursion (the so-called $\beta$-deformation of the topological recursion) is being developed with similar properties~\cite{CEMq2, CEMq1}.

Eventually, it remains to compare this construction with other approaches (Frobenius mani\-folds~\cite{Dubrovin} and construction of integrable hierarchies of topological type~\cite{DubZh}, Poisson bracket structures, Segal--Wilson formalism in the Grassmannian \cite{Segalwilson}, etc.), study its consequences and better understand the underlying geometry.

\appendix

\section[Proof of Conjecture~\ref{consj} up to $o(1/N)$]{Proof of Conjecture~\ref{consj} up to $\boldsymbol{o(1/N)}$}
\label{apphirota1surN}

\begin{proposition}
$\psi(z_1,z_2)$ is self-replicating at least up to $o(1/N)$,
\begin{gather*}
\frac{1}{N} \delta_{z}\psi(z_1,z_2)+\psi(z_1,z)\psi(z,z_2) = o(1/N).
\end{gather*}
\end{proposition}

\begin{proof}
Let us start from $\psi_{12}=\psi(z_1,z_2)$ written as
\begin{gather*}
\psi_{12} = \frac{\ee{N\int_2^1 Y\dd X}}{E_{12}} \frac{\Theta_{12}}{\Theta} \left\{ 1+\frac{1}{N}\hat\psi_{12}+o(1/N)\right\},
\end{gather*}
with
\begin{gather*}
\hat\psi_{12}
 =  \int_2^1 \omega_1^{(1)} + \frac{1}{6}\int_2^1\int_2^1\int_2^1 \om_3^{(0)} + \frac{1}{2} \frac{\Theta''_{12}}{\Theta_{12}}  \oint\oint\int_2^1 \om_3^{(0)}  + \frac{1}{2} \frac{\Theta'_{12}}{\Theta_{12}}  \oint \int_2^1\int_2^1 \om_3^{(0)}  \nonumber \\
\hphantom{\hat\psi_{12}=}{}
 + \left( \frac{\Theta_{12}'}{\Theta_{12}}-\frac{\Theta'}{\Theta}\right) F'_1  + \frac{1}{6} \left( \frac{\Theta_{12}'''}{\Theta_{12}}-\frac{\Theta'''}{\Theta}\right) F'''_0,
\end{gather*}
where, to shorten notations,  $1$ means $z_1$, $2$ means~$z_2$,  $\oint$ means the contour integral around $\bcycle$-cycles (indices are understood in tensor notations, i.e.\ contracted with the indices of derivatives of $\Theta$), and $\Theta_{12}$ means
\begin{gather*}
\Theta_{12} = \Theta\big(\mathbf{w}_0 +2{\rm i}\pi(\mathbf{u}(z_1)-\mathbf{u}(z_2))\big),\qquad \Theta=\Theta(\mathbf{w}_0).
\end{gather*}
We remind that (equation~\eqref{W30ex})
\begin{gather*}
\om_3^{(0)}(z_0,z_1,z_2) = \sum_{i} \Res_{z \rightarrow a_i} \frac{B(z_0,z)B(z_1,z)B(z_2,z)}{\dd X(z)\,\dd Y(z)},
\end{gather*}
and by special geometry
\begin{gather*}
F_0''' = \oint_{\mathcal{B}}\oint_{\mathcal{B}}\oint_{\mathcal{B}} \om_3^{(0)} =  \sum_{i} \Res_{z \rightarrow a_i} \frac{\big(\dd\mathbf{v}(z)\big)^3}{\dd X(z) \dd Y(z)},
\end{gather*}
where we have set
\begin{gather*}
\dd \mathbf{v}(z) = 2{\rm i}\pi \dd \mathbf{u}(z) = \oint_{\mathcal{B}} B(z,\cdot).
\end{gather*}
The expression for $\hat\psi_{12}$ is thus
\begin{gather*}
\hat\psi_{12}   =   \int_2^1 \omega_1^{(1)} + \frac{1}{6}\int_2^1\int_2^1\int_2^1 \om_3^{(0)} + \frac{1}{2} \frac{\Theta''_{12}}{\Theta_{12}}\oint\oint\int_2^1 \om_3^{(0)} + \frac{1}{2}\left(\frac{\Theta'_{12}}{\Theta_{12}}\right)\oint \int_2^1\int_2^1 \om_3^{(0)}  \nonumber \\
\hphantom{\hat\psi_{12}   =}{}
 + \left( \frac{\Theta_{12}'}{\Theta_{12}}-\frac{\Theta'}{\Theta}\right) F'_1
 + \frac{1}{6} \left(\frac{\Theta_{12}'''}{\Theta_{12}}-\frac{\Theta'''}{\Theta}\right) F'''_0 \nonumber \\
\hphantom{\hat\psi_{12}}{}
=
 \int_2^1 \omega_1^{(1)} + \left( \frac{\Theta_{12}'}{\Theta_{12}}-\frac{\Theta'}{\Theta}\right)
 F'_1  + \sum_{i} \Res_{z \rightarrow a_i} \frac{1}{\dd X(z) \dd Y(z)} \Bigg\{\frac{1}{6}\big(\dd S_{12}(z)\big)^3  \\
\hphantom{\hat\psi_{12}   =}{}
+ \frac{1}{2}\left(\frac{\Theta'_{12}}{\Theta_{12}}\right)\dd\mathbf{v}(z)\big(\dd S_{12}(z)\big)^2  \nonumber \\
\hphantom{\hat\psi_{12}   =}{}
   + \frac{1}{2}\left(\frac{\Theta''_{12}}{\Theta_{12}}\right)\big(\dd\mathbf{v}(z)\big)^2 \dd S_{12}(z) + \frac{1}{6} \left(\frac{\Theta_{12}'''}{\Theta_{12}}-\frac{\Theta'''}{\Theta}\right)\big(\dd \mathbf{v}(z)\big)^3 \Bigg\}.
\end{gather*}
We need to apply the insertion operator $\frac{1}{N} \delta_z$ to $\hat\psi_{12}$, and obtain the result up to~$o(1)$. Only the variation of $NF_0'$ appearing in the Theta functions contributes
to this order
\begin{gather*}
\frac{1}{N} \delta_z \hat\psi_{12}   =    \Bigg[\left(\frac{\Theta_{12}'}{\Theta_{12}}-\frac{\Theta'}{\Theta}\right)' F'_1  + \sum_{i}\Res_{z' \rightarrow a_i} \frac{1}{\dd X(z') \dd Y(z')} \Bigg\{\frac{1}{2}\left(\frac{\Theta'_{12}}{\Theta_{12}}\right)'\dd\mathbf{v}(z')\big(\dd S_{12}(z')\big)^2   \\
\hphantom{\frac{1}{N} \delta_z \hat\psi_{12}   =}{}
 + \frac{1}{2}\left(\frac{\Theta''_{12}}{\Theta_{12}}\right)'\big(\dd\mathbf{v}(z')\big)^2 \dd S_{12}(z') + \frac{1}{6}\left(\frac{\Theta_{12}'''}{\Theta_{12}}-\frac{\Theta'''}{\Theta}\right)'\big(\dd\mathbf{v}(z')\big)^3 \Bigg\}\Bigg]\dd\mathbf{v}(z) + o(1).
\end{gather*}
This allows us to compute
\begin{gather*}
  \frac{1}{N} \delta_z \ln\psi_{12}
  =   \dd S_{12}(z)  + \left(\frac{\Theta_{12}'}{\Theta_{12}}-\frac{\Theta'}{\Theta}\right)\dd\mathbf{v}(z) + \frac{1}{2N}\int_{2}^1\int_{2}^1 \om_3^{(0)}(z,\cdot,\cdot)  \\
\hphantom{\frac{1}{N} \delta_z \ln\psi_{12} =}{}
+ \frac{1}{N}\left( \frac{\Theta'_{12}}{\Theta_{12}}\right) \oint\int_2^1 \om_3^{(0)}(z,\cdot,\cdot)  + \frac{1}{2N}{\left( \frac{\Theta''_{12}}{\Theta_{12}}-\frac{\Theta''}{\Theta}\right)}\oint\oint \om_3^{(0)}(z,\cdot,\cdot)\\
\hphantom{\frac{1}{N} \delta_z \ln\psi_{12} =}{}
 + \frac{1}{N^2} \delta_z \hat\psi_{12} + o(1/N)   \\
\hphantom{\frac{1}{N} \delta_z \ln\psi_{12}}{}
 =   \dd S_{12}(z)  + \dd\mathbf{v}(z) \left( \frac{\Theta_{12}'}{\Theta_{12}}-\frac{\Theta'}{\Theta}\right)  + \frac{1}{N} \sum_{i} \Res_{z' \rightarrow a_i} \frac{B(z',z)}{\dd X(z') \dd Y(z')} \Bigg\{\frac{1}{2}\big(\dd S_{12}(z)\big)^2   \\
\hphantom{\frac{1}{N} \delta_z \ln\psi_{12} =}{}
  + \left(\frac{\Theta_{12}'}{\Theta_{12}}\right)\dd\mathbf{v}(z')\dd S_{12}(z')   + \frac{1}{2}{\left(\frac{\Theta''_{12}}{\Theta_{12}}-\frac{\Theta''}{\Theta}\right)}\big(\dd\mathbf{v}(z')\big)^2 \Bigg\}\! + \frac{1}{N^2} \delta_z \hat\psi_{12}  + o(1/N) \nonumber \\
\hphantom{\frac{1}{N} \delta_z \ln\psi_{12}}{}
 =   \dd S_{12}(z)  + \left(\frac{\Theta_{12}'}{\Theta_{12}}-\frac{\Theta'}{\Theta}\right)\dd\mathbf{v}(z) + \left(\frac{\Theta_{12}'}{\Theta_{12}}-\frac{\Theta'}{\Theta}\right)'F'_1 \dd\mathbf{v}(z)    \\
\hphantom{\frac{1}{N} \delta_z \ln\psi_{12}=}{}
  + \frac{1}{N} \sum_{i} \Res_{z' \rightarrow a_i} \frac{1}{\dd X(z') \dd Y(z')} \Bigg\{\frac{1}{2}\big(\dd S_{12}(z')\big)^2B(z',z) \\
\hphantom{\frac{1}{N} \delta_z \ln\psi_{12}=}{}
+ \left( \frac{\Theta_{12}'}{\Theta_{12}}\right)\dd\mathbf{v}(z')\dd S_{12}(z')B(z',z)  + \frac{1}{2} {\left(\frac{\Theta''_{12}}{\Theta_{12}}-\frac{\Theta''}{\Theta}\right)}\big(\dd\mathbf{v}(z')\big)^2B(z',z)   \\
\hphantom{\frac{1}{N} \delta_z \ln\psi_{12}=}{}
 + \frac{1}{2}\left(\frac{\Theta''_{12}}{\Theta_{12}}\right)'\big(\dd\mathbf{v}(z')\big)^2\dd\mathbf{v}(z')\dd S_{12}(z)  + \frac{1}{2}\left(\frac{\Theta'_{12}}{\Theta_{12}}\right)'\dd\mathbf{v}(z')\big(\dd S_{12}(z')\big)^2\dd\mathbf{v}(z)
 \\
\hphantom{\frac{1}{N} \delta_z \ln\psi_{12}=}{}
 + \frac{1}{6}\left( \frac{\Theta_{12}'''}{\Theta_{12}}-\frac{\Theta'''}{\Theta}\right)'\big(\dd\mathbf{v}(z')\big)^3\dd\mathbf{v}(z)\Bigg\}
     + o(1/N).
\end{gather*}
We may transform the f\/irst line using the ref\/ined duality equation established in Proposition~\ref{thYBsp}, which is a consequence of the Fay identity satisf\/ied by the Theta function of the spectral curve. It can be rephrased as
\begin{gather}
\label{grad}
\dd S_{12}(z)  + \left( \frac{\Theta_{12}'}{\Theta_{12}}-\frac{\Theta'}{\Theta}\right)\dd\mathbf{v}(z) = -\frac{E_{12}}{E_{1z}E_{z2}} \frac{\Theta_{1z}\Theta_{z2}}{\Theta \Theta_{12}}.
\end{gather}
On the other hand we have
\begin{gather*}
  \frac{\psi_{1z} \psi_{z2}}{\psi_{12}}
 =  \frac{E_{12}}{E_{1z}E_{z2}} \frac{\Theta_{1z}\Theta_{z2}}{\Theta \Theta_{12}}\left( 1+\frac{1}{N}\big(\hat \psi_{1z}+\hat \psi_{z2}-\hat \psi_{12}\big)+ o(1/N) \right) \nonumber \\
\hphantom{\frac{\psi_{1z} \psi_{z2}}{\psi_{12}}}{}
 =  \frac{E_{12}}{E_{1z}E_{z2}} \frac{\Theta_{1z}\Theta_{z2}}{\Theta \Theta_{12}}\Bigg(1+\frac{1}{N}
 \left[\frac{\Theta_{1z}'}{\Theta_{1z}}+\frac{\Theta_{z2}'}{\Theta_{z2}}-\frac{\Theta_{12}'}{\Theta_{12}}-\frac{\Theta'}{\Theta}\right] F'_1  \\
\hphantom{\frac{\psi_{1z} \psi_{z2}}{\psi_{12}}=}{}
 + \frac{1}{N} \sum_{i} \Res_{z' \rightarrow a_i} \frac{1}{\dd X(z') \dd Y(z')}\Bigg\{\frac{1}{2}\Bigg(\frac{\Theta'_{1z}}{\Theta_{1z}}\big(\dd S_{1z}(z')\big)^2+\frac{\Theta'_{z2}}{\Theta_{z2}}\big(\dd S_{z2}(z')\big)^2
 \\
\hphantom{\frac{\psi_{1z} \psi_{z2}}{\psi_{12}}=}{}
 -\frac{\Theta'_{12}}{\Theta_{12}}\big(\dd S_{12}(z')\big)^2\Bigg)\dd\mathbf{v}(z')  + \frac{1}{2}\Bigg(\frac{\Theta''_{1z}}{\Theta_{1z}}\dd S_{1z}(z')+\frac{\Theta''_{z2}}{\Theta_{z2}}\dd S_{z2}(z')\\
\hphantom{\frac{\psi_{1z} \psi_{z2}}{\psi_{12}}=}{}
 -\frac{\Theta''_{12}}{\Theta_{12}}\dd S_{12}(z')\Bigg)\big(\dd\mathbf{v}(z')\big)^2
 + \frac{1}{6}\left(\frac{\Theta_{1z}'''}{\Theta_{1z}}+\frac{\Theta_{z2}'''}{\Theta_{z2}}
 -\frac{\Theta_{12}'''}{\Theta_{12}}-\frac{\Theta'''}{\Theta}\right)\big(\dd\mathbf{v}(z')\big)^3 \Bigg\}\\
\hphantom{\frac{\psi_{1z} \psi_{z2}}{\psi_{12}}=}{}
  -\frac{1}{2}\dd S_{1z}(z')\dd S_{z2}(z')\dd S_{12}(z') +  o(1/N)\Bigg).
\end{gather*}
Let us now compute
\begin{gather}
  \frac{1}{N} \delta_z \ln\psi_{12}+\frac{\psi_{1z} \psi_{z2}}{\psi_{12}}
 =   \Bigg\{\left(\frac{\Theta_{12}'}{\Theta_{12}}-\frac{\Theta'}{\Theta}\right)'\dd\mathbf{v}(z)\nonumber\\
 \qquad{}
   + \frac{E_{12}}{E_{1z}E_{z2}}\frac{\Theta_{1z}\Theta_{z2}}{\Theta \Theta_{12}}
 \left(\frac{\Theta_{1z}'}{\Theta_{1z}}+\frac{\Theta_{z2}'}{\Theta_{z2}}-\frac{\Theta_{12}'}{\Theta_{12}}-\frac{\Theta'}{\Theta}\right)\Bigg\}F'_1  \nonumber \\
 \qquad{}
  + \sum_{i} \Res_{z' \rightarrow a_i} \frac{1}{\dd X(z') \dd Y(z')}\Bigg\{
\frac{1}{2}\big(\dd S_{12}(z')\big)^2B(z',z)  + \left(\frac{\Theta_{12}'}{\Theta_{12}}\right)\dd\mathbf{v}(z')\dd S_{12}(z')B(z',z)\nonumber \\
\qquad{}
  + \frac{1}{2}{\left( \frac{\Theta''_{12}}{\Theta_{12}}-\frac{\Theta''}{\Theta}\right)}\big(\dd\mathbf{v}(z')\big)^2B(z',z) + \frac{1}{2}\left(\frac{\Theta''_{12}}{\Theta_{12}}\right)'\big(\dd\mathbf{v}(z')\big)^2\dd S_{12}(z')\dd\mathbf{v}(z) \nonumber \\
\qquad{}
  + \frac{1}{2}\left(\frac{\Theta'_{12}}{\Theta_{12}}\right)'\dd\mathbf{v}(z')\big(\dd S_{12}(z')\big)^2\dd\mathbf{v}(z) + \frac{1}{6}\left(\frac{\Theta_{12}'''}{\Theta_{12}}-\frac{\Theta'''}{\Theta}\right)'\big(\dd\mathbf{v}(z')\big)^3\dd\mathbf{v}(z) \nonumber \\
\qquad{}
  + \frac{E_{12}}{E_{1z}E_{z2}}  \frac{\Theta_{1z}\Theta_{z2}}{\Theta \Theta_{12}}
 \Bigg[\frac{1}{2}\left(\frac{\Theta'_{1z}}{\Theta_{1z}}\big(\dd S_{1z}(z')\big)^2+\frac{\Theta'_{z2}}{\Theta_{z2}}\big(\dd S_{z2}(z')\big)^2-\frac{\Theta'_{12}}{\Theta_{12}}\big(\dd S_{12}(z')\big)^2\right)\dd\mathbf{v}(z') \nonumber \\
 \qquad{}
  + \frac{1}{2}\left(\frac{\Theta''_{1z}}{\Theta_{1z}}\dd S_{1z}(z')+\frac{\Theta''_{z2}}{\Theta_{z2}}\dd S_{z2}(z')-\frac{\Theta''_{12}}{\Theta_{12}}\dd S_{12}(z')\right)\big(\dd\mathbf{v}(z')\big)^2 \nonumber \\
  \qquad{}
  + \frac{1}{6}\left( \frac{\Theta_{1z}'''}{\Theta_{1z}}+\frac{\Theta_{z2}'''}{\Theta_{z2}}-\frac{\Theta_{12}'''}{\Theta_{12}}
  -\frac{\Theta'''}{\Theta}\right)\big(\dd\mathbf{v}(z')\big)^3  -\frac{1}{2}\dd S_{1z}(z')\dd S_{z2}(z')\dd S_{12}(z')\Bigg]\Bigg\} \nonumber \\
  \qquad{}
+ o(1/N). \label{A12}
\end{gather}
The coef\/f\/icient of $F'_1$ vanishes, as we can see by computing the gradient of equation~\eqref{grad} to leading order in $N$ (recall that $N$ enters in the def\/inition of our $\Theta$ through the point $\mathbf{w}_0 = NF_0'$)
\begin{gather}
\label{A111}
\left(\frac{\Theta'_{12}}{\Theta_{12}} - \frac{\Theta'}{\Theta}\right)'\dd\mathbf{v}(z)
 -\frac{E_{12}}{E_{1z}E_{z2}}\frac{\Theta_{1z}\Theta_{2z}}{\Theta\Theta_{12}}\left(\frac{\Theta_{1z}'}{\Theta_{1z}} + \frac{\Theta'_{z2}}{\Theta_{z2}} - \frac{\Theta'_{12}}{\Theta_{12}} - \frac{\Theta'}{\Theta}\right).
\end{gather}
This identity can also be seen as a consequence of Fay identity.

Let us now study the residue term of equation~\eqref{A12}, which we write
\begin{gather*}
 \sum_{i} \Res_{z' \rightarrow a_i} \frac{H_{12}(z',z)}{\dd X(z') \dd Y(z')}.
\end{gather*}
First, notice that by construction, $H_{12}(z',z)$ is a meromorphic 1-form in the variable $z$, which means that it has trivial monodromy when~$z$ goes around a non-trivial cycle.
It may have simple poles at $z = z_1$ or $z = z_2$ coming from the ratio of prime forms, but the expression in~$[\cdots]$ vanish when $z = z_1$ or $z_2$, so $H_{12}(z,z')$ is actually regular at $z = z_1$ or $z = z_2$. It may also have a~singularity at $z = z'$ coming from the term $[\cdots]$, which is at~most a double pole. To leading order when $z \rightarrow z'$, we f\/ind
\begin{gather*}
H_{12}(z',z)   =   \frac{1}{2}\dd S_{12}(z')^2  + \left( \frac{\Theta_{12}'}{\Theta_{12}}\right)\dd\mathbf{v}(z')\dd S_{12}(z') + \frac{1}{2}{\left( \frac{\Theta''_{12}}{\Theta_{12}}-\frac{\Theta''}{\Theta}\right)}\big(\dd\mathbf{v}(z')\big)^2  \\
\hphantom{H_{12}(z',z)   =}{}
  \frac{E_{12}}{E_{1z'}E_{z'2}}\frac{\Theta_{1z'}\Theta_{z'2}}{\Theta\Theta_{12}}\left\{\frac{1}{2}\left(\frac{\Theta'_{1z'}}{\Theta_{1z}} + \frac{\Theta'_{z'2}}{\Theta_{z'2}}\right)\dd\mathbf{v}(z') + \frac{1}{2}\dd S_{12}(z')\right\} + O\big(\xi_{z'}^{-1}(z)\big),
\end{gather*}
where $\xi_{z'}$ is our notation for a local coordinate centered at $z'$. We use again equation~\eqref{grad} and f\/ind
\begin{gather*}
H_{12}(z',z)   =   \frac{1}{2}\dd S_{12}(z')^2  + \left( \frac{\Theta_{12}'}{\Theta_{12}}\right)\dd\mathbf{v}(z')\dd S_{12}(z') + \frac{1}{2}{\left( \frac{\Theta''_{12}}{\Theta_{12}}-\frac{\Theta''}{\Theta}\right)}\big(\dd\mathbf{v}(z')\big)^2   \\
\hphantom{H_{12}(z',z)   =}{}
  - \left(\dd S_{12}(z') + \Big(\frac{\Theta_{12}'}{\Theta_{12}} - \frac{\Theta'}{\Theta}\Big)\dd\mathbf{v}(z)\right)\left\{\frac{1}{2}\left(\frac{\Theta'_{1z'}}{\Theta_{1z}} + \frac{\Theta'_{z'2}}{\Theta_{z'2}}\right)\dd\mathbf{v}(z') + \frac{1}{2}\dd S_{12}(z')\right\}  \\
\hphantom{H_{12}(z',z)   =}{}
  + O\big(\xi_{z'}^{-1}(z)\big)   \\
\hphantom{H_{12}(z',z) }{}
  =   \Bigg\{-\dd S_{12}(z')\left(-\frac{\Theta'_{12}}{\Theta_{12}} - \frac{\Theta'}{\Theta} + \frac{\Theta'_{1z'}}{\Theta_{1z'}} + \frac{\Theta'_{z'2}}{\Theta_{z'2}}\right)   \\
\hphantom{H_{12}(z',z)   =}{}
  - \left(\frac{\Theta'_{12}}{\Theta_{12}} - \frac{\Theta'}{\Theta}\right)\left(\frac{\Theta'_{1z'}}{\Theta_{1z'}} + \frac{\Theta'_{z'2}}{\Theta_{z'2}}\right)\dd\mathbf{v}(z') + \left(\frac{\Theta_{12}''}{\Theta_{12}} - \frac{\Theta''}{\Theta}\right)\dd\mathbf{v}(z')\Bigg\}\frac{\dd\mathbf{v}(z')}{2}     \\
\hphantom{H_{12}(z',z)   =}{}
  + O\big(\xi_{z'}^{-1}(z)\big).
\end{gather*}
We can rearrange the terms
\begin{gather*}
H_{12}(z',z)   =   \left\{-\left[\dd S_{12}(z') + \Big(\frac{\Theta'_{12}}{\Theta_{12}} - \frac{\Theta'}{\Theta}\Big)\dd\mathbf{v}(z')\right]\left(\frac{\Theta'_{1z'}}{\Theta_{1z'}} + \frac{\Theta'_{z'2}}{\Theta_{z'2}}-\frac{\Theta'_{12}}{\Theta_{12}} - \frac{\Theta'}{\Theta} \right) \right.   \\
  \left.
\hphantom{H_{12}(z',z)   =}{}
   + \left(\frac{\Theta'_{12}}{\Theta_{12}} - \frac{\Theta'}{\Theta}\right)'\dd\mathbf{v}(z')\right\}\frac{\dd \mathbf{v}(z')}{2} + O\big(\xi_{z'}^{-1}(z)\big),
\end{gather*}
and according to equation~\eqref{A111}, we see that $H_{12}(z',z) \in O\big(\xi^{-1}_{z'}(z)\big)$. Therefore, $H_{12}(z',z)$ is a meromorphic function whose only singularity is a pole at~most simple at $z = z'$. But a~meromorphic function cannot have a single simple pole, so $H_{12}(z,z')$ must be holomorphic, and we can write it
\begin{gather*}
H_{12}(z,z') = h_{12}(z') \dd\mathbf{v}(z).
\end{gather*}
Since the prefactor $h_{12}(z')$ is independent of $z$, we may compute it by specializing to~$z = z_1$ in~$H_{12}(z,z_1)$ def\/ined from equation~\eqref{A12}. Doing so, we obtain
\begin{gather*}
h_{12}(z') \dd\mathbf{v}(z_1)   =   \frac{1}{2}\big(\dd S_{12}(z')\big)^2B(z',z_1) + \left(\frac{\Theta'_{12}}{\Theta_{12}}\right)\dd\mathbf{v}(z')\dd S_{12}(z')B(z',z_1)     \\
\hphantom{h_{12}(z') \dd\mathbf{v}(z_1)   =}{}
    + \frac{1}{2}\left(\frac{\Theta''_{12}}{\Theta_{12}} - \frac{\Theta''}{\Theta}\right)\big(\dd\mathbf{v}(z')\big)^2B(z',z_1) + \frac{1}{2}\left(\frac{\Theta_{12}''}{\Theta_{12}}\right)'\big(\dd\mathbf{v}(z')\big)^3\dd\mathbf{v}(z_1)   \\
\hphantom{h_{12}(z') \dd\mathbf{v}(z_1)   =}{}
+
    \frac{1}{2}\left(\frac{\Theta'_{12}}{\Theta_{12}}\right)'\dd\mathbf{v}(z')\big(\dd S_{12}(z')\big)^2\dd\mathbf{v}(z_1) + \frac{1}{6}\left(\frac{\Theta'''_{12}}{\Theta_{12}} - \frac{\Theta'''}{\Theta}\right)'\big(\dd\mathbf{v}(z')\big)^3\dd\mathbf{v}(z_1)   \\
\hphantom{h_{12}(z') \dd\mathbf{v}(z_1)   =}{}
    - \dd_{z = z_1}\Bigg[\frac{1}{2}\left(\frac{\Theta'_{1z}}{\Theta_{1z}}\big(\dd S_{1z}(z')\big)^2 + \frac{\Theta'_{z2}}{\Theta_{z2}}\big(\dd S_{z2}(z')\big)^2 - \frac{\Theta'_{12}}{\Theta_{12}}\big(\dd S_{12}(z')\big)^2\right)\dd\mathbf{v}(z')  \\
\hphantom{h_{12}(z') \dd\mathbf{v}(z_1)   =}{}
     + \frac{1}{2}\left(\frac{\Theta''_{1z}}{\Theta_{1z}}\dd S_{1z}(z') + \frac{\Theta''_{z2}}{\Theta_{z2}}\dd S_{z2}(z')-\frac{\Theta''_{12}}{\Theta_{12}}\dd S_{12}(z')\right)\big(\dd\mathbf{v}(z')\big)^2  \\
\hphantom{h_{12}(z') \dd\mathbf{v}(z_1)   =}{}
      + \frac{1}{6}\left( \frac{\Theta_{1z}'''}{\Theta_{1z}}+\frac{\Theta_{z2}'''}{\Theta_{z2}}-\frac{\Theta_{12}'''}{\Theta_{12}}-\frac{\Theta'''}{\Theta}\right)
      \big(\dd\mathbf{v}(z')\big)^3  -\frac{1}{2}\dd S_{1z}(z')\dd S_{z2}(z')\dd S_{12}(z')\Bigg]
\\
\hphantom{h_{12}(z') \dd\mathbf{v}(z_1) }{}
  =   \frac{1}{2}\big(\dd S_{12}(z')\big)^2B(z',z_1) + \left(\frac{\Theta'_{12}}{\Theta_{12}}\right)\dd\mathbf{v}(z')\dd S_{12}(z')B(z',z_1)
   \\
\hphantom{h_{12}(z') \dd\mathbf{v}(z_1)   =}{}
    + \frac{1}{2}\left(\frac{\Theta''_{12}}{\Theta_{12}} - \frac{\Theta''}{\Theta}\right)\big(\dd\mathbf{v}(z')\big)^2B(z',z_1) + \frac{1}{2}\left(\frac{\Theta_{12}''}{\Theta_{12}}\right)'\big(\dd\mathbf{v}(z')\big)^3\dd\mathbf{v}(z_1)   \\
\hphantom{h_{12}(z') \dd\mathbf{v}(z_1)   =}{}
+
    \frac{1}{2}\left(\frac{\Theta'_{12}}{\Theta_{12}}\right)'\dd\mathbf{v}(z')\big(\dd S_{12}(z')\big)^2\dd\mathbf{v}(z_1) + \frac{1}{6}\left(\frac{\Theta'''_{12}}{\Theta_{12}} - \frac{\Theta'''}{\Theta}\right)'\big(\dd\mathbf{v}(z')\big)^3\dd\mathbf{v}(z_1)
     \\
\hphantom{h_{12}(z') \dd\mathbf{v}(z_1)   =}{}
    -\Bigg[\left(\frac{\Theta'_{12}}{\Theta_{12}}\right)B(z_1,z')\dd\mathbf{v}(z')\dd S_{12}(z') + \frac{1}{2}\left(\frac{\Theta'_{12}}{\Theta_{12}}\right)'\dd\mathbf{v}(z')\dd\mathbf{v}(z_1)\big(\dd S_{12}(z')\big)^2
     \\
\hphantom{h_{12}(z') \dd\mathbf{v}(z_1)   =}{}
+
    \frac{1}{2}\left(\frac{\Theta_{12}''}{\Theta_{12}} - \frac{\Theta''}{\Theta}\right)\big(\dd\mathbf{v}(z')\big)^2B(z_1,z') + \frac{1}{2}\left(\frac{\Theta''_{12}}{\Theta_{12}}\right)'\!\big(\dd\mathbf{v}(z')\big)^2\dd\mathbf{v}(z_1)\dd S_{12}(z')   \\
\hphantom{h_{12}(z') \dd\mathbf{v}(z_1)   =}{}
    + \frac{1}{6}\left(\frac{\Theta'''_{12}}{\Theta_{12}} - \frac{\Theta'''}{\Theta}\right)'\big(\mathbf{d}v(z')\big)^3\dd\mathbf{v}(z_1) + \frac{1}{2}B(z_1,z')\big(\dd S_{12}(z')\big)^2\Bigg]   =   0.
\end{gather*}
All the terms eventually cancel each other. Thus $H_{12}(z,z') = 0$, and coming back to equation~\eqref{A12}, this proves
\begin{gather*}
\frac{1}{N} \delta_{z} \ln \psi(z_1,z_2) + \frac{\psi(z_1,z)\psi(z,z_2)}{\psi(z_1,z_2)} = o(1/N).  \tag*{\qed}
\end{gather*}
\renewcommand{\qed}{}
\end{proof}

\subsection*{Acknowledgments}
We thank O.~Babelon, M.~Berg\`ere, M.~Bertola, B.~Dubrovin, D.~Korotkin, M.~Mulase,  J.M.~Mu\-{\~n}oz Porras, N.~Orantin, F.~Plaza Martin, E.~Previato, A.~Raimondo, B.~Safnuk for fruitful discussions, T.~Grava and S.~Romano for enlightening discussions concerning dispersionless hie\-rar\-chies, their dispersive deformations and the role of Whitham equations, and I.~Krichever for careful reading, valuable discussions and for pointing out references. This work is partly supported by the ANR project Grandes Matrices Al\'eatoires ANR-08-BLAN-0311-01, by the European Science Foundation through the Misgam program, by the Qu\'ebec government with the FQRNT, by the Fonds Europ\'een S16905 (UE7~-~CONFRA), by the Swiss NSF (no 200021-43434) and the ERC AG CONFRA. B.E.\ thanks the CERN, and G.B.\ thanks the SISSA for their hospitality while this work was pursued.

\pdfbookmark[1]{References}{ref}

\LastPageEnding

\end{document}